\documentclass[reprint, groupedaddress, amsmath, amssymb, aps, prl, floatfix]{revtex4-2}

\usepackage{braket}
\usepackage{printlen} 
\usepackage{amsthm}
\usepackage{physics}

\usepackage{graphicx}
\graphicspath{{supp_mat_figures/}}

\usepackage{dcolumn}
\usepackage{bm}
\usepackage[hidelinks]{hyperref}
\usepackage{tikz}
\usepackage{dsfont}
\usetikzlibrary{quantikz}
\usepackage{siunitx}

\usepackage{lipsum} 

\usepackage[labelformat=simple]{subcaption}

\usepackage{ragged2e}
\DeclareCaptionJustification{justified}{\justifying}
\usepackage{printlen}

\usepackage{etoolbox}

\makeatletter
\def\maketitle{
\@author@finish
\title@column\titleblock@produce
\suppressfloats[t]}
\makeatother

\newcounter{PRLsections}
\renewcommand{\thePRLsections}{\Roman{PRLsections}}
\DeclareRobustCommand{\PRLsec}[2]{%
    \begin{center}
        \medskip
        \refstepcounter{PRLsections}%
        \addcontentsline{toc}{section}{\thePRLsections.\space#1}
        \textbf{\thePRLsections.\quad\label{#2} \uppercase{#1}}
    \end{center}
}

\newcounter{PRLsubsections}[PRLsections]
\renewcommand{\thePRLsubsections}{\Alph{PRLsubsections}}
\DeclareRobustCommand{\subsec}[2]{%
    \iftoggle{arXiv}{ 
        \begin{center}
            \medskip
            \refstepcounter{PRLsubsections}%
            \textbf{\thePRLsubsections.\quad\label{#2} #1} 
        \end{center}
    }{ \subsection{#1} \label{#2} }
}
\makeatletter
\renewcommand{\p@PRLsubsections}{\thePRLsections.\,}
\makeatother

\usepackage{wasysym}

\usepackage[OT2,T1]{fontenc}
\DeclareSymbolFont{cyrletters}{OT2}{wncyr}{m}{n}
\DeclareMathSymbol{\Sha}{\mathalpha}{cyrletters}{"58}

\usepackage[innercaption]{sidecap}
\sidecaptionvpos{figure}{c}

\DeclareSymbolFont{yhlargesymbols}{OMX}{yhex}{m}{n} 
\DeclareMathAccent{\widehat}{\mathord}{yhlargesymbols}{"62}

\usepackage{multirow}
\usepackage{array}
\newcolumntype{L}[1]{>{\raggedright\let\newline\\\arraybackslash\hspace{0pt}}m{#1}}
\newcolumntype{C}[1]{>{\centering\let\newline\\\arraybackslash\hspace{0pt}}m{#1}}
\newcolumntype{R}[1]{>{\raggedleft\let\newline\\\arraybackslash\hspace{0pt}}m{#1}}

\makeatletter
\newcommand*{\balancecolsandclearpage}{%
  \close@column@grid
  \cleardoublepage
  \twocolumngrid
}
\makeatother

\makeatletter
\newenvironment{SuppMat}
 {
 }
\makeatother

\newcommand{\n}[1]{\mathrm{#1}}
\newcommand{\cl}[1]{\mathcal{#1}}
\newcommand{\bb}[1]{\mathbb{#1}}
\newcommand{\h}[1]{\hat{#1}}

\newcommand{\q}{\hat{q}}
\newcommand{\p}{\hat{p}}
\newcommand{\s}{\hat{\sigma}}

\newcommand{\be}{\begin{equation}} 
\newcommand{\ee}{\end{equation}}

\newcommand{\CNOT}{\mathtt{CNOT}}
\newcommand{\CZ}{\mathtt{CZ}}
\newcommand{\ve}{\textbf}

\newcommand{\overbar}[1]{\mkern 1.5mu\overline{\mkern-1.5mu#1\mkern-1.5mu}\mkern 1.5mu}

\newtoggle{arXiv} 
\toggletrue{arXiv}      

\iftoggle{arXiv}{\usepackage{eczoo}}{}

\newcommand{\mycolor}{black}
\newcommand{\mycolorbis}{black}

\begin{document}
\preprint{APS/123-QED}


\title{Two-qubit operations for finite-energy Gottesman-Kitaev-Preskill encodings}

\author{Ivan Rojkov}
  \email{irojkov@phys.ethz.ch}
\author{Paul Moser R\"oggla}
\author{Martin Wagener}
\author{Moritz~Fontbot\'{e}-Schmidt}
\author{Stephan Welte}
\author{Jonathan Home}
\author{Florentin Reiter}
  \email{freiter@phys.ethz.ch}
\affiliation{
    Institute for Quantum Electronics,
    ETH Z\"{u}rich, Otto-Stern-Weg 1, 8093 Z\"{u}rich, Switzerland
    }
\affiliation{Quantum Center, ETH Zürich, 8093 Zürich, Switzerland}

\date{\today}


\begin{abstract}
\noindent
We present techniques for performing two-qubit gates on Gottesman-Kitaev-Preskill (GKP) codes with finite energy, and find that operations designed for ideal infinite-energy codes create undesired entanglement when applied to physically realistic states. We demonstrate that this can be mitigated  using recently developed local error-correction protocols, and evaluate the resulting performance. We also propose energy-conserving finite-energy gate implementations which largely avoid the need for further correction.
\end{abstract}

\maketitle

The realization of a fault-tolerant quantum computer requires the implementation of a universal set of gates performed on error-corrected encoded logical qubits~\cite{gottesman_introduction_2009}. Encoding involves the redundant use of a larger Hilbert space, which is often obtained by mapping information across multiple physical systems.~An alternative is to examine systems with an extended Hilbert space such as harmonic oscillators, of which bosonic codes~\iftoggle{arXiv}{\eczoo{oscillators}\nocite{eczoo_oscillators}}{\cite{eczoo_oscillators}} are a prominent example. One candidate set of bosonic codes are the  Gottesman-Kitaev-Preskill (GKP) codes~\cite{GKP_original_paper}, in which quantum error-correction has recently been demonstrated in both superconducting circuits~\cite{campagne-ibarcq_quantum_2020,sivak_real_time_2023} and trapped ions~\cite{fluhmann_encoding_2019,fluhmann_direct_2020,de_neeve_error_2022} using a single oscillator. In order to embed this encoding into a larger system~\cite{menicucci_universal_2006,fukui_high-threshold_2018,vuillot_quantum_2019,noh_fault-tolerant_2020,noh_low-overhead_2022}, gates between multiple encoded qubits will be required. While multi-qubit gate schemes have been proposed~\cite{GKP_original_paper,menicucci_fault-tolerant_2014,terhal_towards_2020,tzitrin_progress_2020,walshe_continuous-variable_2020}, these consider the action on ``ideal'' infinite-energy GKP states, with the effect on experimentally realizable finite-energy states treated as a tolerable source of error. However, recent theoretical~\cite{royer_stabilization_2020,hastrup_improved_2021} and experimental~\cite{de_neeve_error_2022} works have shown that single-qubit operations can be designed for finite-energy states, which asks the question whether similar strategies can be taken for multi-qubit gates.

In this Letter, we present two approaches to tackle this problem. First, we examine the effect of ideal, i.e., infinite-energy, two-qubit operations on finite-energy GKP states, and show that although the gate operation leads to significant distortion of the states, this is of a form which is correctable by finite-energy error-correction protocols. Second, we introduce direct finite-energy gates which preserve the energy of the states, and thus avoid the need for correction steps. These components serve as a foundation for integrating finite-energy GKP states into larger-scale quantum computing systems, providing a path towards fault-tolerant processing of quantum information.

GKP encodings have a characteristic grid-like structure in phase space and are defined through displacement operators. The logical Pauli operators and the stabilizers for a square code are defined by ${\h{X} = e^{i\p\sqrt{\pi}}}$, ${\h{Z} = e^{-i\q\sqrt{\pi}}}$ and ${\h{S}_x = e^{i\p2\sqrt{\pi}}}$, ${\h{S}_z = e^{-i\q2\sqrt{\pi}}}$, respectively~\footnote{For other code geometries, the analysis and results will be similar after linearly transforming the quadratures.}. The simultaneous eigenstates of these operators are the ideal GKP codewords ${\ket{\mu}_\n{I} = \sum_{s} \ket{q = (2s + \mu) \sqrt{\pi}}}$ where ${\mu\in\{0,1\}}$, ${s\in\mathbb{Z}}$ and the subscript ``I'' stands for \textit{ideal}. Since each component of the superposition is an infinitely squeezed state, ideal codewords have an infinite norm and are thus not physical. A finite-energy version of these states can be constructed using a Gaussian phase-space envelope centered at the origin~\cite{GKP_original_paper,menicucci_fault-tolerant_2014,motes_encoding_2017,noh_quantum_2019,royer_stabilization_2020}. Mathematically, this can be realized by introducing an envelope operator ${\h{E}_\Delta = e^{-\Delta^2\h{n}}}$, where ${\h{n}=\frac{1}{2}(\q^2+\p^2)}$ is the number operator and $\Delta$ parameterizes the size of the code states in phase space. A finite-energy GKP state is then expressed by ${\ket{\psi}_{\!\Delta}\propto\h{E}_\Delta\ket{\psi}_\n{I}}$ and can be thought of as a superposition of periodically-spaced, finitely-squeezed states weighted according to an overall Gaussian envelope. The states and their marginal distributions, ${P(x)=|\langle x\!\ket{0}_\Delta\!|^2}$ with $x\in\{q,p\}$, are thus characterized by two parameters; the peak's standard deviation $\Delta_\n{peak}$ and the inverse of the Gaussian envelope's standard deviation $\Delta_\n{envl}$. For a pure state $\ket{\psi}_{\!\Delta}$, these are both equal to $\Delta$. 

\begin{figure*}[ht!]
    \centering
    \includegraphics{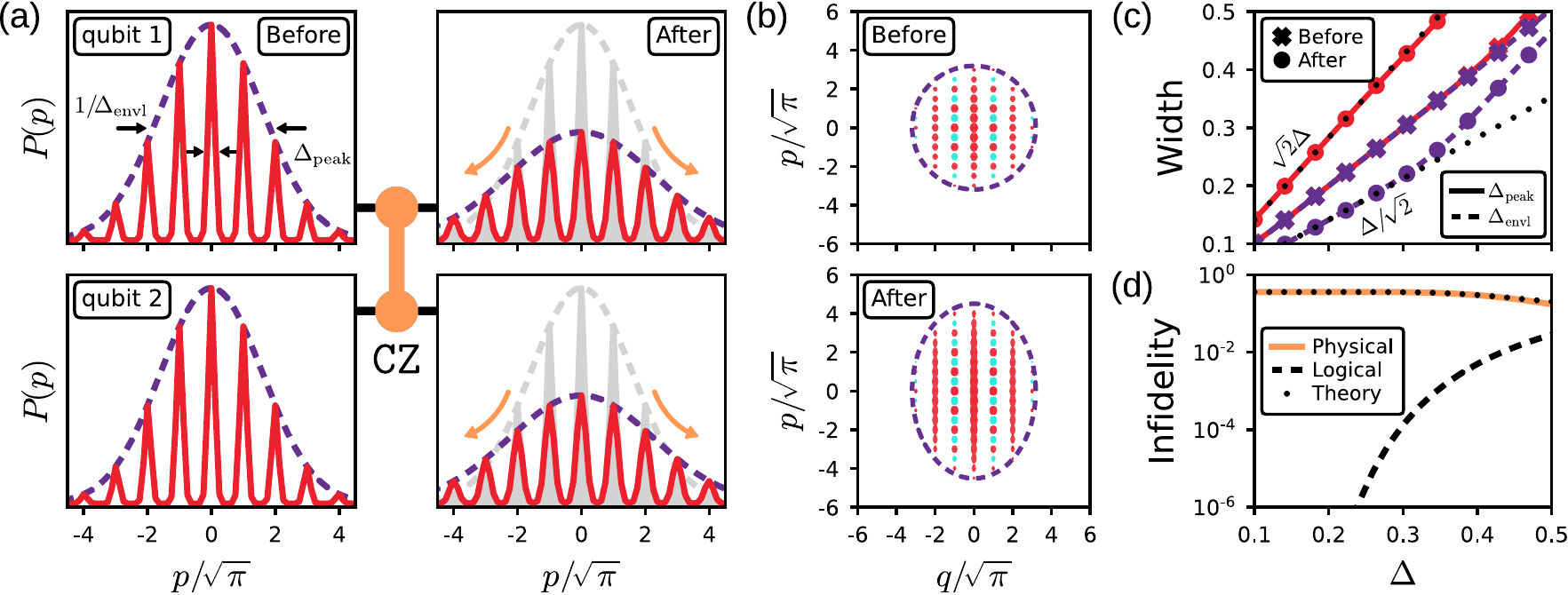}    
    {\phantomsubcaption\label{fig:effect_explanation}}
    {\phantomsubcaption\label{fig:wigner_before_after}}
    {\phantomsubcaption\label{fig:std_dev_delta}}
    {\phantomsubcaption\label{fig:fidelity_delta}}
    \caption{Finite-energy effects in two GKP qubit operations. (a) Momentum marginal distribution $P(p)$ of both oscillators starting in the state $\ket{0}_{\!\Delta}$ before (left) and after (right) the $\CZ$ gate. The output distributions get broadened as the operation corresponds to a continuous set of displacements that spreads each oscillator's wave function conditioned on the position of the other one. (b) Wigner quasiprobability distribution, showing broadening only in the p quadrature. (c) Peak widths as a function of input width, showing clearly the linear relation. (d) The physical and logical infidelity between the input and output states as a function of the energy parameter $\Delta$.}
    \label{fig:fig1}
\end{figure*}


Two-qubit entangling gates are essential operations for universal quantum computation~\cite{bremner_practical_2002}. For GKP codes, such gates can be realized using quadrature--quadrature coupling Hamiltonians that are equivalent to each other up to local phase-space transformations. Here we focus on the controlled $\h{Z}$ gate, a two-qubit operation expressed as ${\CZ = e^{i\q_1\q_2}}$~\cite{GKP_original_paper,menicucci_fault-tolerant_2014}, where the indices $1$ and $2$ denote the two oscillators. The action of $\CZ$ on position eigenstates, $\ket{q_1=m_1\sqrt{\pi}}\ket{q_2=m_2\sqrt{\pi}}$, is to add in a prefactor $e^{-i \pi m_1 m_2}$ with $m_1$, $m_2$ being either odd or even integers. If the input state is $\ket{1}_\n{I}\ket{1}_\n{I}$ the system acquires an overall phase of $\pi$, whereas all other input states acquire a multiple of $2\pi$. 

Due to their limited extent in phase space, finite-energy GKP states are not translationally invariant and thus the action of the CZ gate described above produces distortion of the underlying states. Consider the situation depicted in~Fig.~\ref{fig:effect_explanation}; from the perspective of the second subsystem the gate corresponds to a series of displacements ${\sum_{q_1}\!\bra{q_1}\CZ\ket{q_1} = \sum_{q_1}\!e^{i q_1\q_2} \equiv \sum_{q_1}\!\h{D}_2(i\,q_1/\sqrt{2})}$. Each of these displacements shifts both the individual Gaussian peaks and the envelope of the state. The final subsystem state can thus be regarded as a Gaussian mixture model, i.e., a weighted sum of Gaussian functions with a broader envelope and peak width. This distortion occurs only in the $p$ quadrature (cf. Fig.~\ref{fig:wigner_before_after}).

In order to analytically quantify the broadening of the envelope we evaluate the marginal distribution of the first oscillator after tracing out the second one from the state following the gate: ${P(p_1)=\bra{p_1}\n{Tr}_2\big[\CZ\,\rho_\Delta\,\CZ^\dagger\big]\ket{p_1}}$ where ${\rho_\Delta = \ket{00}_{\!\Delta}\!\prescript{}{\Delta\!}{\bra{00}}}$. We perform this calculation using the shifted grid state representation~\cite{GKP_original_paper,glancy_error_2006,ketterer_quantum_2016,terhal_encoding_2016,weigand_generating_2018,matsuura_equivalence_2020} and recognizing that the main additional contributions come from the first nearest neighbor peaks (cf. Supplemental Material~\footnote{See Supplemental Material, which includes Refs.~\cite{mardia_directional_2000,albert_performance_2018,baragiola_all-gaussian_2019,hastrup_measurement-free_2021,weedbrook_gaussian_2012,braunstein_squeezing_2005,zak_finite_1967,mensen_phase-space_2021,lloyd_quantum_1999,bloch_canonical_1962,cariolaro_bloch-messiah_2016,sivak_kerr-free_2019,grimsmo_quantum_2021,haljan_spin-dependent_2005,bennett_mixed-state_1996,wootters_entanglement_1998,gardiner_quantum_2004,hatano_finding_2005,gan_hybrid_2020,shaw_stabilizer_2024,layden_first-order_2022,cvsim_rojkov_2024,bezanson2017julia,kramer2018quantumoptics}, for more detailed derivations of the results presented in the main text and their extension to states with a non-symmetric energy envelope in $q$ and $p$.}). 
This assumption is valid for GKP states with $\Delta\lesssim0.4$ which is consistent with recent experimental realizations~\cite{campagne-ibarcq_quantum_2020,de_neeve_error_2022,sivak_real_time_2023}. The marginal distribution of the target state after the gate retains a finite-energy GKP form but with characteristic parameters being updated to
\begin{equation} \label{eq:variance_peak_and_envelope}
    \Delta_\n{peak}^2 = 2\Delta^2 
    \quad \textnormal{and} \quad
    \Delta_\n{envl}^2 = \frac{\Delta^2}{2}\,.
\end{equation}
Thus both the peak and envelope widths of the marginal distribution after the $\CZ$ gate increase by a factor of $\sqrt{2}$ compared to their initial values. Fig.~\ref{fig:std_dev_delta} shows the comparison between these input and output state parameters. Eq.~\eqref{eq:variance_peak_and_envelope} is accurate to $\cl{O}(\Delta^6)$, but further improvements can be made by including contributions from subsequent neighboring peaks using the same method. Analytical expressions for the position marginal distributions and the purity of each subsystem are discussed further in the Supplemental Material~\cite{Note2}.

An alternative measure of the quality of a given GKP code are the effective squeezing parameters~\cite{terhal_encoding_2016,duivenvoorden_single-mode_2017,weigand_generating_2018}, which are defined as ${\sigma_{x/z}^2 = \frac{1}{\pi} \log{\big(|\text{Tr}[\,\hat{S}_{x/z}\rho\,]\,|^{-2}\big)}}$ and quantify the closeness of a system's state $\rho$ to the unit eigenstates of the code stabilizers. In ideal GKP states, both effective squeezing parameters are $0$, while for pure finite-energy states such as $\ket{0}_{\!\Delta}$, $\sigma_{x/z} = \Delta$. After the $\CZ$ gate is applied, we find that the effective squeezing parameters of each subsystem read $\sigma_{x} = \sqrt{2}\Delta$ and $\sigma_{z} = \Delta$. The former expression is consistent with the peak and envelope broadening in $P(p_1)$ of Eq.~\eqref{eq:variance_peak_and_envelope}. The expression for $\sigma_{z}$ indicates that the marginal distribution in the position space will be unaffected by the gate.

The main consequence of these finite-energy modifications is the lowering of the physical overlap fidelity between the input and desired output states, ${F=|\!\bra{\psi_\n{out}}\CZ\ket{\psi_\n{in}}\!|^2}$. As an example, we derive this quantity for the input state $\ket{00}_{\!\Delta}$ using as above the shifted grid state method and the first nearest neighbor assumption, obtaining

\begin{equation} \label{eq:fidelity_formula}
    F \approx \frac{16}{25} \frac{\left( 1 + 4\,e^{-\frac{\pi}{5\Delta^2}}\right)^2}
    {\left(1+2\,e^{-\frac{\pi }{\Delta^2}}\right)^4 \left(1+2\,e^{-\frac{\pi}{4\Delta^2}}\right)^4}\,.
\end{equation}
Fig.~\ref{fig:fidelity_delta} shows that this expression agrees with the numerically evaluated overlap using state vector simulations. Eq.~\eqref{eq:fidelity_formula} provides an accurate approximation of the true fidelity up to $\cl{O}(\Delta^3)$ (a higher-order formula and the fidelity for other states can be found in SM~\cite{Note2}). In the limit ${\Delta\rightarrow0}$ the fidelity approaches a finite value ${F_\n{max}=16/25=0.64}$ which is independent of the input state. Despite converging to ideal codewords, GKP states with ${\Delta^2\ll1}$ still possess finite-width peaks and envelope which remain susceptible to broadening due to the two-qubit interaction. The logical fidelity which for the state $\ket{00}_{\!\Delta}$ is accessible by integrating its position marginal distribution over $[-\sqrt{\pi}/2;\sqrt{\pi}/2)+2\sqrt{\pi}\bb{Z}$ remains close to unity~\cite{glancy_error_2006,tzitrin_progress_2020,pantaleoni_modular_2020}.

The $\CZ$ gate operation considered above can be realized exactly using two beamsplitters and a single-layer of single-mode squeezers~\cite{terhal_towards_2020,tzitrin_progress_2020}. We find that it is also possible to achieve the same interaction using an alternative decomposition consisting of two squeezing operations and only one application of the beamsplitter. This is given by ${\CZ(\theta,r) = \h{S}^{\otimes2}(r)\h{B}_A(\theta)\h{S}^{\otimes2}(-r)}$ with ${\h{S}^{\otimes2}(r)=\h{S}(r) \otimes \h{S}(r)}$ and ${\h{S}(r) = e^{i\frac{1}{2}r(\q_j\p_j + \p_j\q_j)}}$ representing the squeezing operation on mode $j$, while ${\h{B}_A(\theta) = e^{i\theta (\q_1\q_2 + \p_1\p_2)}}$ is the anti-symmetric beamsplitter transformation. This chain of operations can then be written as
\begin{equation}\label{eq:approximateCZ_exp}
    \CZ(\theta,r) = e^{i\theta(e^{2r}\q_1\q_2 + e^{-2r} \p_1\p_2)}.
\end{equation}
The ideal desired gate is obtained when ${\theta = e^{-2r}}$ and ${r\rightarrow\infty}$, which makes this decomposition a convergent but approximate realization of $\CZ$. The Hilbert-Schmidt distance between the symplectic representations of $\CZ(\theta,r)$ and $\CZ$ scales as $\sqrt{3}\,e^{-4r}$ which constitutes a deviation with respect to the ideal operator norm that is below $1\%$ at ${r>1.06}$. In practice, we find that the overlap fidelity $F$ is above $0.6$ for ${r\geq0.75}$ and at ${r=1.0}$ has less than $0.7\%$ error relative to ${F_\n{max}}$. The approximate decomposition in Eq.~\eqref{eq:approximateCZ_exp} has the advantage of requiring a weaker bilinear interaction than previously proposed schemes~\cite{terhal_towards_2020,tzitrin_progress_2020}, allowing for a flexible selection of the beamsplitter coupling strength complying with a fault-tolerant concatenation of GKP and surface codes, aiding in reducing the accumulation of errors during the gate time~\cite{noh_fault-tolerant_2020,Note2}. The limitation of this scheme is that it maintains squeezed quadratures for an extended duration, thereby enhancing the susceptibility of the GKP code to minor deviations. This decomposition as well as the previously proposed ones induce the same finite-energy effects as the ideal operation.


\begin{figure}[t]
    \centering
    {\phantomsubcaption\label{fig:qec_circuit}}
    {\phantomsubcaption\label{fig:qec_fidelity}}
    {\phantomsubcaption\label{fig:qec_infidelity_diff_states}}
    \includegraphics{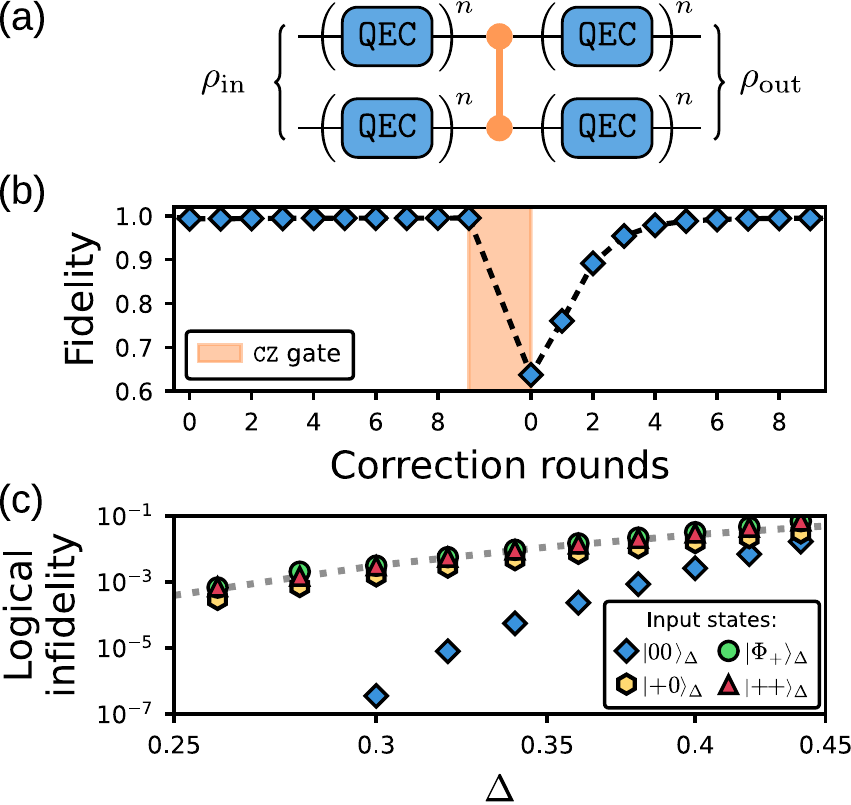}
    \caption{Correction of finite-energy effects. (a) The circuit representation of the error-corrected $\CZ$ gate. The system is initialized in a state ${\rho_\n{in}=\ket{\psi_\n{in}}_{\!\Delta}\!\prescript{}{\Delta\!}{\bra{\psi_\n{in}}}}$ and undergoes $n$ stabilization cycles. After ${n=9}$ rounds, the ideal $\CZ$ gate is applied, followed by ${n=9}$ additional rounds of correction. (b) Fidelity $F$ evaluated after each QEC round with $\ket{\psi_\n{in}}_{\!\Delta}=\ket{00}_{\!\Delta}$ and $\Delta=0.3$.\, (c) The logical infidelity as a function of $\Delta$ evaluated as the difference of overlap fidelities at the 9$^\n{th}$ round of QEC before and after~$\CZ$. Before the gate, the overlap fidelity is obtained with $F(\rho,\ket{\psi_\n{in}}_{\!\Delta})$, whereas after $\CZ$ using $F(\rho,\ket{\psi_\n{out}}_{\!\Delta})$ where $\ket{\psi_\n{out}}_{\!\Delta}$ is the desired output state which for each input state corresponds to $\ket{00}_{\!\Delta}$, $\ket{+0}_{\!\Delta}$, $\ket{\Phi_{-}}_{\!\Delta}$ or ${\left(\ket{0+}_{\!\Delta}\!+\ket{1-}_{\!\Delta}\right)\!/\!\sqrt{2}}$, respectively. The dotted curve represents the analytical result from~Eq.~\eqref{eq:diff_logical_fidelity}.}
    \label{fig:fig2}
\end{figure}

A first solution against finite-energy effects is to correct them locally, given that despite the impact of these effects on the state of the oscillators, the $\CZ$ gate effectively executes the intended logical operation. Using recently demonstrated quantum error correction (QEC) protocols~\cite{royer_stabilization_2020,de_neeve_error_2022,sivak_real_time_2023}, we show that this works. Fig.~\ref{fig:qec_circuit} shows the protocol for the error-corrected $\CZ$ gate. We initialize the system in a pure state ${\rho_\n{in}=\ket{\psi_\n{in}}_{\!\Delta}\!\prescript{}{\Delta\!}{\bra{\psi_\n{in}}}}$ and perform a series of finite-energy stabilization cycles that correct for small displacements in position and momentum and for deformations in the state's energy envelope. Halfway through the series of QEC rounds we perform the $\CZ$ gate and resume the stabilization. 
{\color{\mycolorbis}After each correction round before the gate, we evaluate the overlap fidelity as ${F(\rho,\ket{\psi_\n{in}}_{\!\Delta})}$, whereas for those after the $\CZ$ we use ${F(\rho,\ket{\psi_\n{out}}_{\!\Delta})}$ with ${F(\rho,\ket{\psi})=\expval{\rho}{\psi}}$ and ${\ket{\psi_\n{out}}_{\!\Delta}\propto\h{E}_\Delta\CZ\ket{\psi_\n{in}}_\n{I}}$ being the desired output state.}
As an example, Fig.~\ref{fig:qec_fidelity} illustrates this quantity for $\ket{\psi_\n{in}}_{\!\Delta}=\ket{00}_{\!\Delta}=\ket{\psi_\n{out}}_{\!\Delta}$ as a function of the correction round. As anticipated, the gate lowers the fidelity, but after a few rounds of error correction the fidelity recovers. 

{\color{\mycolorbis}To quantify the logical infidelity we evaluate ${F(\rho_b,\ket{\psi_\n{in}}_{\!\Delta})-F(\rho_\n{out},\ket{\psi_\n{out}}_{\!\Delta})}$ with $\rho_b$ and $\rho_\n{out}$ being the states before the gate and after the entire protocol.}
We observe that this difference is finite (cf.~Fig.~\ref{fig:qec_infidelity_diff_states}). This is explained by the distortion in phase space of both oscillators' state that then increases the probability that the finite-energy stabilization procedure misinterprets $\ket{0}_{\!\Delta}$ for $\ket{1}_{\!\Delta}$ (and vice-versa). The states with logical coherences, e.g. $\ket{+}_{\!\Delta}$, are the most affected by this distortion because their information is primarily stored in the momentum quadrature. Therefore, we observe that the logical infidelity for states such as $\ket{+0}_{\!\Delta}$, $\ket{++}_{\!\Delta}$, or the Bell state $\ket{\Phi_{+}}_{\!\Delta}\propto\ket{00}_{\!\Delta}+\ket{11}_{\!\Delta}$ is several orders of magnitude higher than for the computational states. We can approach the infidelity for those states using some ideal decoders~\cite{Note2} or by integrating appropriate marginal distributions. Taking $\ket{+0}_{\!\Delta}$ as the analytically simplest example, we construct $P(p_1)$ using peaks and envelope widths derived in Eq.~\eqref{eq:variance_peak_and_envelope} and integrate it over the domain $[-\sqrt{\pi}/2;\sqrt{\pi}/2)+2\sqrt{\pi}\bb{Z}$ to obtain 
\be \label{eq:diff_logical_fidelity}
    1-F_\n{logic} \approx \frac{2\sqrt{2}\Delta }{\pi} \, e^{-\frac{\pi }{8 \Delta ^2}} \left(1- \frac{4\Delta^2}{\pi}\right)
\ee
which is accurate to $\cl{O}(\Delta^5)$. This expression is illustrated by the dashed curve in Fig.~\ref{fig:qec_infidelity_diff_states}. Despite this undesired behavior, the infidelity of the stabilized $\CZ$ gate for states with $\Delta\leq0.34$ is below $1\%$, a value that has been shown to be sufficient for a fault-tolerant concatenation of GKP codes with discrete variable encodings~\cite{noh_low-overhead_2022}. The performance of the error-corrected gate can be enhanced by improving the recovery procedure using reshaping of GKP states into a rectangular lattice~\cite{royer_stabilization_2020} or post-selection based on correlation in syndrome outcomes between multiple rounds of QEC~\cite{sivak_real_time_2023} (cf. Supp. Mat.~\cite{Note2}).



An alternative to local error correction is to use a finite-energy version of the gate which minimally distorts physical GKP states~\cite{menicucci_fault-tolerant_2014,royer_stabilization_2020, de_neeve_error_2022}. The finite-energy form of the $\CZ$ gate is the nonunitary interaction $\CZ_\Delta = \h{E}_\Delta\CZ\h{E}_\Delta^{-1} = e^{i\q_1\q_2 - \Delta^2 ( \q_1\p_2 + \p_1\q_2 ) }$ whose implementation requires us to couple the two oscillators to some auxiliary system. Inspired by a collisional model of dissipation~\cite{ciccarello_collision_2017} {\color{\mycolor}and a block-encoding of the nonunitary operation~\cite{gilyen_quantum_2019}}, we approximate $\CZ_\Delta$ by a unitary operation $e^{i\q_1\q_2\s_x - i \Delta^2 ( \q_1\p_2 + \p_1\q_2 )\s_y}$ followed by a reset of the auxiliary spin. The Trotterized verions this operation is realizable through spin-conditioned beamsplitters and/or squeezing~\cite{Note2}. Unfortunately, this engineered dissipative dynamics is subject to approximation errors. As a consequence, the fidelity that one reaches with the 1st order Trotter formula is $0.80$ when ${\Delta^2\ll1}$, an improvement of only $0.16$ compared to the ideal $\CZ$ gate. {\color{\mycolor}Higher order formulas entail a significant increase in the number of operations and complexity of the gate}.

\begin{figure}[t]
    \centering
    {\phantomsubcaption\label{fig:qutrit_circuit}}
    {\phantomsubcaption\label{fig:qutrit_infidelity_diff_states}}
    \includegraphics{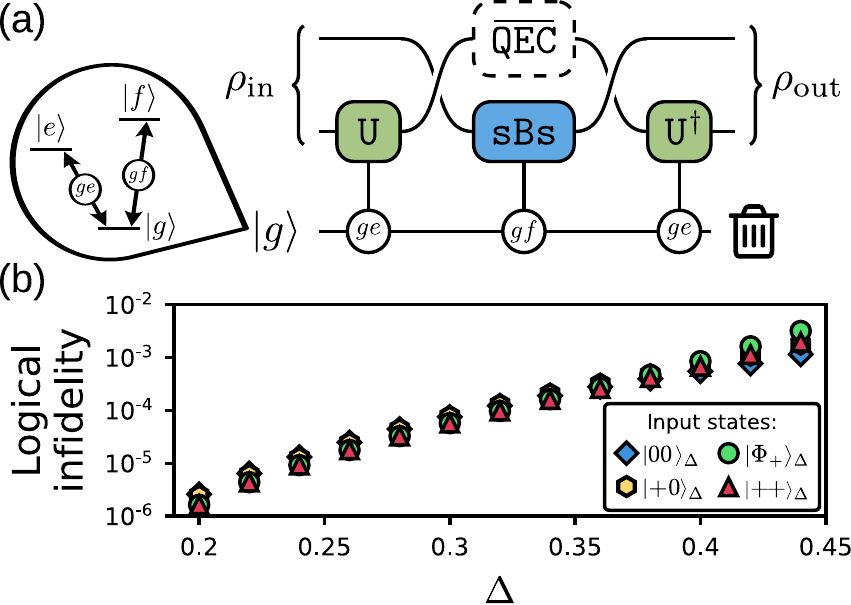}
    \caption{Two-qubit gate mediated by an auxiliary three-level system. (a) Circuit representation of the protocol. The system is initialized in a state $\rho_\n{in}$. The auxiliary system starting in $\ket{g}$ is used to retrieve the logical information of one of the oscillator using $\mathtt{CU}$ after what the two oscillators are swapped using a 50:50 beamsplitter. $\mathtt{sBs}$ represents a finite-energy stabilization which also effects a Pauli $Z$ operation if the qutrit is in $\ket{g}$. The circuit is then reversed in order to disentangle the oscillators from the auxiliary system. (b) The logical infidelity evaluated in a similar fashion as in Fig.~\ref{fig:qec_infidelity_diff_states}.}
    \label{fig:fig3}
\end{figure}


{\color{\mycolor}
To achieve higher fidelities, we propose a third approach that involves employing an auxiliary three-level system coupled to only one of the oscillators. In this hybrid discrete-continuous variable system we use the qutrit to facilitate quantum information exchange between the two oscillators. To entangle two GKP states, we first query the logical information of the initial oscillator using the auxiliary system, then swap bosonic states and apply an operation on the second oscillator conditioned on the auxiliary state. Finally, we swap the bosonic states back and disentangle the first oscillator from the auxiliary system. This procedure is illustrated in Fig.~\ref{fig:qutrit_circuit}. Since the swap via {\color{\mycolorbis}two} 50:50 beamsplitter is intrinsically finite-energy, the scheme preserves the states' energy as long as the conditional $\mathtt{CU}$ and $\mathtt{sBs}$ are finite energy. The former operation reads $\mathtt{CU}=e^{-i l \q\s_y^{(ge)}}\,e^{-i l \Delta^2\p\s_x^{(ge)}}$ with ${l=\sqrt{\pi}/2}$ and is identical to the unitary used for the readout of finite-energy GKP states proposed in Ref.~\cite{hastrup_improved_2021,de_neeve_error_2022}. With $\s_x^{(ge)}$ and $\s_y^{(ge)}$ being Pauli operators between states $\ket{g}$ and $\ket{e}$ of the qutrit, this operation entangles those states with $\ket{0}_{\!\Delta}$ and $\ket{1}_{\!\Delta}$, respectively. The second operation is the so-called small-Big-small circuit $\mathtt{sBs} = e^{-i l \Delta^2 \p\s_y^{(gf)}}\,e^{i l \q\s_x^{(gf)}}\,e^{-i l \Delta^2 \p\s_y^{(gf)}}$ which has been used to perform measurement-free error correction of the position quadrature of GKP states~\cite{royer_stabilization_2020,de_neeve_error_2022}. On top of the corrective power $\mathtt{sBs}$ has the property to apply a logical Pauli $Z$ that we use to add a phase to the oscillator state if the auxiliary system is in $\ket{g}$. The final operation is the inverse of the first one followed by a reset of the qutrit. The logical infidelity of this scheme shown in Fig.~\ref{fig:qutrit_infidelity_diff_states} is consistent across states and lower than that of the error-corrected gate.
{\color{\mycolorbis} For comparison for $\Delta\leq0.34$ the fidelity is below $0.01\%$.}
The main source of this infidelity stems from $\mathtt{CU}$ and $\mathtt{CU}^\dagger$, which can be mitigated by correcting the first oscillator state in between the two swaps. After $\mathtt{CU}$, the GKP state is stabilized not by $(\hat{S}_x,\hat{S}_z)$ but by $(-\hat{S}_x,\hat{S}_z)$. Fortunately, autonomous error correction protocol can be adapted to this new stabilizer set (represented as $\overbar{\mathtt{QEC}}$ in Fig.~\ref{fig:qutrit_circuit})~\cite{Note2}.
}


All the required elements, such as beamsplitters and squeezers, for realizing both ideal and finite-energy two GKP qubit operations have been successfully demonstrated in experimental setups utilizing trapped ions \cite{leibfriedTrappedIonQuantumSimulator2002,brown_coupled_2011,gorman_two-mode_2014,burd_quantum_2019,sutherland_universal_2021,hou_coherently_2022,katz_n-body_2022,katz_programmable_2023,shapira_robust_2023} and superconducting microwave cavities \cite{baust_tunable_2015,pfaff_controlled_2017,collodo_observation_2019,frattini_optimizing_2018,gao_programmable_2018,hillmann_universal_2020,chapman_high_2023,pietikainen_controlled_2022}. In supplementary material~\cite{Note2}, we provide details on experimental requirements for a trapped ion system.

{\color{\mycolor}
Our work addresses a critical issue in continuous-variable quantum information processing by proposing protocols to address the fidelity loss issue in two-qubit gates between GKP qubits. To this end we use of quantum error correction, introduce a direct construction of finite-energy two-qubit gate, and propose a protocol relying on an auxiliary three-level system to mediate the logical information between oscillators. Our proposed solutions can be made fault-tolerant:
This could be addressed through the use of biased-noise auxiliary systems~\cite{grimsmo_quantum_2021} or multiple discrete-variable systems, offering the potential for fault tolerance through the utilization of flag-qubits, detection or correction codes, or path-independent gate technique~\cite{shi_fault_tolerant_2019,hastrup_universal_2022,ma_path_independent_2020}. While this has been suggested using bosonic biased systems such as Cat codes, further work is necessary to investigate how to incorporate these methods into multi-qubit GKP gates offering a path towards fault tolerance using hybrid discrete-continuous variable codes~\cite{andersen_hybrid_2015}. Another important question for later study is how the finite-energy effects of ideal two-qubit gates together with standard continuous-variable noise processes, such as photon loss and dephasing, modify the thresholds of GKP-based encodings~\cite{fukui_high-threshold_2018,vuillot_quantum_2019,noh_fault-tolerant_2020,noh_low-overhead_2022}.}~We believe that our findings will contribute to the use of GKP codes for fault-tolerant quantum computation using dissipative QEC~\cite{reiter_dissipative_2017,de_neeve_error_2022} as well as for applications such as error-corrected quantum sensing~\cite{duivenvoorden_single-mode_2017} that is bias-free~\cite{rojkov_bias_2022}.

\medskip 
{\color{\mycolorbis}This paper is dedicated to the memory of our dear colleague and friend Martin Wagener, who passed away during the peer-review of the manuscript.}
I.R. thanks Daniel Weigand for the extensive and helpful discussion on the shifted grid state method, Brennan de Neeve for numerous discussions on GKP codes and Elias Zapusek for insightful questions throughout the project.
This work was supported as a part of NCCR QSIT, a National Centre of Competence (or Excellence) in Research, funded by the Swiss National Science Foundation (SNSF) grant no. 51NF40-185902. I.R. and F.R. acknowledge financial support via the SNSF Ambizione grant no. PZ00P2$\_$186040. M.W. acknowledges support via the SNSF research grant no. 200020$\_$179147. S.W. acknowledges financial support via the SNSF Swiss Postdoctoral Fellowship grant no. TMPFP2$\_$210584.


\bibliography{references}
\balancecolsandclearpage

\iftoggle{arXiv}{\begin{SuppMat}\iftoggle{arXiv}{
    \cleardoublepage 
	\title{Supplementary Material: Two qubit operations \\ for finite-energy Gottesman-Kitaev-Preskill encodings}
    \maketitle
}{}
\onecolumngrid
\setcounter{equation}{0}

In this Supplemental Material, we give detailed derivations of formulas and provide justifications to assertions from the main text. In particular, we present a more general form of the finite-energy grid states and discuss their representation using the shifted grid state basis. Then, we derive the position marginal distribution, the purity and the overlap fidelity of the gate. Section~\ref{sec:gate_decompositions} treats thoroughly the decomposition of the $\CZ$ and $\CNOT$ gates into linear optical elements, whereas in Section~\ref{sec:ft_threshold} we discuss their fault-tolerant aspect using experimental evidences. The error correction of the undesired effects is the topic of Section~\ref{sec:error_correction}. Finally, we discuss in more depth the finite-energy form of the entangling operations, present their first implementation and numerical simulation results.

\iftoggle{arXiv}{ \PRLsec{Notation}{sec:notation} }{ \section{Notation} \label{sec:notation} }

In this section, the relevant quantities and their notation are introduced. We are working with harmonic oscillators in phase-space. The quadrature operators $\q$ and $\p$ are defined via the creation $\hat{a}^{\dagger}$  and annihilation operator $\hat{a}$ as 
\be\label{eq:quadrature_operators}
    \q = \frac{1}{\sqrt{2}} (\hat{a} + \hat{a}^{\dagger}) 
    \qquad \text{and} \qquad
    \p = \frac{i}{\sqrt{2}} (\hat{a}^{\dagger} - \hat{a}).
\ee
These operators obey the canonical commutation relation $[\q,\p] = i$. The number operator determining the energy of the oscillator state is therefore defined as ${\hat{n}:=\hat{a}^{\dagger}\hat{a} = \frac{1}{2}(\q^2 + \p^2)}$. 

The two-qubit gates discussed further below are in essence conditional displacements in phase-space. A general one-mode displacement by an amount $\alpha \in \bb{C}$ in terms of the quadrature operators is defined as 
\begin{equation} \label{eq:displacement_operator}
    \hat{D}(\alpha) = e^{-i\sqrt{2}\,\big(\text{Re}(\alpha)\,\p - \text{Im}(\alpha)\,\q\big)}.
\end{equation}
A displacement of magnitude $l\in\bb{R}$ strictly along the $q$-axis corresponds to $\hat{D}(l/\sqrt{2})$, whereas a displacement along $p$ by the same amount is given by $\hat{D}(-il/\sqrt{2})$. These displacements play a crucial role in the GKP encodings, as we will show in the next section.

Other important transformations in phase-space are the squeezing and beamsplitter interactions. The former one is a single-mode operation, which is given by
\begin{equation}\label{eq:squeezing_operator}
    \hat{S}(r) = e^{i \frac{1}{2} r\,(\q\p + \p\q)}. 
\end{equation}
The squeezing operator enhances the resolution of a state in one quadrature and lowers it in the other, as in the Heisenberg picture the quadrature operators are transformed as $\q \to e^r \q$ and $\p \to e^{-r}\p$. On the other hand, the beamsplitter interaction is a two-mode operation realized via a bilinear coupling Hamiltonian that effectively describes an interferometer between the two oscillators. It is not uniquely defined. For our purpose, we rely on two definitions
\be \label{eq:beamsplitter_operators}
    \hat{B}_A(\theta) = e^{i\theta(\q_1\q_2 + \p_1\p_2)} 
    \qquad \text{and} \qquad
    \hat{B}_S(\theta) = e^{i\theta(\q_1\p_2 - \p_1\q_2)}.
\ee

Finally, since normal as well as wrapped normal distributions~\cite{mardia_directional_2000} will be repeatedly utilized in the following sections, we state below the convention that we are heeding
\be \label{eq:normal_and_wrapped_norm}
\begin{split}
    \mathsf{pdf}_X(x) &= \frac{1}{\sqrt{\pi\sigma^2}} e^{\frac{1}{\sigma^2}(x-\mu)^2} \quad\quad\text{where}\quad X\sim\cl{N}\big[\mu,\sigma\big]\\
    \mathsf{pdf}_X(x) &= \frac{1}{\sqrt{\pi\sigma^2}} \sum_{k\in\bb{Z}} e^{\frac{1}{\sigma^2}(x-\mu+ 2\pi k)^2} \quad\text{where}\quad X\sim\cl{WN}\big[\mu,\sigma\big].
\end{split}
\ee
Here, $\mathsf{pdf}_X$ denotes the probability density function of the random variable $X$ following a normal~$\cl{N}$ or wrapped normal~$\cl{WN}$ distribution with a mean $\mu$ and a standard deviation $\sigma$.

\iftoggle{arXiv}{ \PRLsec{Ideal GKP codes}{sec:ideal_codes} 
}{ \section{Ideal GKP codes} \label{sec:ideal_codes} }

\subsec{Stabilizer operators}{sec:stab_op}

As outlined in the main text, the ideal grid states consist of an infinite and translation-invariant superposition of position $\q$ or momentum $\p$ eigenstates. These eigenstates are in fact oscillator's ground states that are infinitely squeezed in the corresponding direction. There exist several variations of GKP encodings which differ from each other by their lattice type. To show this we follow the derivation from \citet{royer_stabilization_2020}. Consider the generalized quadrature operators defined by
\be \label{eq:general_q_p}
    \h{Q} = \alpha_1 \q + \beta_1 \p \qquad \text{and} \qquad \h{P} = \beta_2 \q + \alpha_2 \p
\ee
or alternatively with
\be 
    \begin{pmatrix} \h{Q} \\ \h{P} \end{pmatrix}
    =
    \begin{pmatrix} \alpha_1 & \beta_1 \\ \beta_2 & \alpha_2 \end{pmatrix}
    \begin{pmatrix} \q \\ \p \end{pmatrix}
    =:
    L
    \begin{pmatrix} \q \\ \p \end{pmatrix}
\ee
where $\alpha_i$, $\beta_i\,\in\mathbb{R}$ and $L$ represents the lattice matrix. Note that these operators do not satisfy the usual commutation relation. With the motivation of having translation-invariant codes, we construct the stabilizer group generators as
\be \label{eq:general_stab}
    \h{T}_x = e^{i\h{Q}} \qquad \text{and} \qquad \h{T}_z = e^{i\h{P}}
\ee
which then commute as $\h{T}_x\h{T}_z=e^{i\vartheta}\,\h{T}_z\h{T}_x$ with $\vartheta=\alpha_1\alpha_2-\beta_1\beta_2\equiv\n{det}(L)$. Thus to have a translation invariant code one requires that $\vartheta=2\pi d$ where $d\in\mathbb{Z}$ is the dimension of the logical subspace. For GKP states encoding a qubit (i.e., $d=2$), the two choices of lattices that are the most popular in the literature~\cite{GKP_original_paper,albert_performance_2018,fukui_high-threshold_2018,baragiola_all-gaussian_2019,fluhmann_encoding_2019, campagne-ibarcq_quantum_2020,royer_stabilization_2020,hastrup_measurement-free_2021,de_neeve_error_2022} are the square and hexagonal ones that read
\be 
   L_{\,\scalebox{0.85}{$\Square$}} = 2\sqrt{\pi} \,
   \begin{pmatrix} 0 & 1 \\ -1 & 0 \end{pmatrix}
   \qquad \text{and} \qquad
   L_{\,\scalebox{0.85}{$\hexagon$}} = 2\sqrt{\frac{2\pi}{\sqrt{3}}} \,
   \begin{pmatrix} 0 & 1 \\ -\sin\left(\frac{\pi}{3}\right) & \cos\left(\frac{\pi}{3}\right) \end{pmatrix}.
\ee
In this work, we focus exclusively on square grids. However, the following derivations can be performed with generalized quadrature operators and similar finite-energy effects would be observed for other code lattices. For us, the stabilizers are thus defined as in the main text
\be \label{eq:square_stab_ops}
    \h{S}_x := \,\h{T}_x\, = e^{i2\sqrt{\pi}\,\p} \qquad\qquad 
    \h{S}_z := \,\h{T}_z\, = e^{-i2\sqrt{\pi}\,\q}\,.
\ee

\subsec{Logical Pauli operators}{sec:logical_operators}

The logical Pauli operators generate a lattice dual to the code one~\cite{GKP_original_paper} and are given by the square root of the code stabilizers $\h{T}_x$ and $\h{T}_z$. In our case, they are expressed as
\be \label{eq:square_logic_ops}
    \h{X} := \sqrt{\h{T}_x} = e^{i\p\sqrt{2}} \qquad\qquad 
    \h{Z} := \sqrt{\h{T}_z} = e^{-i\q\sqrt{2}}\,.
\ee
One can straightforwardly check that they satisfy the commutation relations that are required by a stabilizer code, namely $\big[\h{X},\h{S}\big]=0=\big[\h{Z},\h{S}\big]\,\,\,\forall\,S\in\{S_x,S_z\}$ and $\h{X}\,\h{Z}=-\h{Z}\,\h{X}$.

\subsec{Computational basis}{sec:comput_basis}

We can now proceed with the logical codewords. The $\ket{0}_\n{I}$ and $\ket{1}_\n{I}$ states which effectively represent the $+1$ and $-1$ eigenstates of the Pauli $\h{Z}$ operator can for arbitrary GKP encodings be formulated as~\cite{albert_performance_2018,royer_stabilization_2020} 
\be \label{eq:general_com_basis}
    \ket{c}_\n{I} = \frac{1+(-1)^c\,\h{Z}}{2}\,\sum_{j,k\in\mathbb{Z}}\,\h{T}_x^j\,\h{T}_z^k\,\ket{q=0}\,,
\ee
where $\ket{q=0}$ is the eigenstate of $\q$ associated to the $0$ eigenvalue. Hence, with the stabilizers defined in Eq.~\eqref{eq:square_stab_ops} these states have the following form
\be \label{eq:ideal_com_basis_q}
    \ket{0}_\n{I} = \sum_{s\in\mathbb{Z}}\,\ket{q=2s\sqrt{\pi}}\qquad\text{and}\qquad\ket{1}_\n{I} = \sum_{s\in\mathbb{Z}}\,\ket{q=(2s+1)\sqrt{\pi}}\,.
\ee
Here, we express them in the position eigenstate basis (\textit{eigenbasis}), but we can easily transform them into their momentum representation using the Fourier transform that relates the real and the dual spaces. Indeed, the transform allows to convert a position wave function $\psi(q):=\langle{q}\ket{\psi}$ into its momentum form $\psi(p):=\langle{p}\ket{\psi}$,
\be \label{eq:fourier_transform}
    \cl{F}\big[\psi(q)\big](p) := \frac{1}{\sqrt{2\pi}} \int_{-\infty}^{\infty} \psi(q)\,e^{-i q p} \n{d}q \equiv \psi(p) \,.
\ee
Since the position wave function of the states in Eq.~\eqref{eq:ideal_com_basis_q} are simply forests of equally-spaced Dirac delta functions, also known as Dirac combs, their Fourier transform is again a forest of delta functions but with a different spacing. We can write it as $\cl{F}\big[\Sha_{T}(q)\big](p)=\Sha_{2\pi/T}(p)$ where $\Sha_{T}(x)=\sum_s \delta(x+s\,T)$ characterizes a Dirac comb with a period~$T$. As a result, in the momentum eigenbasis the computational states of a square GKP code are 
\be \label{eq:ideal_com_basis_p}
    \ket{0}_\n{I} = \sum_{s\in\mathbb{Z}}\,\ket{p=s\sqrt{\pi}}\qquad\text{and}\qquad\ket{1}_\n{I} = \sum_{s\in\mathbb{Z}}\,e^{is\pi}\ket{p=s\sqrt{\pi}}\,,
\ee
where the additional phase in $\ket{1}_\n{I}$ arises from the $\sqrt{\pi}$--shift of the grid compared to the $\ket{0}_\n{I}$ state. 
Due to the fact that $\cl{F}\big[\delta(q-y)\big](p)=\delta(p-y)$ and $\cl{F}\big[\delta(p-y)\big](q)=-\delta(q-y)$ with $y\in\bb{R}$, one can alternatively define the Fourier transform as a unitary operator $\hat{F}$. Indeed, these expressions establish that $\h{F}\q=\p\h{F}$ and $\q\h{F}=-\h{F}\p$. Thus, in terms of the two quadratures it reads~\cite{GKP_original_paper,weedbrook_gaussian_2012}
\be \label{eq:fourier_operator}
    \h{F} := e^{i\frac{\pi}{4}(\q^2 + \p^2)} \,,
\ee
effectively describing a ${\pi/2}$\,--\,rotation of the phase space or a quarter-cycle evolution under the oscillator Hamiltonian. This tells us that the $q$\,-- and $p$\,--\,representations of a quantum state are unitarily equivalent, meaning that the action of $\q$ on $\psi(q)$ is identical to $\p$ on $\psi(p)$. One can therefore focus exclusively on the representation that is the most convenient in a given situation. This is also the reason why the comparison between the $\CZ$ and $\CNOT$ operations that we will make in the later sections is fair and accurate.

More generally, arbitrary rotations of phase-space are represented by the following unitary
\be \label{eq:rotation_operator}
    \h{R}(\phi) := e^{i\frac{\phi}{2}(\q^2 + \p^2)} \,.
\ee
This operation that is easily realizable by changing the reference frame of the oscillator's rotating picture will be useful in our discussion about the unitary decompositions of the two-qubit gates for GKP states (cf. Section~\ref{sec:gate_decompositions}).

Finally, we emphasize that the states given in Eqs.~\eqref{eq:ideal_com_basis_q} and~\eqref{eq:ideal_com_basis_p} are not normalizable because the position and momentum eigenstates that they are made of do not live in the space of physical quantum states $L_2(\bb{R})$.

\subsec{Shifted grid state basis}{sec:shifted_grid_basis}

Despite their non-physical nature, the grid states can be used to define a complete and orthonormal basis of phase space. Let $\ket{\psi,u,v}$ be a square GKP state $\ket{\psi}_\n{I}$ that has been consecutively displaced in momentum and position by $v$ and $u$, respectively, namely $\ket{\psi,u,v} := e^{iu\p}e^{iv\q}\ket{\psi}_\n{I}$. Then, for the computational zero state defined previously this so-called shifted grid state will have the form
\be \label{eq:shifte_grid_states}
    \ket{u,v} \equiv \ket{0,u,v} = e^{-iu\p}e^{-iv\q}\ket{0}_\n{I} =
    \sum_{s\in\mathbb{Z}} e^{-iv2\sqrt{\pi}s} \, \ket{q=2\sqrt{\pi}s+u} = 
    \sum_{s\in\mathbb{Z}} e^{-iu(\sqrt{\pi}s+v)} \, \ket{p=\sqrt{\pi}s+v}\,.
\ee
One can notice that the variables parametrizing these states can be bounded to $u\in[-\sqrt{\pi};\sqrt{\pi})$ and $v\in[-\sqrt{\pi}/2;\sqrt{\pi}/2)$ thanks to some (pseudo) periodicity of the grid states in position and and momentum representations
\be \label{eq:periodicity_u_v}
    \ket{u\pm2\sqrt{\pi},v} = e^{\pm i v 2\sqrt{\pi}} \ket{u,v} 
    \qquad\text{and}\qquad
    \ket{u,v\pm\sqrt{\pi}} = \ket{u,v}\,.
\ee
Moreover, these states are orthogonal~\cite{weigand_generating_2018}
\be \label{eq:sgs_orthonormal}
\begin{split}
    \langle{u_1,v_1}\ket{u_2,v_2} &= 
    \sum_{s_1,s_2\in\mathbb{Z}} e^{i\,2\sqrt{\pi}(v_1s_1-v_2s_2)} \, \langle{q_1=2\sqrt{\pi}s_1+u_1} \ket{q_2=2\sqrt{\pi}s_2+u_2} = \\
    &=\sum_{s_1,s_2\in\mathbb{Z}} \, e^{i\,2\sqrt{\pi}(v_1s_1-v_2s_2)} \, \delta\left({2\sqrt{\pi}(s_1-s_2) + (u_1-u_2)}\right) = \\
    &= \frac{1}{2\sqrt{\pi}}\,\, \sum_{s_1,s_2\in\mathbb{Z}} \, e^{i\,2\sqrt{\pi}(v_1s_1-v_2s_2)} \, \delta\left({(s_1-s_2) + \frac{u_1-u_2}{2\sqrt{\pi}}}\right) = \\
    &= \frac{1}{2\sqrt{\pi}}\,\, \sum_{s\in\mathbb{Z}} \, e^{i\,2\sqrt{\pi}(v_1-v_2)s} \, \delta\left({\frac{u_1-u_2}{2\sqrt{\pi}}}\right) = 
    \frac{1}{2\sqrt{\pi}}\,\, \delta\Big(2\sqrt{\pi}(v_1-v_2)\Big) \, \delta\left(\frac{u_1-u_2}{2\sqrt{\pi}}\right) = \\
    &= \frac{1}{2\sqrt{\pi}}\,\, \delta\big({v_1-v_2}\big) \, \delta\big({u_1-u_2}\big) = \cl{N}_{u,v}^2\,\, \delta\big({v_1-v_2}\big) \, \delta\big({u_1-u_2}\big)\,.
\end{split}
\ee
In this derivation, we use the fact that the position eigenstates are orthonormal and that due to the domain of $u_1$ and $u_2$ the unique non-trivial solution is $s_1=s_2\equiv s$. It also shows that these shifted grid states can be rescaled by $\cl{N}_{u,v}$ such that they become orthonormal. With this in mind, let $\cl{B}_{u,v}$ be the set of shifted grid states defined~in~Eq.~\eqref{eq:shifte_grid_states}, i.e., $\cl{B}_{u,v} = \big\{ \ket{u,v} \,\,|\,\, u\in[-\sqrt{\pi};\sqrt{\pi})\,\,\text{and}\,\,v\in[-\sqrt{\pi}/2;\sqrt{\pi}/2) \big\}$. Then, it constitutes a complete orthogonal basis of the oscillator's Hilbert space given that
\be \label{eq:sgs_complete}
\begin{split}
    \int\displaylimits_{-\sqrt{\pi}}^{\sqrt{\pi}}\!\n{d}u\int\displaylimits_{-\sqrt{\pi}/2}^{\sqrt{\pi}/2}\!\n{d}v \,\, \ket{u,v}\!\!\bra{u,v}q=x\rangle &= 
    \int\displaylimits_{-\sqrt{\pi}}^{\sqrt{\pi}}\!\n{d}u\int\displaylimits_{-\sqrt{\pi}/2}^{\sqrt{\pi}/2}\!\n{d}v \,\,  
    \sum_{s\in\mathbb{Z}} e^{iv2\sqrt{\pi}s} \, \delta\big(x-(2\sqrt{\pi}s+u)\big) \, \ket{u,v} =
    \\
    &= \int\displaylimits_{-\sqrt{\pi}}^{\sqrt{\pi}}\!\n{d}u\int\displaylimits_{-\sqrt{\pi}/2}^{\sqrt{\pi}/2}\!\n{d}v \,\,  
    \sum_{s\in\mathbb{Z}} e^{iv2\sqrt{\pi}s} \, \delta\big(2\sqrt{\pi}(s_x-s)+(u_x-u)\big) \, \ket{u,v} = \\
    &= \int\displaylimits_{-\sqrt{\pi}/2}^{\sqrt{\pi}/2}\!\n{d}v \,\,  
    e^{-iv2\sqrt{\pi}s_x} \, \ket{u_x,v} = \int\displaylimits_{-\sqrt{\pi}/2}^{\sqrt{\pi}/2}\!\n{d}v \,\, \sum_{s\in\mathbb{Z}}
    \,  e^{iv2\sqrt{\pi}(s_x-s)} \, \ket{q=2\sqrt{\pi}s+u_x} = \\
    &= \sum_{s\in\mathbb{Z}}
    \,  \delta_{s_x,s} \, \ket{q=2\sqrt{\pi}s+u_x} = \ket{q=2\sqrt{\pi}s_x+u_x} = \ket{q=x} \,.
\end{split}
\ee
Derivations in Eqs.~\eqref{eq:sgs_orthonormal} and~\eqref{eq:sgs_complete} were inspired by the proofs of Lemma A1 and A2 from Ref.~\cite{weigand_generating_2018}. We can thus represent an arbitrary state in phase space using these states. Crucially, the choice of $\ket{u,v}$ as a basis state is not essential and may be replaced by $\ket{\psi,u,v}$ with the input GKP state being $\ket{+}_\n{I}$ (the~$+1$ eigenstate of $\h{X}$) as in Ref.~\cite{weigand_generating_2018} or any other infinite-energy grid state. These new bases will be equivalently orthogonal/orthonormal and complete as $\cl{B}_{u,v}$.

We must stress that the shifted grid state basis have been first introduced in 1967 by~\citet{zak_finite_1967} as the $kq$ representation. This has been then revived in the original GKP proposal~\cite{GKP_original_paper} as the ``error wave functions'' and used for analytical calculations in Refs.~\cite{glancy_error_2006,ketterer_quantum_2016,terhal_encoding_2016,weigand_generating_2018}.

\iftoggle{arXiv}{ \PRLsec{Finite-energy GKP codes}{sec:finite_codes} 
}{ \section{Finite-energy GKP codes} \label{sec:finite_codes} }

\subsec{Envelope operators}{sec:envl_operator}

As we mentioned above and in the main text, the ideal grid states are nonphysical due to the fact that they have an infinite norm and thus their wave function is not in $L_2(\bb{R})$. To rectify this, one must limit these states in phase space effectively making their energy finite. Mathematically, speaking this can be done using the envelope operator~\cite{menicucci_fault-tolerant_2014,noh_quantum_2019,royer_stabilization_2020}, also known as the embedded-error operator~\cite{GKP_original_paper,motes_encoding_2017}. Its most general form reads
\be \label{eq:energy_op}
    \h{E}_{\Delta,\kappa} = \exp\left(-\frac{\Delta^2}{2}\q^2 - \frac{\kappa^2}{2}\p^2 \right)
\ee
which is a non-unitary operator that can be approximated by the product of $e^{-\frac{\Delta^2}{2}\q^2}$ and $e^{-\frac{\kappa^2}{2}\p^2}$ in the assumption of $\Delta^2,\kappa^2\ll1$. The latter operators correspond to some Gaussian modulation of the wave function in position and momentum, respectively, and explain how they restrict the ideal GKP states in phase space.

Moreover, the definition in Eq.~\eqref{eq:energy_op} can be reduced to the operator used in the main text with $\kappa=\Delta$, i.e.,~${\h{E}_{\Delta}\equiv \h{E}_{\Delta,\Delta}}$, in which case one can reformulate it using the number operator $\h{n}=\frac{1}{2}(\q^2+\p^2)$. Thus, $\h{E}_{\Delta}$ would correspond to some Gaussian function in the energy basis (a.k.a.~Fock basis) and explain how it bounds the energy of the grid states presented earlier.

\subsec{Finite-energy wave functions}{sec:FE_wave_function}

The finite-energy version of an ideal GKP state $\ket{\psi}_\n{I}$ is given by $\ket{\psi}_{\Delta,\kappa}=\cl{N}_{\Delta,\kappa}\h{E}_{\Delta,\kappa}\ket{\psi}_\n{I}$ with $\cl{N}_{\Delta,\kappa}$ being the normalization constant. Hence, the computational basis states given in Eqs.~\eqref{eq:ideal_com_basis_q} and~\eqref{eq:ideal_com_basis_p} will under the action of the envelope operator be expressed as
\be \label{eq:finite_com_basis}
\begin{split}
    \ket{0}_{\Delta,\kappa} &= \cl{N}_{\Delta,\kappa}
    \int\displaylimits_{-\infty}^{\infty}\!\n{d}q \,\, \sum_{s \in \mathbb{Z}} \,\,
    e^{-\frac{1}{2}\kappa^2 (2s\sqrt{\pi})^2} \, e^{-\frac{1}{2\Delta^2}(q-2s\sqrt{\pi})^2}\,\ket{q} = \\
    &= \cl{N}_{\Delta,\kappa}'
    \int\displaylimits_{-\infty}^{\infty}\!\n{d}p \,\, \sum_{s \in \mathbb{Z}} \,\,
    e^{-\frac{1}{2}\Delta^2p^2}\,e^{-\frac{1}{2\kappa^2}(p-s\sqrt{\pi})^2}\,\ket{p}
\end{split}
\ee
(for the $\ket{1}_I$ replace the index $2s$ by $2s+1$) in the position and momentum basis, respectively. The normalization factors $\cl{N}_{\Delta,\kappa}$ and $\cl{N}_{\Delta,\kappa}'$ can be obtained from computing $\prescript{}{\Delta,\kappa}\langle0\!\ket{0}_{\Delta,\kappa}$. In the limit of $\Delta^2,\kappa^2\ll1$, the former one can be approximated by $\cl{N}_{\Delta,\kappa}^{\,2}\approx\frac{\kappa}{\Delta}\frac{1}{\pi}$. For an exact solution, we refer the reader to Refs.~\cite{ketterer_quantum_2016,royer_stabilization_2020,matsuura_equivalence_2020}. 

From these expressions, we notice that in position space the width of the Gaussian envelope is determined using the $\kappa$ parameter whereas the second exponential depending on $\Delta$ broadens individual peaks that in the ideal case were Dirac delta functions. Conversely, the wave function in the momentum space has the envelope and peaks parameterized by $\Delta$ and $\kappa$, respectively. This swapping of the roles is related to the Fourier transform that links the two representations.

Strictly speaking, Eq.~\eqref{eq:finite_com_basis} is not derived using the envelope operator. Indeed, there exists various representations for finite-energy GKP states among which some using Jacobi theta functions~\cite{mensen_phase-space_2021}, Hermit polynomials, coherent-state basis~\cite{albert_performance_2018} or displaced and squeezed vacuum states~\cite{GKP_original_paper}. The expression in Eq.~\eqref{eq:finite_com_basis} corresponds to the latter representation since $\h{E}_{\Delta,\kappa}\ket{\psi}_\n{I}$ would give a wave function (in terms of Hermit polynomials) with less explicit roles of $\kappa$ and $\Delta$. However, all these representations are equivalent under the desired condition $\Delta^2,\kappa^2\ll1$~\cite{albert_performance_2018,matsuura_equivalence_2020}. 

\subsec{Shifted grid state representation}{sec:shited_grid_repr}

Using the shifted grid state basis described in Section~\ref{sec:ideal_codes}, any oscillator's state can be represented as
\be
    \ket{\psi}=\int\displaylimits_{-\sqrt{\pi}}^{\sqrt{\pi}}\!\n{d}u\int\displaylimits_{-\sqrt{\pi}/2}^{\sqrt{\pi}/2}\!\n{d}v \,\, 
    \psi(u,v)\,\ket{u,v}.
\ee
where $\psi(u,v)$ is the wave function in this basis. In principle, the wave function has an arbitrary form as long as it satisfies the normalization requirement $\iint\n{d}u\n{d}v\,|\psi(u,v)|^2=1$. However, a certain class of probability distributions called wrapped distributions is particularly useful in this situation. Coming from the field of directional statistics these distributions are used to characterize random variables with some spatial (e.g. angles) or temporal (e.g. time periods) periodicity~\cite{mardia_directional_2000}. Given the $2\sqrt{\pi}$ periodicity of the grid states (cf. Eqs.~\eqref{eq:ideal_com_basis_q},~\eqref{eq:ideal_com_basis_p} and~\eqref{eq:finite_com_basis}) wrapped distributions are thus a suitable representation of their wave functions. In this work, we consider that the wave function of $\ket{0}_{\Delta,\kappa}$ is separable with respect to the two degrees of freedom and reads
\be \label{eq:shifted_grid_repr}
    \psi_0(u,v)=f_\Delta(u)\,\,g_\kappa(v)=
    \sum_{s\in\bb{Z}} \frac{1}{(\pi\Delta^2)^{1/4}} e^{-\frac{1}{2\Delta^2}(u+2\sqrt{\pi}s)^2}
    \,\,
    \sum_{t\in\bb{Z}} \frac{1}{(\pi\kappa^2)^{1/4}} e^{-\frac{1}{2\kappa^2}(v+\sqrt{\pi}t)^2}\,.
\ee
Here, $f_\Delta$ and $g_\kappa$ have been derived from the square root of the probability density function of two wrapped normal distributions~$\cl{WN}\big[0,\Delta\big]$ and $\cl{WN}\big[0,\kappa\big]$ centered at the origin and with standard deviations $\Delta$ and $\kappa$ (cf. Eq.~\eqref{eq:normal_and_wrapped_norm}). However, this approximation is valid only for small enough $\Delta$ and $\kappa$ parameters. We will see in Section~\ref{sec:physical_fidelity} that for larger values the normalization terms $(\pi\Delta^2)^{-1/4}$ and $(\pi\kappa^2)^{-1/4}$ will require some corrections. We must also mention that depending on the desired properties and considered states, alternative wrapped probability distributions can be used to express $\psi_0(u,v)$ such as bivariate wrapped normal or Von Mises distributions~\cite{ketterer_quantum_2016,weigand_generating_2018}.

This representation of $\ket{0}_{\Delta,\kappa}$ is identically equivalent to its position and momentum wave functions given in Eq.~\eqref{eq:finite_com_basis},
\be \label{eq:sgs_position_wf}
\begin{split}
    \psi_0(q=2\sqrt{\pi}s+u) &= \langle{q=2\sqrt{\pi}s+u}\ket{0}_{\Delta,\kappa} 
     = \int\!\n{d}u'\!\int\!\n{d}v' \,\, f_\Delta(u')\,g_\kappa(v') \,\langle{q=2\sqrt{\pi}s+u}\ket{u',v'}\propto \\
    &\propto \int\!\n{d}u'\!\int\!\n{d}v' \,\, f_\Delta(u')\,g_\kappa(v') \, \sum_{s'\in\bb{Z}} e^{-iv'2\sqrt{\pi}s'} \, \delta\big(2\sqrt{\pi}(s'-s)+(u'-u)\big)\propto \\
    &\propto \int\!\n{d}u'\!\int\!\n{d}v' \,\, f_\Delta(u')\,g_\kappa(v')\,e^{-iv'2\sqrt{\pi}s}\,\delta\big(u'-u\big) \propto \\
    &\propto f_\Delta(u)\,\int\!\n{d}\tilde{v} \,\, e^{-i\tilde{v}s}\,\,\mathsf{pdf}_{\tilde{V}}(\tilde{v})
    =f_\Delta(u)\,\,\Big\langle e^{-i \tilde{V} s} \Big\rangle= f_\Delta(u)\,e^{-\frac{1}{2}(2\sqrt{\pi}\kappa)^2 s^2} \propto\\
    &\propto e^{-\frac{1}{2}\kappa^2(2\sqrt{\pi}s)^2}\,\sum_{t\in\bb{Z}} e^{-\frac{1}{2\Delta^2}(u+2\sqrt{\pi}t)^2} 
    \approx e^{-\frac{1}{2}\kappa^2(2\sqrt{\pi}s)^2} e^{-\frac{1}{2\Delta^2}(q-2\sqrt{\pi}s)^2}\,. 
\end{split}
\ee
Here, from the second to the third line we used the similar approach as in Eqs.~\eqref{eq:sgs_orthonormal} and~\eqref{eq:sgs_complete}. On top of these, in the fourth line we made the change of variable $\tilde{v}\mapsto2\sqrt{\pi}v'$ which allows us to transform the given integral into the first moment of the random variable $\tilde{V}\sim\cl{WN}\big[0,2\sqrt{\pi}\kappa\big]$. The last equality approximates the sum of Gaussian functions by neglecting all the terms for which $t\neq0$. This approximation is valid as long as these functions are sharp enough, i.e., when $\Delta^2\ll1$. It is straightforward to notice that one will get a similar result when $\ket{0}_{\Delta,\kappa}$ is replaced by the expression in Eq.~\eqref{eq:finite_com_basis}. If one wants to obtain the distribution for any arbitrary $q$, it is necessary to disentangle the dependence of $q$ on $s$ by including a summation over the index. 

\iftoggle{arXiv}{ \PRLsec{Gate operations for two GKP qubits}{sec:ideal_operations}
}{ \section{Gate operations for two GKP qubits}\label{sec:ideal_operations} }

\subsec{Symplectic and Gaussian operations}{sec:symplec_gaussian_op}

As we will see below the $\CZ$ gate that we consider in the main text is a symplectic and Gaussian operation. A gate is said to be symplectic if it preserves the symplectic form of the phase space~\cite{weedbrook_gaussian_2012}. In other words, it preserves the canonical commutation relations of the quadratures $\q$ and $\p$. If one defines $\mathbf{\h{x}}=\big(\q_1,\p_1,\q_2,\p_2,\ldots,\q_n,\p_n\big)$ as the vector of position and momentum operators of $n$ oscillators, then the commutation relation dictates that $\big[\h{x}_i,\h{x}_j\big]=2i\,\Omega_{ij}$ where $\Omega = \bigoplus_i^n \omega$ is known as the symplectic form of phase space with $\omega$ being a 2x2 skew-symmetric matrix defined as $(\omega)_{1,2}=-(\omega)_{2,1}=1$. 

Another important definition from the field of Gaussian quantum information (i.e., quantum information that uses bosonic systems) is the notion of Gaussian states and transformations. The former is a subset $\cl{S}_\n{G}$ of all physical states in phase space that have the particularity of having a Gaussian Wigner quasiprobability distribution. A unitary $U$ is said to be Gaussian if it leaves this subset invariant, i.e., $\big\{ U\ket{\Phi}\,\n{with}\,\ket{\Phi}\in\cl{S}_\n{G}\big\}=\cl{S}_\n{G}$. Gaussian operations are generated with Hamiltonians that are quadratic in the position and momentum operators which in the Heisenberg picture corresponds to a linear map~\cite{weedbrook_gaussian_2012}
\be \label{eq:gaussian_op_heisenberg}
    \mathbf{\h{x}} \mapsto \h{U}^\dagger\mathbf{\h{x}}\h{U} = S_{\h{U}}\mathbf{\h{x}}+d_{\h{U}}
    \qquad \text{with} \qquad
    S_{\h{U}}\in \bb{R}^{2n\times2n} \quad \text{and}\quad d_{\h{U}}\in \bb{R}^{2n}\,.
\ee
These two parameters completely characterize unitary Gaussian operations. Thus, a $n$-mode transformation $\big(S_{\h{U}},d_{\h{U}}\big)$ is said to be symplectic if and only if $S_{\h{U}}\,\Omega\,S_{\h{U}}^\intercal = \Omega$.

Examples of such operations are for instance the squeezing and Fourier operators given in Eqs.~\eqref{eq:squeezing_operator} and~\eqref{eq:fourier_operator}, respectively. They are both one mode interactions and are defined by the transformations
\be \label{eq:squeez_fourier_heis}
    S_{\h{S}(r)} = 
    \begin{pmatrix}
        e^{r} & 0 \\
        0 & e^{-r}
    \end{pmatrix}
    \qquad\qquad
    S_{\h{F}} = 
    \begin{pmatrix}
        0 & -1 \\
        1 & 0
    \end{pmatrix}\equiv\omega\,.
\ee
From a direct calculation, it follows that ${S_{\h{S}(r)}\,\Omega\,S_{\h{S}(r)}^{\intercal} = \Omega = S_{\h{F}}\,\Omega\,S_{\h{F}}^{\intercal}}$. As we will see below $\CZ$ and $\CNOT$ gates are also members of the symplectic Gaussian operations.

GKP states do not belong to the subset of Gaussian states since parts of their Wigner distributions are negative (pure Gaussian states have no negative components in their Wigner distribution~\cite{weedbrook_gaussian_2012}). However, it has been proven that symplectic gates are a necessary requirement for universal quantum computation with continuous variable systems~\cite{lloyd_quantum_1999,menicucci_universal_2006}. Moreover, symplectic gates can be implemented using linear optical elements, such as phase shifters, squeezers and beamsplitters~\cite{lloyd_quantum_1999}, and are therefore relevant for experimental schemes.

\subsec{$\CZ$ gate}{subsec:cz_gate}

The logical $\CZ$ gate is a two mode entangling operation, given by 
\begin{equation}\label{eq:CZ_operator}
    \CZ = e^{i \q_1\q_2}.
\end{equation}
Its transformation matrix in phase space can be obtained as for the squeezing and Fourier operations in Eq.~\eqref{eq:squeez_fourier_heis} by conjugating the two quadratures $\q$ and $\p$ with the unitary above. We obtain
\begin{equation}\label{eq:CZ_phase_space}
    \begin{pmatrix}
    \q_1 \\ \p_1 \\ \q_2 \\ \p_2 
    \end{pmatrix} \mapsto
    S_\CZ
    \begin{pmatrix}
    \q_1 \\ \p_1 \\ \q_2 \\ \p_2 
    \end{pmatrix}
    =
    \begin{pmatrix}
        1 & 0 & 0 & 0 \\
        0 & 1 & -1 & 0 \\
        0 & 0 & 1 & 0 \\
        -1 & 0 & 0 & 1 
    \end{pmatrix}   
    \begin{pmatrix}
    \q_1 \\ \p_1 \\ \q_2 \\ \p_2 
    \end{pmatrix}.
\end{equation}
One can verify that it satisfies the symplectic condition $S_{\CZ}\,\Omega\,S_{\CZ}^{\intercal} = \Omega$ and that the Hamiltonian generating this transformation is at most quadratic in $\q$ and $\p$. Thus, $\CZ$ belongs to symplectic Gaussian operations.

The action on the ideal computational basis states presented in Eq.~\eqref{eq:ideal_com_basis_q} can be understood, if we consider the significance of this operation for the target state given that the control one is in a position eigenstate
\begin{equation}
    \bra{q_1}\CZ\ket{q_1} = e^{i q_1 \q_2} = \h{D}\left(i\frac{q_1}{\sqrt{2}}\right)
\end{equation}
with $\h{D}$ being the displacement operator defined in Eq.~\eqref{eq:displacement_operator}. Since the ideal computational basis state are superpositions of eigenstates located at either even or odd multiples of $\sqrt{\pi}$, only the input state $\ket{1}_\n{I}\ket{1}_\n{I}$ will acquire an overall phase of $\pi$ (the product of two odd multiples of $\sqrt{\pi}$ is an odd multiple of $\pi$), whereas all the other states will acquire an overall phase of $2\pi$. It thus reproduces the behavior of a $\CZ$ gate acting on a qubit.

\subsec{$\CNOT$ gate}{subsec:cnot_gate}

The $\CNOT$ gate is a two-qubit gate which performs a logical $\hat{X}$ operation on the target qubit if the control qubit is in state $\ket{1}$. This translates in the GKP setting to be a conditional displacement operator. It is defined by a quadrature--quadrature coupling Hamiltonian
\begin{equation}\label{eq:CNOT_operator}
    \CNOT = e^{i\q_1 \p_2}\,.
\end{equation}
It is equivalent to a $\CZ$ operation upon a Fourier transform of the target mode (in practice apply a Fourier gate from Eq.~\eqref{eq:fourier_operator} on the target system before and after the $\CZ$ operation). Therefore, the discussion about finite energy effects in the main text is equivalent for both gates, as long as the right perturbed quadratures are specified. The phase-space representation is obtained as above and yields
\begin{equation}\label{eq:CNOT_phase_space}
    \begin{pmatrix}
    \q_1 \\ \p_1 \\ \q_2 \\ \p_2 
    \end{pmatrix} 
    \mapsto S_\CNOT
    \begin{pmatrix}
    \q_1 \\ \p_1 \\ \q_2 \\ \p_2 
    \end{pmatrix}
    =
    \begin{pmatrix}
        1 & 0 & 0 & 0  \\
        0 & 1 & 0 & -1 \\
        1 & 0 & 1 & 0  \\
        0 & 0 & 0 & 1 
    \end{pmatrix}   
    \begin{pmatrix}
    \q_1 \\ \p_1 \\ \q_2 \\ \p_2 
    \end{pmatrix}.
\end{equation}
As for the $\CZ$ gate, notice that if we assume that the control system is in a position eigenstate, the gate embodies a displacement along the $q$--axis for the target state (i.e., with index 2)
\begin{equation}
    \bra{q_1}\CNOT\ket{q_1} = e^{i q_1 \p_2} = \h{D}\left(-\frac{q_1}{\sqrt{2}}\right) \,.
\end{equation}
Hence, the $\CNOT$ gate is a translation along $q_2$ by an amount $q_1/\sqrt{2}$. If the target state is in a superposition of even multiples of $\sqrt{\pi}$ position eigenstates and the control state in a superposition of odd ones, each term in the target state is shifted by an odd multiple, the resulting state will be in an odd superposition (the sum of an even and odd integers is odd). Using this knowledge, we can apply the gate to the ideal computational basis states which are superposition of position eigenstates located at even and odd multiples of $\sqrt{\pi}$. We find that
\begin{equation}
    \begin{matrix}
    \ket{00}_\n{I} \mapsto \ket{00}_\n{I} \\
    \ket{01}_\n{I} \mapsto \ket{01}_\n{I} \\
    \ket{10}_\n{I} \mapsto \ket{11}_\n{I} \\
    \ket{11}_\n{I} \mapsto \ket{10}_\n{I} 
    \end{matrix} 
    \qquad \text{or alternatively} \qquad
    \begin{bmatrix}
    \ket{00}_\n{I} \\ \ket{01}_\n{I} \\ \ket{10}_\n{I} \\ \ket{11}_\n{I} 
    \end{bmatrix} 
    \mapsto
    \begin{bmatrix}
    1 & 0 & 0 & 0 \\
    0 & 1 & 0 & 0 \\
    0 & 0 & 0 & 1 \\
    0 & 0 & 1 & 0
    \end{bmatrix}
    \begin{bmatrix}
    \ket{00}_\n{I} \\ \ket{01}_\n{I} \\ \ket{10}_\n{I} \\ \ket{11}_\n{I} 
    \end{bmatrix}, 
\end{equation}
which is precisely the action that a controlled NOT operation should have on a two-level system. Here and from now on, we denote $\ket{\psi_1\psi_2}_\alpha:=\ket{\psi_1}_\alpha\ket{\psi_2}_\alpha$ with the first state being the state of the control system and the second the state of the target one, while $\alpha$ indicates ideal or finite-energy version of these states.

\subsec{$\CZ$ and $\CNOT$ action on shifted grid states}{subsec:cz_cnot_gates_shift_grid_state}

Given the definition of shifted grid states in Eq.~\eqref{eq:shifte_grid_states} we compute how the $\CZ$ gate transforms a pair of such states
\be \label{eq:sgs_CZ}
\begin{split}
    \CZ \ket{u_1,v_1} \ket{u_2,v_2} &= \CZ\,e^{-iu_1\p_1}e^{-iv_1\q_1}e^{-iu_2\p_2}e^{-iv_2\q_2}\, \ket{00}_\n{I} = \\
    &= \CZ\,e^{-iu_1\p_1}\,\CZ^{\dagger}\CZ\,e^{-iv_1\q_1}\,\CZ^{\dagger}\CZ\,e^{-iu_2\p_2}\,\CZ^{\dagger}\CZ\,e^{-iv_2\q_2}\,\CZ^{\dagger}\CZ\, \ket{00}_\n{I} =\\
    &= e^{-iu_1\p_1}\,e^{-i(v_1-u_2)\q_1}\,e^{-iu_2\p_2}\,e^{u_1u_2[\q_2,\p_2]}\,e^{-i(v_2-u_1)\q_2}\,\ket{00}_\n{I} =\\
    &= e^{iu_1u_2}\,\ket{u_1,v_1-u_2} \ket{u_2,v_2-u_1}
\end{split}
\ee
where we make use of Eq.~\eqref{eq:CZ_phase_space} as well as the Baker-Campbell-Hausdorf (BCH) formula for commuting $e^{-iu_1\q_2}$ and $e^{-iu_2\p_2}$. The end result shows us that the $\CZ$ gate converts a pair of shifted grid states into another pair of shifted grid states that has as expected a global phase given by the product of position shifts of both grid states,i.e., ~$u_1u_2$. However, they are also mutually dependent. The new pair is such that the shift in the momentum space of the first state~$v_1$ is modified by an amount that corresponds to a shift in the position space of the second state~$u_2$. Moreover, we notice that the gate acts symmetrically on the second state. This property is also observed in the $\CNOT$ action
\be 
    \CNOT \ket{u_1,v_1} \ket{u_2,v_2} = \ket{u_1,v_1+v_2} \ket{u_2-u_1,v_2}\,.
\ee
Indeed, now the position shift in the target state get enhanced by the position shift of the control one, whereas the momentum shift of the control state get a contribution from the momentum shift of the target. Hence, if in Fig.~1(a) we would have plotted the momentum marginal distribution of the $\ket{00}_{\Delta}$ after the $\CNOT$ gate instead of the $\CZ$ one, we would have observed that the finite-energy effects arise only in the control system and none in the target one.

\subsec{Controlled Pauli gates}{subsec:controlled_paulis}

It is well known that a set of single-qubit rotations combined with a unique entangling two-qubit gate is sufficient for universal quantum computation~\cite{bremner_practical_2002}. For GKP qubits, single qubit rotations can be achieved through gate teleportation technique proposed in~\citet{campagne-ibarcq_quantum_2020} and demonstrated experimentally in \citet{fluhmann_encoding_2019}. Thus, $\CZ$ and $\CNOT$ gates presented above are both sufficient for universality. Nonetheless, the general form of a controlled Pauli gate for a square GKP code reads~\cite{grimsmo_quantum_2021}
\be
    \mathtt{CPAULI}_{AB} = e^{i \h{a}_1 \h{b}_2}
\ee
where $A$ and $B$ set the Pauli bases for the control and target qubits, respectively, and $\h{a},\h{b}\in\{\q,\p,\q-\p\}$ correspond to their associated quadrature operator. In other words, this unitary performs the logical Pauli gate $\h{B}$ on the target qubit if and only if the control one is a $-1$ eigenstate of the operator $\h{A}$. Most important is the fact that $\mathtt{CPAULI}_{AB}$ is once again realized using a quadrature--quadrature coupling and constitutes a displacement of one subsystem conditioned on some phase-space coordinate of the other. Therefore, due to the lack of translation symmetry of finite-energy GKP states, all controlled Pauli gates will induce equivalent undesired effects.

\iftoggle{arXiv}{ \PRLsec{Marginal distributions}{sec:marginal_distributions}
}{ \section{Marginal distributions} \label{sec:marginal_distributions} }

The first finite-energy effect that is considered in the main text is the peaks and envelope broadening of the momentum marginal distributions of both subspaces' states after the application of the $\CZ$ gate onto $\ket{00}_\Delta$. It is however worth mentioning that the analysis that follows also concerns the broadening that other controlled Pauli gates induce on marginal distributions along the corresponding disturbed quadratures. The particular choice of $\CZ$ is motivated by the numerical methods used in this work (cf. Section~\ref{sec:num_method}). Moreover, the amount of broadening is independent of the input state. Indeed, the marginal distributions do not reveal the relative phase of a state, consequently the amount of broadening will be the same for $\ket{11}_\Delta$ or $\ket{++}_\Delta$ and only the position of their peaks in phase space will differ from the case presented below.

Let us start by observing the action of the $\CZ$ gate on the shifted-grid state representation of $\ket{00}_{\Delta,\kappa}$ (cf. Eq.~\eqref{eq:shifted_grid_repr})
\be \label{eq:cz_action_on_00}
\begin{split}
    \ket{\Psi} = \CZ \ket{00}_{\Delta,\kappa} &= \CZ 
    \iint\!\n{d}u_1\n{d}v_1\iint\!\n{d}u_2\n{d}v_2 \,\, 
    \psi_0(u_1,v_1)\psi_0(u_2,v_2) \ket{u_1,v_1} \ket{u_2,v_2} =\\
    &= \iint\!\n{d}A_1\iint\!\n{d}A_2 \,\, 
    e^{iu_1u_2}\,\psi_0(u_1,v_1)\,\psi_0(u_2,v_2) \ket{u_1,v_1-u_2} \ket{u_2,v_2-u_1}
\end{split}
\ee
where for conciseness we omit the limits of the integrals and simplify $\n{d}u_i\n{d}v_i$ by $\n{d}A_i$ ($A$ for area). This result follows directly from Eq.~\eqref{eq:sgs_CZ}. We now proceed with the tracing out of the second system
\be 
\begin{split}
    \n{Tr}_2\big[\ket{\Psi}\!\!\bra{\Psi}\big] &= 
    \iint\!\n{d}u'_2\n{d}v'_2 \,\,
    \Big|\!\bra{u'_2,v'_2} \Psi\rangle\Big|^2 = \\
    &= \iint\!\n{d}A'_2 \,\, \left|
    \iint\!\n{d}A_1\iint\!\n{d}A_2 \,\, 
    \Psi(u_1,v_1,u_2,v_2)\,\delta(u_2-u'_2)\,\delta(v_2-u_1-v'_2) \ket{u_1,v_1-u_2}
    \right|^2 = \\
    &= \iint\!\n{d}A'_2 \,\, \left|
    \iint\!\n{d}A_1 \,\, 
    \Psi(u_1,v_1,u'_2,v'_2+u_1) \ket{u_1,v_1-u'_2}
    \right|^2
\end{split}
\ee
where we assumed the shifted grid states being orthonormal and $\Psi(u_1,v_1,u_2,v_2):=e^{iu_1u_2}\,\psi_0(u_1,v_1)\,\psi_0(u_2,v_2)$ is the wave function in the shifted grid state basis. The momentum marginal distribution is thus given by
\be \label{eq:marg_integral_0}
\begin{split}
    P(\tilde{p}_1)&=\bra{\tilde{p}_1}\n{Tr}_2\big[\ket{\Psi}\!\!\bra{\Psi}\big]\ket{\tilde{p}_1} 
    = \iint\!\n{d}A'_2 \,\, \Bigg| \iint\!\n{d}A_1 \,\, 
    \Psi(u_1,v_1,u'_2,v'_2+u_1)\,\langle \tilde{p}_1\!\ket{u_1,v_1-u'_2} \Bigg|^2 = \\
    &= \cl{N}_{u,v}^{-1} \iint\!\n{d}A'_2 \,\, \left| \iint\!\n{d}A_1 \,\, 
    \Psi(u_1,v_1,u'_2,v'_2+u_1) \sum_{s\in\bb{Z}} e^{iu_1(\sqrt{\pi}s+v_1-u'_2)}
    \delta\!\left(\tilde{p}_1-\sqrt{\pi}s-v_1+u'_2\right)
    \right|^2 = \\
    &= \cl{N}_{u,v}^{-1} \iint\!\n{d}A'_2 \,\, \Bigg| \iint\!\n{d}A_1 \,\, 
    \Psi(u_1,v_1,u'_2,v'_2+u_1)\,e^{iu_1(\sqrt{\pi}\tilde{s}+v_1-u'_2)}\,
    \frac{1}{\sqrt{\pi}}\,\delta\!\left(\tilde{v}_1-v_1+u'_2\right) 
    \Bigg|^2 = \\
    &= \frac{1}{\pi}\cl{N}_{u,v}^{-1} \iint\!\n{d}A'_2 \,\, \Bigg| \varint\!\n{d}u_1 \,\, 
    e^{iu_1u'_2}\,\psi_0(u_1,\tilde{v}_1+u'_2)\,\psi_0(u'_2,v'_2+u_1)\,e^{iu_1\tilde{p}_1}\,
    \Bigg|^2 = \\
    &= \frac{1}{\pi}\cl{N}_{u,v}^{-1} \iint\!\n{d}A'_2 \,\, 
    f^2_\Delta(u'_2)\,g^2_\kappa(\tilde{v}_1+u'_2)\,
    \Bigg| \varint\!\n{d}u_1\,\,e^{iu_1(u'_2+\tilde{p}_1)}\,f_\Delta(u_1)\,g_\kappa(v'_2+u_1)\, \Bigg|^2 \,.
\end{split}
\ee
Here, we used that $\psi_0(u,v)=f_\Delta(u)\,\,g_\kappa(v)$ and the fact that these are real-valued functions. The term $\cl{N}_{u,v}$ comes from the othonormality condition of the shifted grid states (cf. Eq.~\eqref{eq:sgs_orthonormal}). The solution of the inner most integral reads
\be \label{eq:marg_integral_1}
\begin{split}
    &\varint\!\n{d}u_1\,\,e^{iu_1(u'_2+\tilde{p}_1)}\,f_\Delta(u_1)\,g_\kappa(v'_2+u_1)=\\
    &=\frac{1}{(\pi\Delta^2)^{1/4}}\,\frac{1}{(\pi\kappa^2)^{1/4}} \, \sum_{s,t\in\bb{Z}}\, 
    \varint\!\n{d}u_1\,\,e^{iu_1(u'_2+\tilde{p}_1)}
    \,e^{-\frac{1}{2\Delta^2}(u_1+2\sqrt{\pi}s)^2}\,e^{-\frac{1}{2\kappa^2}(v'_2+u_1+\sqrt{\pi}t)^2}\approx \\
    &\approx \cl{N}_1 \,\,
    e^{-\frac{2i\Delta^2}{2(\Delta^2+\kappa^2)}(\tilde{p}_1+u'_2)v'_2 } \,
    e^{-\frac{\Delta^2\kappa^2}{2(\Delta^2+\kappa^2)}(\tilde{p}_1+u'_2)^2 } \,
    \sum_{t\in\bb{Z}} e^{-\frac{1}{2(\Delta^2+\kappa^2)} (v'_2 + \sqrt{\pi} t)^2}\,
    e^{ \frac{2i\Delta^2}{2(\Delta^2+\kappa^2)} \sqrt{\pi}t(\tilde{p}_1+u'_2)}
\end{split}
\ee
where we summarized the normalization terms by $\cl{N}_1$. In this derivation we have taken into consideration only the zeroth order term of the first sum, i.e.,~$s=0$, which is a sensible assumption thanks to the fact that $f_\Delta$ has a twice as large periodicity compared to $g_\kappa$ and that $\Delta^2,\kappa^2\ll1$. To be more specific, the neglected terms become relevant only for $\Delta,\kappa > 0.6$. The absolute value squared is then well approximated by 
\be 
    \bigg| \varint\!\n{d}u_1\,\,e^{iu_1(u'_2+\tilde{p}_1)}\,f_\Delta(u_1)\,g_\kappa(v'_2+u_1) \bigg|^2 
    \approx \cl{N}_1^2 \,\, e^{-\frac{\Delta^2\kappa^2}{\Delta^2+\kappa^2} (\tilde{p}_1+u'_2)^2} \,
    \sum_{t\in\bb{Z}} e^{-\frac{1}{\Delta^2+\kappa^2} (v'_2 + \sqrt{\pi} t)^2 }\,
\ee
which is strictly speaking an upper bound of the left-hand side expression. However, the assumption that $\Delta,\kappa<0.6$ makes the cross term elements arising from the sum irrelevant. In order to have an analytical expression for the momentum marginal distribution $P(\tilde{p}_1)$ we must solve two more integrals. Firstly,
\be \label{eq:marg_integral_2}
    \varint\!\n{d}v'_2\,\,\sum_{t\in\bb{Z}} e^{-\frac{1}{\Delta^2+\kappa^2} (v'_2 + \sqrt{\pi} t)^2 } = \sqrt{\pi}\sqrt{\Delta^2+\kappa^2}
\ee
which is trivial since it is equivalent (up to a proportionality coefficient) to integrating the probability density function of a wrapped normal distribution $\cl{WN}\big[0,\frac{1}{2}\sqrt{\pi}(\Delta^2+\kappa^2)\big]$ over $[-\pi;\pi)$. The second one reads
\be \label{eq:marg_integral_3}
\begin{split}
    &\varint\!\n{d}u'_2\,\,f^2_\Delta(u'_2)\,g^2_\kappa(\tilde{v}_1+u'_2)\,
    e^{-\frac{\Delta^2\kappa^2}{\Delta^2+\kappa^2} (\tilde{p}_1+u'_2)^2} \approx \\
    &\approx\frac{1}{(\pi\Delta^2)^{1/2}}\,\frac{1}{(\pi\kappa^2)^{1/2}} \sum_{s,t\in\bb{Z}}\, 
    \varint\!\n{d}u'_2\,\,
    e^{-\frac{1}{\Delta^2}(u'_2+2\sqrt{\pi}s)^2}\,
    e^{-\frac{1}{\kappa^2}(\tilde{v}_1+u'_2+\sqrt{\pi}t)^2}\,
    e^{-\frac{\Delta^2\kappa^2}{\Delta^2+\kappa^2} (\tilde{p}_1+u'_2)^2} \approx \\
    &\approx \cl{N}_2^2 \,\, 
    e^{-\frac{\Delta^2+\kappa^2}{(\Delta^2+\kappa^2)^2+\Delta^4\kappa^4} (\tilde{p}_1-\sqrt{\pi}\tilde{s})^2}
    \, e^{-\frac{\Delta^2\kappa^4}{(\Delta^2+\kappa^2)^2+\Delta^4\kappa^4} \tilde{p}_1^2}
    \, e^{-\frac{\Delta^4\kappa^2}{(\Delta^2+\kappa^2)^2+\Delta^4\kappa^4} (\sqrt{\pi}\tilde{s})^2}\,
\end{split}
\ee
where we replace $\tilde{v}_1=\tilde{p}_1-\tilde{s}\sqrt{\pi}$ and store as before the normalization constant in $\cl{N}_2^2$. To obtain this result we first approximate the square of the sums (arising from $f^2_\Delta$ and $g^2_\kappa$) by the sum of squares, this is once again reasonable for $\Delta,\kappa<0.6$ as the cross terms become negligible. The second approximation is similar to Eq.~\eqref{eq:marg_integral_1}, we set $s=0$. However, due to third exponential in the integrand, we can also consider solely the term $t=0$.

The normalization constants from Eqs.~\eqref{eq:marg_integral_1} and~\eqref{eq:marg_integral_3} have the following form
\be \label{eq:marg_norm_terms}
    \cl{N}_1^2 = \frac{2\Delta\kappa}{\Delta^2+\kappa^2} 
    \qquad \text{and} \qquad 
    \cl{N}_2^2 =\frac{1}{\sqrt{\pi}} \sqrt{\frac{\Delta^2+\kappa^2}{(\Delta^2+\kappa^2)^2+\Delta^4\kappa^4}} \,\,.
\ee
Finally, combining Eqs.~\eqref{eq:marg_integral_2},~\eqref{eq:marg_integral_3} and~\eqref{eq:marg_norm_terms} with \eqref{eq:marg_integral_0} gives us the following expression for the momentum marginal distribution of the first/control system
\be 
    P(\tilde{p}_1) = \frac{1}{\sqrt{\pi}} \frac{4\Delta\kappa}{\sqrt{(\Delta^2+\kappa^2)^2+\Delta^4\kappa^4}} \,
    e^{-\frac{\Delta^2+\kappa^2}{(\Delta^2+\kappa^2)^2+\Delta^4\kappa^4} (\tilde{p}_1-\sqrt{\pi}\tilde{s})^2}
    \, e^{-\frac{\Delta^2\kappa^4}{(\Delta^2+\kappa^2)^2+\Delta^4\kappa^4} \tilde{p}_1^2}
    \, e^{-\frac{\Delta^4\kappa^2}{(\Delta^2+\kappa^2)^2+\Delta^4\kappa^4} (\sqrt{\pi}\tilde{s})^2}\,.
\ee
As for Eq.~\eqref{eq:sgs_position_wf}, we can write this distribution for an arbitrary $p_1$ by adding a sum over the $s$ index which unravels $\tilde{p}_1$ and $\tilde{s}$ dependency. Moreover, one can combine the second and third terms in this expression by replacing $\tilde{p}_1$ by $\sqrt{\pi}\tilde{s}$ as the contribution from the variable $\tilde{u}_1$ is minor. Thus,
\be \label{eq:marginal_final}
    P(p_1) = \cl{N} \, \sum_{s\in\bb{Z}}\,\,
    e^{-\frac{\Delta^2+\kappa^2}{(\Delta^2+\kappa^2)^2+\Delta^4\kappa^4} (p_1-\sqrt{\pi}s)^2}\, 
    e^{-\frac{\Delta^2\kappa^2(\kappa^2+\Delta^2)}{(\Delta^2+\kappa^2)^2+\Delta^4\kappa^4} (\sqrt{\pi}s)^2}
\ee
where we can easily identify the peak’s variance and the inverse of the envelope’s variance
\be \label{eq:delta_peak_envl_full}
    \Delta_\n{peak}^2 = \frac{(\Delta^2+\kappa^2)^2+\Delta^4\kappa^4}{\Delta^2+\kappa^2}
    \qquad \textnormal{and} \qquad
    \Delta_\n{envl}^2 = \frac{\Delta^2\kappa^2(\kappa^2+\Delta^2)}{(\Delta^2+\kappa^2)^2+\Delta^4\kappa^4}\,.
\ee
Setting $\kappa=\Delta$ and taking their Taylor expansion for $\Delta^2\ll1$ leads to the results from the main text.

In Figure~\ref{fig:delta_envl_and_peaks} we show the correspondence between these analytical formula and the simulation data. Here, we fit using the least-squares method the following function to the rescaled marginal distributions  
\be \label{eq:fit_function}
    \mathsf{f}\big[\Delta_\n{fit},\kappa_\n{fit}\big](x,\n{parity}) = \sum_{s=-50}^{50} \,\,
    e^{-\kappa_\n{fit}^2 (\n{parity}\,s\sqrt{\pi})^2} \, e^{-\frac{1}{\Delta_\n{fit}^2}(x-\n{parity}\,s\sqrt{\pi})^2}
\ee
with $\Delta_\n{fit}$ and $\kappa_\n{fit}$ being the fitting parameters, $x$ representing the $q$ or $p$ coordinates and $\n{parity}\in\{1,2\}$ setting the spacing between the peaks to be $\sqrt{\pi}$ or $2\sqrt{\pi}$. We fit it to the $q$- and $p$-marginals before and after the application of the $\CZ$ gate for input states $\ket{00}_{\Delta,\kappa}$ with different $\Delta$ and $\kappa$ (denoted by $\Delta_\n{sim}$). For the momentum marginal distributions, we observe a good agreement between the theory and the simulations until $\Delta_\n{sim}<0.45$. To explain the discrepancy for smaller GKP states it would require taking into account contributions of the neighboring peaks in Eqs.~\eqref{eq:marg_integral_1},~\eqref{eq:marg_integral_2} and~\eqref{eq:marg_integral_3}. However, above this value the states loose their encoding power. Indeed, the logical states are less distinguishable and their overlap becomes larger. Moreover, most of the experimental realizations work with codes of size $\Delta,\kappa<0.4$, i.e., in the parameter regions where the above formulas approximate well the desired quantities.

\begin{figure}[t!]
    \centering
    \includegraphics{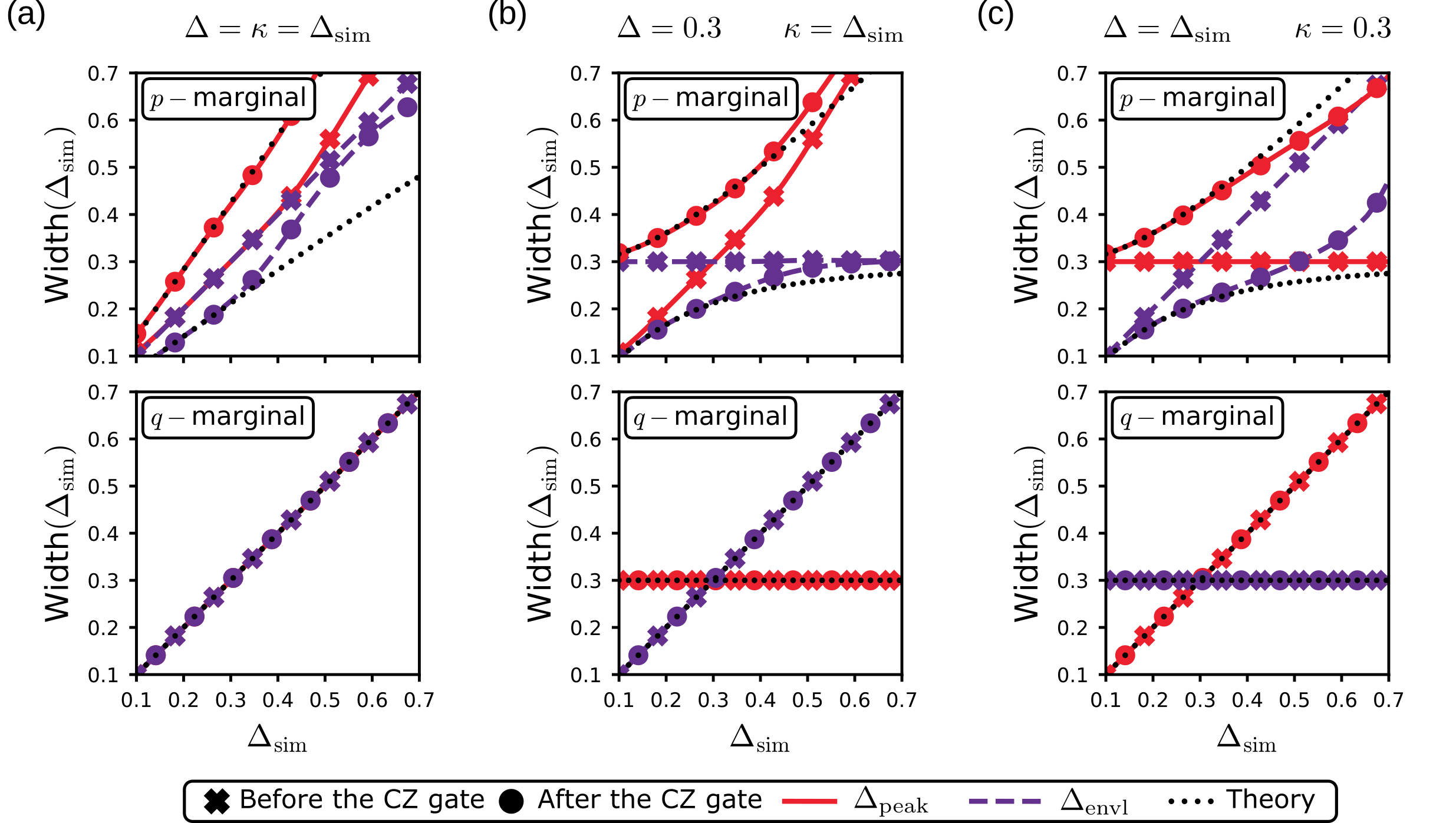}
    {\phantomsubcaption\label{fig:delta_envl_and_peaks_equal_kappa_delta}}
    {\phantomsubcaption\label{fig:delta_envl_and_peaks_fixed_delta}}
    {\phantomsubcaption\label{fig:delta_envl_and_peaks_fixed_kappa}}
    \caption{Broadening of the peaks and envelope widths of the $p$- and $q$-marginals after the $\CZ$ gate. The two widths are characterized by $\Delta_\n{peak}$ and $1/\Delta_\n{envl}$, respectively. The considered input state is $\ket{00}_{\Delta,\kappa}$ (cf. Eq~\eqref{eq:finite_com_basis}) with different $\Delta$ and $\kappa$. The simulated data have been obtained by fitting Eq.~\eqref{eq:fit_function} to the marginal distributions (cf. Section~\ref{sec:num_method}) using the least-squares method.}
    \label{fig:delta_envl_and_peaks}
\end{figure}

For the marginal distributions along the $q$ quadrature, the situation is different. Here, we observe that $\Delta_\n{peak}$ and $\Delta_\n{envl}$ do not vary throughout the process. This behavior can be inferred from the easily derived analytical expression of the distribution
\be
\begin{split}
    P(\tilde{q}_1)&=\bra{\tilde{q}_1}\n{Tr}_2\big[\ket{\Psi}\!\!\bra{\Psi}\big]\ket{\tilde{q}_1} \propto \\
    &\propto \iint\!\n{d}A'_2 \,\, \left| \iint\!\n{d}A_1 \,\, 
    \Psi(u_1,v_1,u'_2,v'_2+u_1) \sum_{s\in\bb{Z}} e^{i(v_1-u'_2)2\sqrt{\pi}s}
    \delta\!\left(\tilde{q}_1-2\sqrt{\pi}s-u_1\right)
    \right|^2 \propto \\
    &= \iint\!\n{d}A'_2 \,\, 
    f^2_\Delta(u'_2)\,g^2_\kappa(v'_2+\tilde{u}_1)\,f^2_\Delta(\tilde{u}_1)
    \Bigg| \varint\!\n{d}v_1\,\,e^{iv_1\,2\sqrt{\pi}\tilde{s}}\,g_\kappa(v_1)\, \Bigg|^2 \,.
\end{split}
\ee
Indeed, we note that the integral over $\n{d}A'_2$ is trivial and the rest leads to the same result as in Eq.~\eqref{eq:sgs_position_wf}. The peculiar characteristic of a two-qubit gate to unequally alter two orthogonal marginal distributions is to the best of our knowledge proper to GKP codes but can in principle appear in other continuous variable encodings.

\iftoggle{arXiv}{ \PRLsec{Effective squeezing parameters}{sec:eff_squeez_param}
}{ \section{Effective squeezing parameters} \label{sec:eff_squeez_param} }

As mentioned in the main text, the effective squeezing parameters represent an alternative quality measure of the grid states. They are defined as 
\begin{equation}\label{eq:eff_squeez_param}
    \sigma_{x/z} = \sqrt{\frac{1}{2\pi} \log{\big(|\text{Tr}\,\hat{S}_{x/z}\rho\,|^{-2}\big)}},
\end{equation}
with $\hat{S}_{x/z}$ being the position and momentum stabilizers of the code (cf. Section~\ref{sec:ideal_codes}). Eq.~\eqref{eq:eff_squeez_param} has been derived from the Holevo phase variance which measures how close $\rho$ is to an eigenstate of a certain operator (in our case $\h{S}_{x/z}$). For a derivation of this measure, we refer the reader to Refs.~\cite{terhal_encoding_2016,duivenvoorden_single-mode_2017,weigand_generating_2018}. The authors of these works show that these quantities reflect well the main parameters of a pure GKP state. Indeed, assuming that $\rho$ is in the state $\ket{\psi}_{\Delta,\kappa}$ it follows straightforwardly that $\sigma_x=\Delta$ and $\sigma_z\approx\kappa$. It is thus tempting to make a parallel between the effective squeezing parameters and $\Delta_\n{peak}$ and $\Delta_\n{envl}$. However, we must emphasize that this result is only valid for pure quantum states, an issue which will be discussed further below.

We now calculate $\sigma_{x/z}$ before and after the $\CZ$ gate. First, recall the partial trace property saying that ${\n{Tr}_1\big(\h{U}\,\n{Tr}_2\rho\big)=\n{Tr}(\h{U}\otimes\h{I})\rho}$ with $\h{U}$ -- a unitary operator acting exclusively on the first system. This means that we can derive the effective squeezing parameters for either the control or target systems directly, i.e., without primarily calculating the partial trace. Let us focus as in the previous section only on the control qubit with the initial state being the computational basis state $\ket{00}_{\Delta,\kappa}$. The stabilizer $\h{S}_z$ commutes with $\CZ$. It follows that the expectation value for $\h{S}_z \otimes \h{I}$ does not change under the application of the $\CZ$ gate, i.e., $\CZ^{\dagger} (\h{S}_z \otimes \h{I})\,\CZ = \h{S}_z \otimes \h{I}$. Therefore,  
\begin{equation}\label{eq:stabilizerSz}
    \n{Tr}\Big( (\h{S}_z \otimes \h{I})\, \CZ \,\rho_{\Delta,\kappa}\, \CZ^\dagger \Big) = 
    \n{Tr}\Big( (\h{S}_z \otimes \h{I})\, \rho_{\Delta,\kappa} \Big) \equiv
    \n{Tr}\Big( \h{S}_z \,\ket{0}\!\!\bra{0}_{\Delta,\kappa} \Big) =
    e^{-\Delta^2 \pi},
\end{equation}
where $\rho_{\Delta,\kappa}:=\ket{00}\!\!\bra{00}_{\Delta,\kappa}$ and the last equality follows straightforwardly from the representation of $\ket{0}_{\Delta,\kappa}$ given in Eq.~\eqref{eq:finite_com_basis} (see~\cite{duivenvoorden_single-mode_2017} for the derivation). Plugging this result into Eq.~\eqref{eq:eff_squeez_param}, we see that the effective squeezing parameter $\sigma_z=\Delta$ which means it remains invariant under application of the $\CZ$ gate. 

On the other hand, the stabilizer $\h{S}_x$ does not commute with the logical $\CZ$ operation. The conjugation of $S_x$ with $\CZ$ can readily be computed with the BCH formula giving $\CZ^{\dagger} (\h{S}_x \otimes \h{I})\,\CZ = \h{S}_x \otimes \h{S}_z$. Using this and the result from Eq.~\eqref{eq:stabilizerSz}, we find that
\begin{equation}\label{eq:stabilizerSx}
    \n{Tr}\Big( (\h{S}_x \otimes \h{I})\, \CZ \,\rho_{\Delta,\kappa}\, \CZ^\dagger \Big) = 
    \n{Tr}\Big( (\h{S}_x \otimes \h{S}_z)\, \rho_{\Delta,\kappa} \Big) \equiv
    \n{Tr}\Big( \h{S}_x \,\ket{0}\!\!\bra{0}_{\Delta,\kappa} \Big)
    \n{Tr}\Big( \h{S}_z \,\ket{0}\!\!\bra{0}_{\Delta,\kappa} \Big) \approx
    e^{-(\kappa^2 + \Delta^2) \pi}
\end{equation}
which in turn gives us
\begin{equation}\label{eq:eff_squeezing_after}
    \sigma_{x} = \sqrt{{\sigma'}_{x}^2 + \Delta^2} \approx \sqrt{\kappa^2 + \Delta^2}
\end{equation}
where the first equality would be the exact solution. Here, $\sigma'_{x}$ is the effective squeezing parameter associated to $\h{S}_x$ prior the $\CZ$ gate. For symmetric GKP states, i.e., $\Delta=\kappa$, we obtain the result form the main text. Note that Eq.~\eqref{eq:eff_squeezing_after} spotlights, that the effective squeezing parameter along the $p$ quadrature of the first system is influenced by the effective squeezing parameter along the $q$ quadrature of the second system.

Figure~\ref{fig:eff_squeez_param} shows the agreement between the simulated data and the analytical derivations for input states $\ket{00}_{\Delta,\kappa}$ with different $\Delta$ and $\kappa$ parameters. We observe that, as expected, the effective squeezing parameter associated to $\h{S}_x$ is left unchanged after the $\CZ$ gate, whereas $\sigma_x$ follows the expression from Eq.~\eqref{eq:eff_squeezing_after}. Interestingly, we note that this expression can be approximately equalized to $\Delta_\n{peak}$ given in Eq.~\eqref{eq:delta_peak_envl_full}. We must warn the reader that this quality measure of GKP states has to be taken carefully, meaning that $\sigma_z=\alpha$ and $\sigma_x=\beta$ do not necessarily correspond to a finite energy state of a form $\ket{\psi}_{\alpha,\beta}$ but can in fact represent a mixed state with such particular expectation values. In Section~\ref{sec:purity} we will indeed see that the traced-out post-$\CZ$ state is not pure.

It is worth emphasizing that even though the results above were derived for a particular input state the outcome is actually state independent thanks to the linearity of the trace and, as already mentioned, to the fact that for any $\ket{\psi}_{\Delta,\kappa}$ the effective parameters are $\sigma_x=\Delta$ and $\sigma_z\approx\kappa$.
\clearpage
\begin{figure}[t!]
    \centering
    \includegraphics{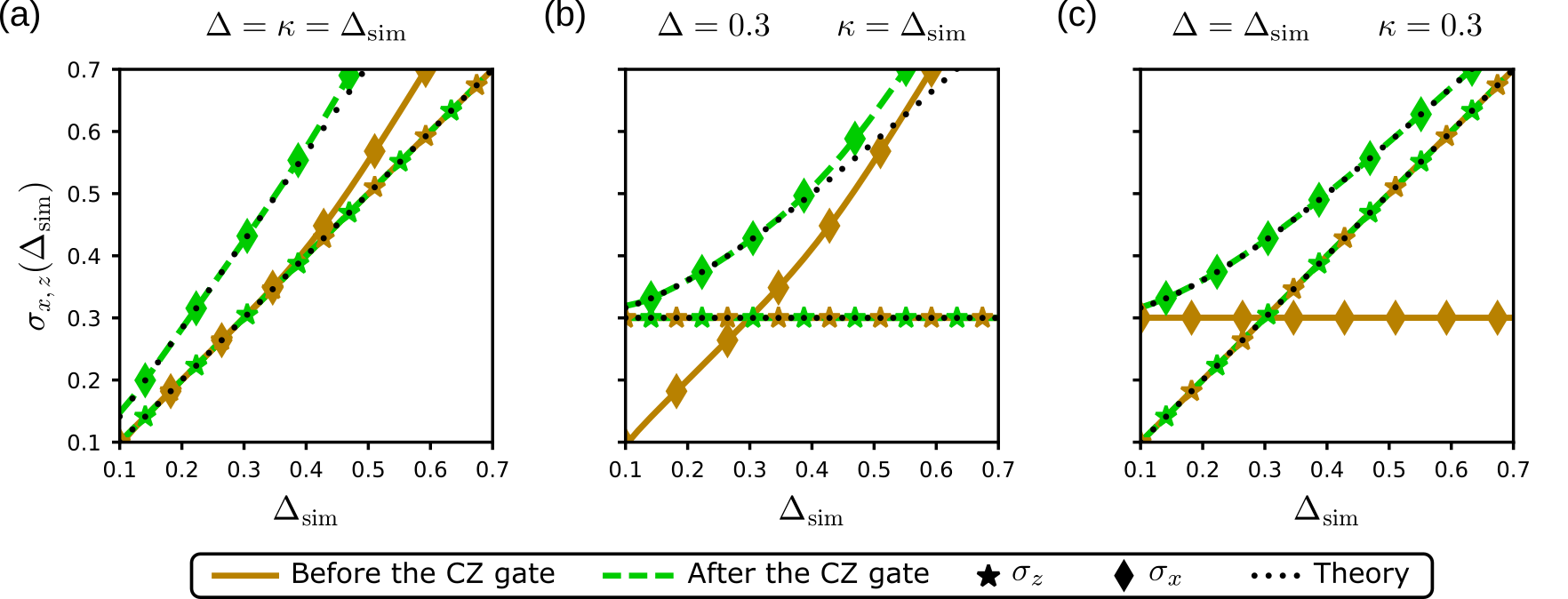}
    {\phantomsubcaption\label{fig:eff_squeez_param_equal_kappa_delta}}
    {\phantomsubcaption\label{fig:eff_squeez_param_fixed_delta}}
    {\phantomsubcaption\label{fig:eff_squeez_param_fixed_kappa}}
    \caption{Effective squeezing parameters $\sigma_{x/z}$ defined in Eq.~\eqref{eq:eff_squeez_param} before and after the application of the $\CZ$ gate. The input state was chosen to be $\ket{00}_{\Delta,\kappa}$ (cf. Eq~\eqref{eq:finite_com_basis}) with different $\Delta$ and $\kappa$ parameters. The two theoretical curves correspond to $\sigma_z=\Delta$ and $\sigma_x=\sqrt{\Delta^2+\kappa^2}$. The simulated data have been obtained by directly evaluating the expectation value of the respective stabilizer operators.}
    \label{fig:eff_squeez_param}
\end{figure}

\iftoggle{arXiv}{ \PRLsec{Physical fidelity}{sec:physical_fidelity}
}{ \section{Physical fidelity} \label{sec:physical_fidelity} }

We will now focus on the physical overlap fidelity of the two-qubit GKP gates. We define the physical overlap fidelity of a gate $\mathtt{G}$ as 
\be \label{eq:phys_fidelity_def}
    F\big(\mathtt{G},\ket{\psi_\n{in}},\ket{\psi_\n{out}}\big)=\big|\!\bra{\psi_\n{out}}\mathtt{G}\ket{\psi_\n{in}}\!\big|^2
\ee
with $\ket{\psi_\n{in}}$ and $\ket{\psi_\n{out}}$ being the input and the expected output states, respectively. This expression can also be seen as the success probability of the gate $\mathtt{G}$ for the input state $\ket{\psi_\n{in}}$. Thus, in what follows we will use it equivalently to fidelity. Moreover, we choose the input and output states to be the computational basis state $\ket{00}_{\Delta,\kappa}$ and the gate $\mathtt{G}$ to be $\CZ$ and abbreviate the quantity of interest simply as $F$. The analysis and result would be identical for the controlled NOT and other controlled Pauli gates thanks to the unitary equivalence of these operations. We discuss the overlap fidelity for other input states at the end of this section.

In an effort to simplify the derivation we split the calculation into a main part and a secondary one. The latter part refers to the normalization term which thus allows us to neglect constant terms in the primary part. Moreover, as for the marginal distributions (cf. Section~\ref{sec:marginal_distributions}), we treat this problem using a shifted grid state basis. We have previously determined the form of the state $\ket{\Psi}$ after the $\CZ$ (cf. Eq.~\eqref{eq:cz_action_on_00}). Therefore,
\be
\begin{split}
    \mathsf{Main} &= \prescript{}{\Delta,\kappa}{\bra{00}}\CZ\ket{00}_{\Delta,\kappa} = 
    \prescript{}{\Delta,\kappa}{\langle{00}\!}\ket{\Psi}= \\
    &= \iint\!\n{d}A_1\iint\!\n{d}A_2\iint\!\n{d}A'_1\iint\!\n{d}A'_2 \,\, 
    e^{iu_1u_2}\,\psi_0(u_1,v_1)\,\psi_0(u_2,v_2)\,\psi_0(u'_1,v'_1)\,\psi_0(u'_2,v'_2) \\
    &\hspace{140pt}\langle{u'_1,v'_1}\!\ket{u_1,v_1-u_2} \langle{u'_2,v'_2}\!\ket{u_2,v_2-u_1} = \\
    &= \iint\!\n{d}A_1\iint\!\n{d}A_2 \,\,
    e^{iu_1u_2}\,\psi_0(u_1,v_1)\,\psi_0(u_2,v_2)\,\psi_0(u_1,v_1+u_2)\,\psi_0(u_2,v_2+u_1) = \\
    &= \iint\!\n{d}A_1\iint\!\n{d}A_2 \,\, 
    e^{iu_1u_2}\,f_\Delta^2(u_1)\,f_\Delta^2(u_2)\,g_\kappa(v_1)g_\kappa(v_1+u_2)\,g_\kappa(v_2)g_\kappa(v_2+u_1).
\end{split}
\ee
Despite being not separable, these integrals can be solved consecutively starting from the $v_1$ and $v_2$ terms. We can notice that both integrals have the structure of a convolution of probability density functions of two wrapped normal distributions. Similar to their unwrapped analogue, these distributions are closed under convolution~\cite{mardia_directional_2000}. Hence,
\be
    h_\kappa(u) := \varint\!\n{d}v\,\, g_\kappa(v) g_\kappa(v+u) =
    (2\pi)(2\sqrt{\pi}\kappa^2)\,\,
    \mathsf{pdf}_{\cl{WN}[0,\sqrt{2}\,2\sqrt{\pi}\kappa]}\!(2\sqrt{\pi}u) = 
    \sqrt{\pi}\kappa \,\, \sum_{t\in\bb{Z}} \,\,e^{-\frac{1}{4\kappa^2}(u+\sqrt{\pi}t)}
\ee
where the pair $(u,v)$ corresponds either to $(u_2,v_1)$ or $(u_1,v_2)$. It is worth reminding the reader that here the functions $f_\Delta$ and $g_\kappa$ do not contain their respective normalization factors $(\sqrt{\pi}\Delta)^{1/4}$ and $(\sqrt{\pi}\Delta)^{1/4}$ as stated in Eq.~\eqref{eq:shifted_grid_repr}. We then proceed with the next integral
\be
\begin{split}
    \varint\!\n{d}u_1\,\, e^{iu_1u_2}\,f_\Delta^2(u_1)\,h_\kappa(u_1) &\approx
    \varint\!\n{d}u_1\,\, \sqrt{\pi}\kappa \,\, e^{iu_1u_2} \, \sum_{s,t\in\bb{Z}} \,\, 
    e^{-\frac{1}{\Delta^2}(u_1+2\sqrt{\pi}s)^2}\,
    e^{-\frac{1}{4\kappa^2}(u_1+\sqrt{\pi}t)} \approx \\
    &\approx \frac{2\pi \Delta\kappa^2}{\sqrt{\Delta^2+4\kappa^2}} \,
    e^{-\frac{\Delta^2\kappa^2}{\Delta^2+4\kappa^2}\,u_2^2}
    \left( 1 +
    e^{\pm\frac{i\sqrt{\pi}\Delta^2}{\Delta^2+4\kappa^2}\,u_2}\,
    e^{\frac{\pi}{\Delta^2+4\kappa^2}}
    \right),
\end{split}
\ee
where we first approximate the square of a sum (arising from $f_\Delta^2$) by a sum of squares (identically to Eq.~\eqref{eq:marg_integral_3}). The second approximation is that after integration we keep only the terms $(s=0,t=0)$ and $(s=0,t=\pm1)$. This assumption translates that the main contribution to the fidelity comes from the overlap of coinciding peaks (of the two grid states) as well as from their first neighboring peaks. This compact result allows us to solve the last integral. For conciseness, we divide it into two pieces:
\be \label{eq:succ_proba_integral_3_0}
\begin{split}
    t=0: \qquad &\varint\!\n{d}u_2\,\, 
    f_\Delta^2(u_2)\,h_\kappa(u_2)\,
    \frac{2\pi \Delta\kappa^2}{\sqrt{\Delta^2+4\kappa^2}}\, 
    e^{-\frac{\Delta^2\kappa^2}{\Delta^2+4\kappa^2}\,u_2^2}
    \approx \\
    &\approx\varint\!\n{d}u_2\,\, 
    \frac{2\pi \Delta\kappa^2}{\sqrt{\Delta^2+4\kappa^2}}\,\sqrt{\pi}\kappa\,
    \sum_{s,r\in\bb{Z}} \,\, 
    e^{-\frac{1}{\Delta^2}(u_2+2\sqrt{\pi}s)^2}\,
    e^{-\frac{1}{4\kappa^2}(u_2+\sqrt{\pi}r)}\, 
    e^{-\frac{\Delta^2\kappa^2}{\Delta^2+4\kappa^2}\,u_2^2} \approx\hspace{58pt}\\
    &\approx 
    \frac{4\pi^2 \Delta^2\kappa^4}{\sqrt{(\Delta^2+4\kappa^2)^2+\Delta^4\kappa^4}}\,
    \left(1+2\,
    e^{-\pi\frac{\Delta^2+4\kappa^2+\Delta^4\kappa^2}{(\Delta^2+4\kappa^2)^2+\Delta^4\kappa^4}}
    \right)
\end{split}
\ee
\be 
\begin{split}
    t=\pm1: \qquad &\varint\!\n{d}u_2\,\, 
    f_\Delta^2(u_2)\,h_\kappa(u_2)\,
    \frac{2\pi \Delta\kappa^2}{\sqrt{\Delta^2+4\kappa^2}}\, 
    e^{-\frac{\Delta^2\kappa^2}{\Delta^2+4\kappa^2}\,u_2^2}\,
    e^{\pm\frac{i\sqrt{\pi}\Delta^2}{\Delta^2+4\kappa^2}\,u_2}\,
    e^{\frac{\pi}{\Delta^2+4\kappa^2}}
    \approx\hspace{50pt} \\
    &\approx 
    \frac{4\pi^2 \Delta^2\kappa^4}{\sqrt{(\Delta^2+4\kappa^2)^2+\Delta^4\kappa^4}}\,
    \left(
    e^{-\pi\frac{\Delta^2+4\kappa^2+\Delta^4\kappa^2}{(\Delta^2+4\kappa^2)^2+\Delta^4\kappa^4}} +
    e^{-\pi\frac{\Delta^2+2i}{2\Delta^2\kappa^2 + i(\Delta^2+4\kappa^2)}} +
    e^{-\pi\frac{\Delta^2-2i}{2\Delta^2\kappa^2 - i(\Delta^2+4\kappa^2)}}
    \right),
\end{split}
\ee
where we made the same approximation for $f_\Delta^2$ and considered as before $s=r=0$ (coincident peaks) and $(s=0,r=1)$ (first neighboring peaks). Thus, the principal part of the fidelity expression reads
\be \label{eq:succ_proba_main}
    \mathsf{Main} \approx \frac{4\pi^2 \Delta^2\kappa^4}{\sqrt{(\Delta^2+4\kappa^2)^2+\Delta^4\kappa^4}}\,
    \left( 1 + 
    4\,e^{-\pi\frac{\Delta^2+4\kappa^2+\Delta^4\kappa^2}{(\Delta^2+4\kappa^2)^2+\Delta^4\kappa^4}} +
    2\,e^{-\pi\frac{\Delta^2+2i}{2\Delta^2\kappa^2 + i(\Delta^2+4\kappa^2)}} +
    2\,e^{-\pi\frac{\Delta^2-2i}{2\Delta^2\kappa^2 - i(\Delta^2+4\kappa^2)}}
    \right).
\ee
It is now important to include the secondary part which as mentioned gathers the normalization terms. In the overlap fidelity, the normalization constant plays an important role since it guarantees a probabilistic interpretation of this quantity. Indeed, $F$ characterizes the probability that the state $\CZ\ket{\psi_\n{in}}$ is identically equal to the state $\ket{\psi_\n{out}}$ (thus the name success probability). It is therefore important to have an accurate expression for this secondary part. We calculate it as following
\be
    \cl{N}_\mathsf{Main} = \prescript{}{\Delta,\kappa}{\bra{00}}{00}\rangle_{\Delta,\kappa} 
    = \iint\!\n{d}A_1\iint\!\n{d}A_2 \,\, 
    \psi_0^2(u_1,v_1)\,\psi^2_0(u_2,v_2) 
    = \iint\!\n{d}A_1\iint\!\n{d}A_2 \,\, 
    f_\Delta^2(u_1)\,f_\Delta^2(u_2)\,g_\kappa^2(v_1)\,g_\kappa^2(v_2).
\ee
These integrals are separable and we can solve them individually. Omitting details, we find that 
\be
    \varint\!\n{d}u\,\, f_\Delta^2(u) \approx \Delta\sqrt{\pi} (1 + 2\,e^{-\frac{\pi}{\Delta^2}})
    \qquad \text{and} \qquad
    \varint\!\n{d}v\,\, g_\kappa^2(v) \approx \kappa\sqrt{\pi} (1 + 2\,e^{-\frac{\pi}{4\kappa^2}}),
\ee
when once again taking into account the first neighboring peaks. Note that in the limit $\Delta,\kappa\ll1$, these expressions converge to the normalization constants of the shifted grid state representation of $\ket{00}_{\Delta,\kappa}$ that we defined in Eq.~\eqref{eq:shifted_grid_repr}. Knowing this the total secondary part equals
\be \label{eq:phys_fidelity_norm}
    \cl{N}_\mathsf{Main} = 
    \Delta^2\kappa^2\pi^2\,\left(1 + 2\,e^{-\frac{\pi}{\Delta^2}}\right)^2 \left(1 + 2\,e^{-\frac{\pi}{4\kappa^2}}\right)^2\,.
\ee
The success probability has at the end the following form
\be \label{eq:phys_fidelity_full}
    F = \left|\frac{\mathsf{Main}}{\cl{N}_\mathsf{Main}}\right|^2 = 
    \left|
    \frac{ 4\kappa^2
    \left( 1 + 
    4\,e^{-\pi\frac{\Delta^2+4\kappa^2+\Delta^4\kappa^2}{(\Delta^2+4\kappa^2)^2+\Delta^4\kappa^4}} +
    2\,e^{-\pi\frac{\Delta^2+2i}{2\Delta^2\kappa^2 + i(\Delta^2+4\kappa^2)}} +
    2\,e^{-\pi\frac{\Delta^2-2i}{2\Delta^2\kappa^2 - i(\Delta^2+4\kappa^2)}}
    \right)
    }{
    \sqrt{(\Delta^2+4\kappa^2)^2+\Delta^4\kappa^4}
    \left(1 + 2\,e^{-\frac{\pi}{\Delta^2}}\right)^2
    \left(1 + 2\,e^{-\frac{\pi}{4\kappa^2}}\right)^2}
    \right|^2\,.
\ee
In the case of a symmetric GKP state, this expression boils down to
\be
    F = 
    \left|
    \frac{4}{\sqrt{25+\Delta^4}} \,
    \frac{1 + 
    4\,e^{-\pi\frac{5+\Delta^4}{\Delta^2(25+\Delta^4)}} +
    2\,e^{-\pi\frac{\Delta^2+2i}{2\Delta^4 + 5i\,\Delta^2 }} +
    2\,e^{-\pi\frac{\Delta^2-2i}{2\Delta^4 - 5i\,\Delta^2 }}
    }{
    \left(1 + 2\,e^{-\frac{\pi}{\Delta^2}}\right)^2
    \left(1 + 2\,e^{-\frac{\pi}{4\Delta^2}}\right)^2}
    \right|^2
\ee
which can then be approximated up to $\cl{O}(\Delta^3)$ by the expression in the main text.

Figure~\ref{fig:fidelity_vs_delta_kappa} illustrates the compatibility of Eq.~\eqref{eq:phys_fidelity_full} with the simulated overlap fidelity. We see that in the free regimes presented there the analytical solution agrees very well with simulations for $\Delta,\kappa<0.6$ which is a larger domain of validity than for the marginal distributions (cf. Section~\ref{sec:marginal_distributions}) where we took more restrictive assumptions. It is however crucial to emphasize that the shifted grid state methodology utilized here enabled us to achieve a high degree of precision in approximating the true fidelity even for energy parameters $\Delta,\,\kappa$ that are significantly larger than those typically found in experimental realizations of GKP states. This demonstrates its potential as a reliable method for analyzing other GKP operations in the future. Figure~\ref{fig:fidelity_vs_delta_kappa} also reveals a characteristic difference between varying only the $\Delta$ or only the $\kappa$ parameters. In the former case the fidelity decreases monotonically for increasing $\Delta$, whereas in the latter case it is instead monotonically increasing for increasing $\kappa$. This behavior can be explained by the fact that throughout the $\CZ$ gate finite-energy effects arise exclusively in the $p$-quadrature of the GKP states, thus narrowing the envelope of this quadrature (i.e., increasing $\kappa$) allows to mitigate the effects and get higher probabilities (see ``$\Delta=0.3 \quad \kappa=\Delta_\mathrm{sim}$'' curve in Figure~\ref{fig:fidelity_vs_delta_kappa}). However, higher values of $\kappa$ also lead to the broadening of the peaks in the $q$-quadrature which consequently makes the logical states less distinguishable and lowers the quality of the encoded information.

\begin{SCfigure}[][t!]
    \centering
    \includegraphics{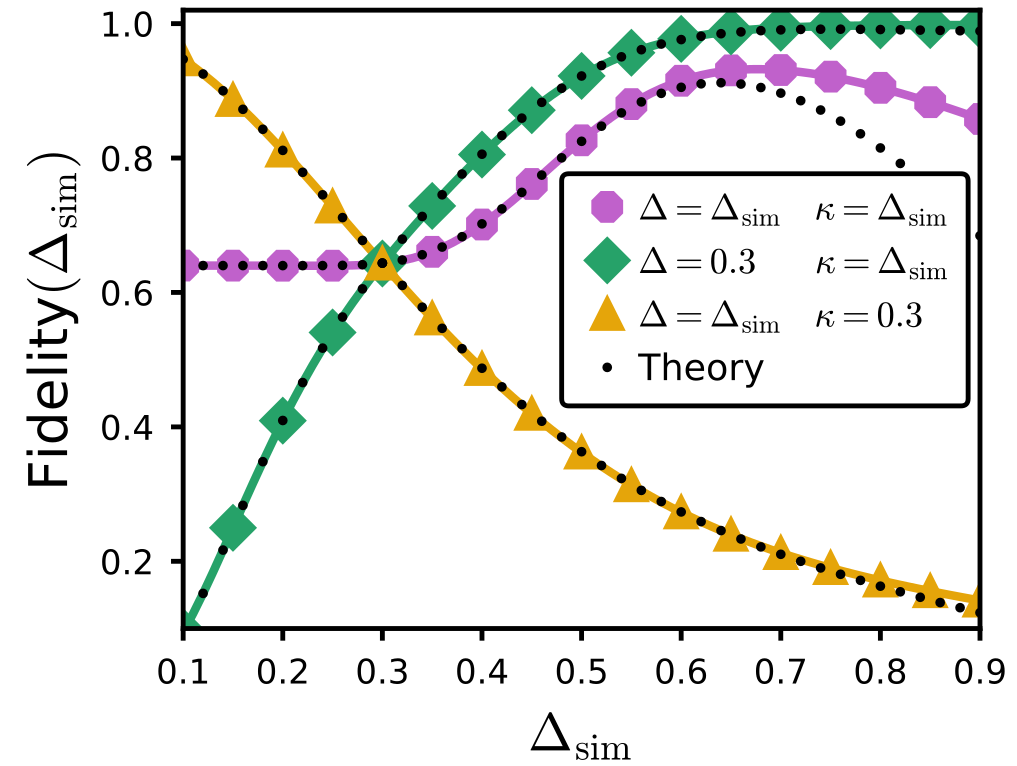}
    \caption{Physical overlap fidelity $F$ as defined in Eq~\eqref{eq:phys_fidelity_def} for input states $\ket{00}_{\Delta,\kappa}$ (cf.~ Eq~\eqref{eq:finite_com_basis}) with different $\Delta$ and $\kappa$ parameters. The dotted curve follow the analytical approximation given in Eq.~\eqref{eq:phys_fidelity_full}.}
    \label{fig:fidelity_vs_delta_kappa}
\end{SCfigure}

\subsec{Fidelity for other input states}{subsec:fidel_other_states}

The overlap fidelity for other states that $\ket{00}_{\Delta,\kappa}$ considered so far can be obtained using the similar procedure but with different input functions than $\psi_0(u,v)$ from Eq.~\eqref{eq:shifted_grid_repr}. However, thanks to the pseudo-periodicity of $\ket{u,v}$ states and to the observation that the wave function of $\ket{1}_{\Delta,\kappa}$ written in the shifted grid state basis reads ${\psi_1(u,v)\equiv\psi_0(u+\sqrt{\pi},v)}$ we infer that the normalization is simply modified by a phase factor
\be
    \cl{N}_1 :=  \prescript{}{\Delta,\kappa}{\bra{1}}{1}\rangle_{\Delta,\kappa} =
    \pi\Delta\kappa\,\left(1 + 2\,e^{-\frac{\pi}{\Delta^2}}\right) \left(1 - 2\,e^{-\frac{\pi}{4\kappa^2}}\right)\,.
\ee
This phase factor will also be found in the rest of the integrals presented above such that the approximated success probability functions for the states $\ket{11}_{\Delta,\kappa}$ and $\ket{01}_{\Delta,\kappa}$ read
\be \label{eq:phys_fidelity_full_11}
    F\big(\CZ,\ket{11}_{\Delta,\kappa},\ket{11}_{\Delta,\kappa}\big) =
    \frac{16\kappa^4}{(\Delta^2+4\kappa^2)^2+\Delta^4\kappa^4}
    \frac{
    \left( 1 - 
    4\,e^{-\pi\frac{\Delta^2+4\kappa^2+\Delta^4\kappa^2}{(\Delta^2+4\kappa^2)^2+\Delta^4\kappa^4}}
    \right)^2
    }{
    \left(1 + 2\,e^{-\frac{\pi}{\Delta^2}}\right)^4
    \left(1 - 2\,e^{-\frac{\pi}{4\kappa^2}}\right)^4} \,,
\ee
\be \label{eq:phys_fidelity_full_01}
    F\big(\CZ,\ket{01}_{\Delta,\kappa},\ket{01}_{\Delta,\kappa}\big) = 
    \frac{16\kappa^4}{(\Delta^2+4\kappa^2)^2+\Delta^4\kappa^4}
    \frac{
    \left( 1 - 
    16\,e^{-\pi\frac{\Delta^2+4\kappa^2+\Delta^4\kappa^2}{(\Delta^2+4\kappa^2)^2+\Delta^4\kappa^4}}
    \right)
    }{
    \left(1 + 2\,e^{-\frac{\pi}{\Delta^2}}\right)^4
    \left(1 - 4\,e^{-\frac{\pi}{2\kappa^2}}\right)^2} \,.
\ee
The function for the $\ket{10}_{\Delta,\kappa}$ state will be identical to Eq.~\eqref{eq:phys_fidelity_full_01}. Note that in these expressions the prefactor is the same as in Eq.~\eqref{eq:phys_fidelity_full} which suggests that for $\Delta^2,\kappa^2\ll1$ this overlap fidelity will be the same for any two-qubit input state $\ket{\psi}_{\Delta,\kappa}$. In the particular case of symmetric GKP states, i.e., $\Delta=\kappa$, this constant will be equal to $16/25$. Figure~~\ref{fig:fidelity_vs_delta_kappa_other_states} compares the analytical expressions in Eqs.~\eqref{eq:phys_fidelity_full_11} and~\eqref{eq:phys_fidelity_full_01} with the simulated overlap fidelity and shows a good agreement for $\Delta_\n{sim}<0.45$ which is beyond the region of interest for $\Delta$ and $\kappa$.

\begin{SCfigure}[][t!]
    \centering
    \includegraphics{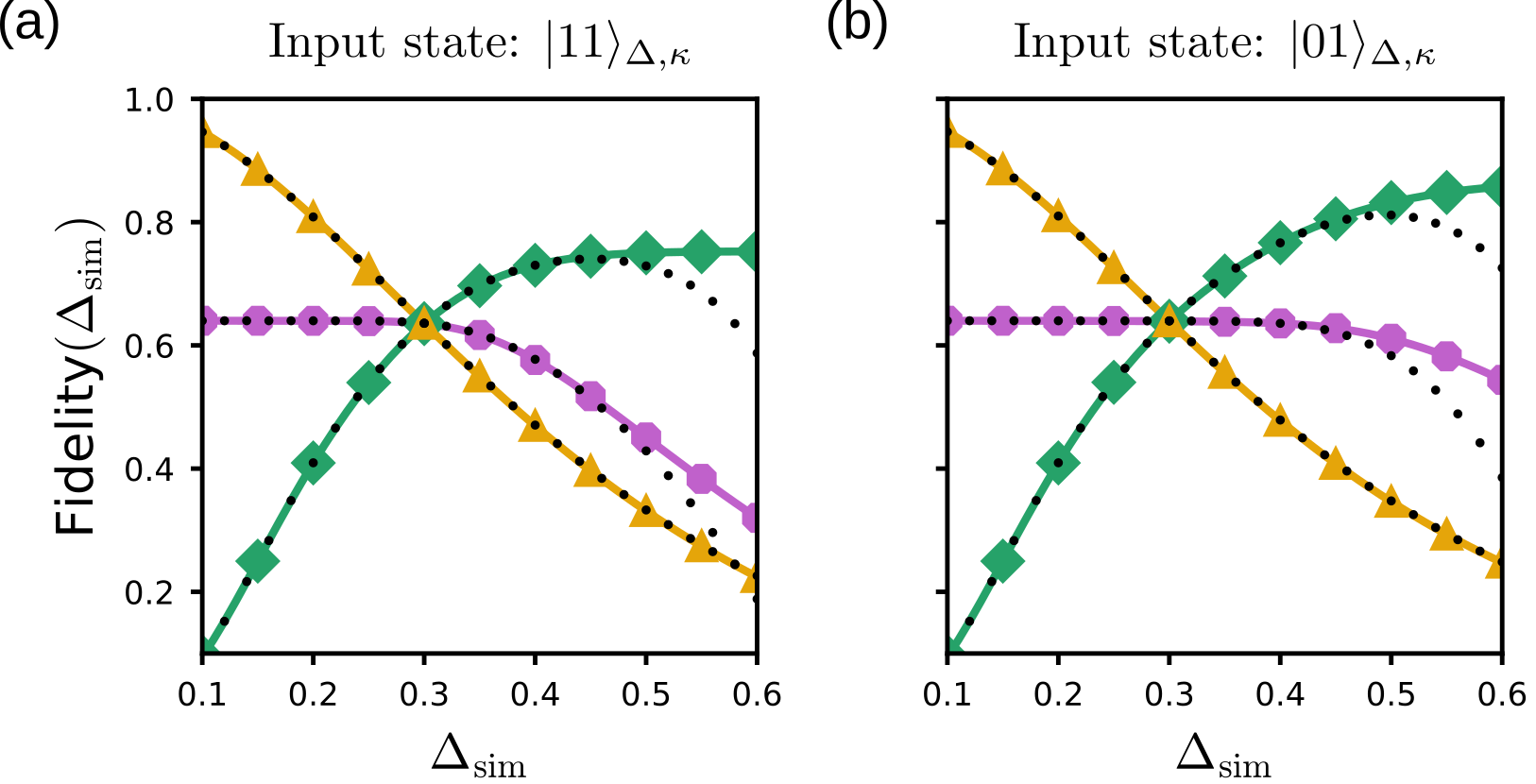}
    \caption{Physical overlap fidelity $F$ as defined in Eq~\eqref{eq:phys_fidelity_def} for input states $\ket{11}_{\Delta,\kappa}$ and $\ket{01}_{\Delta,\kappa}$ with different $\Delta$ and $\kappa$ parameters. The three curves correspond to the three situations presented in Figure~\ref{fig:fidelity_vs_delta_kappa} whereas the dotted curves follow the analytical approximations given in Eqs.~\eqref{eq:phys_fidelity_full_11} and~\eqref{eq:phys_fidelity_full_01}.}
    \label{fig:fidelity_vs_delta_kappa_other_states}
\end{SCfigure}

\iftoggle{arXiv}{ \PRLsec{Purity}{sec:purity} 
}{ \section{Purity} \label{sec:purity} }

As already mentioned previously the purity of the traced-out output state does not follow the behavior that is expected from the standard $\CZ$ operation in two-level systems. The purity of a quantum state $\rho$ is defined as $\mathsf{P}(\rho):=\n{Tr}\big[\rho^2\big]$ and evaluates how mixed the given state is. For a pure state, this measure would be equal to 1 whereas for a maximally mixed state in an $d$ dimensional Hilbert space it would be $1/d$. In our case we are interested in the purity of the traced-out system after the $\CZ$ gate, in other words $\rho=\n{Tr}_2\big[\ket{\Psi}\!\!\bra{\Psi}\big]$ where $\ket{\Psi}=\CZ\ket{00}_{\Delta,\kappa}$.

In the discrete variable case, the $\CZ$ gate does not create any entanglement between the two systems when those are both in the computational state $\ket{0}$, that is the joint state $\ket{00}$ is left invariant under these operations. Thus, the output state is still separable and the state of each of the subsystems remains pure. Conversely, the GKP version of these gates creates some undesired entanglement between the two bosonic modes, hence lowering the purity of each subsystem state.

We can derive the purity in a similar manner to the marginal distributions and to the fidelity presented in Sections~\ref{sec:marginal_distributions} and~\ref{sec:physical_fidelity}, respectively. For the state $\rho=\n{Tr}_2\big[\ket{\Psi}\!\!\bra{\Psi}\big]$ it reads
\be \label{eq:purity_integral}
\begin{split}
    \mathsf{P}\left(\rho\right)
    &=\iint\!\n{d}A_1\iint\!\n{d}A_2\iint\!\n{d}A'_1\iint\!\n{d}A'_2 \\
    &\hspace{20pt}
    e^{i(u_1-u'_1)u_2}\,
    \psi_0(u_1,v_1+u_2)\,\psi_0(u_2,v_2+u_1)\,
    \psi^*_0(u'_1,v'_1+(u_2-u'_2))\,\psi^*_0(u_2,v_2+u'_1)\, \\
    &\hspace{20pt}
    e^{i(u'_1-u_1)u'_2}\,
    \psi_0(u'_1,v'_1)\,\psi_0(u'_2,v'_2+u'_1)\,
    \psi^*_0(u_1,v_1+u'_2)\,\psi^*_0(u'_2,v'_2+u_1)=\\
    &=\iint\!\n{d}A_1\iint\!\n{d}A_2\iint\!\n{d}A'_1\iint\!\n{d}A'_2 \,\, e^{i(u_1-u'_1)(u_2-u'_2)} \\
    &\hspace{20pt}
    f_\Delta^2(u_1)\,f_\Delta^2(u'_1)\,g_\kappa(v_2+u_1)g_\kappa(v_2+u'_1)\,g_\kappa(v'_2+u_1)g_\kappa(v'_2+u'_1)\\
    &\hspace{20pt}
    f_\Delta^2(u_2)\,f_\Delta^2(u'_2)\,g_\kappa(v_1+u_2)g_\kappa(v_1+u'_2)\,g_\kappa(v'_1+(u_1-u'_1))g_\kappa(v'_1) \,,
\end{split}
\ee
where $\psi_0$ is again the wave function of $\ket{\Psi}$ in the shifted grid state representation given in Eq.~\ref{eq:shifted_grid_repr}. As for the overlap fidelity calculation, we note the presence of convolutions in the $g_\kappa$ functions. Using the same assumptions as in the previous sections, we conclude that the purity follows
\be \label{eq:purity_formula}
    \mathsf{P}\!\left(\rho\right) = 
    \frac{ \kappa^2
    \left( 1 + 
    4\,e^{-\frac{\pi}{2}\frac{\Delta^2+\kappa^2+\Delta^4\kappa^2}{(\Delta^2+\kappa^2)^2+\Delta^4\kappa^4}} +
    2\,e^{-\pi\frac{\Delta^2+i}{\Delta^2\kappa^2 + i(\Delta^2+\kappa^2)}} +
    2\,e^{-\pi\frac{\Delta^2-i}{\Delta^2\kappa^2 - i(\Delta^2+\kappa^2)}}
    \right)
    }{
    \sqrt{(\Delta^2+\kappa^2)^2+\Delta^4\kappa^4}
    \left(1 + 2\,e^{-\frac{\pi}{\Delta^2}}\right)^2
    \left(1 + 2\,e^{-\frac{\pi}{4\kappa^2}}\right)^2}
    \,.
\ee
For symmetric GKP states (i.e., $\Delta=\kappa$) and in the limit of large squeezing (i.e., $\Delta,\kappa\rightarrow0$), the purity converges to a value of 1/2. This this also the behavior we observe in the simulations. The results are presented in Figure~\ref{fig:purity_vs_delta_kappa}. We note that the analytical approximation has a smaller validity range. Indeed, Eq.~\eqref{eq:purity_formula} remains valid for $\Delta,\kappa\leq0.3$. This is explained by the fact that the purity depends on the square of the density matrix which enhances the error of the approximation. However, a better analytical approximation can be constructed if one takes into account further neighboring peaks in the infinite sums that compose the shifted grid state representation of $\ket{00}_{\Delta,\kappa}$.

\begin{SCfigure}[][t!]
    \centering
    \includegraphics{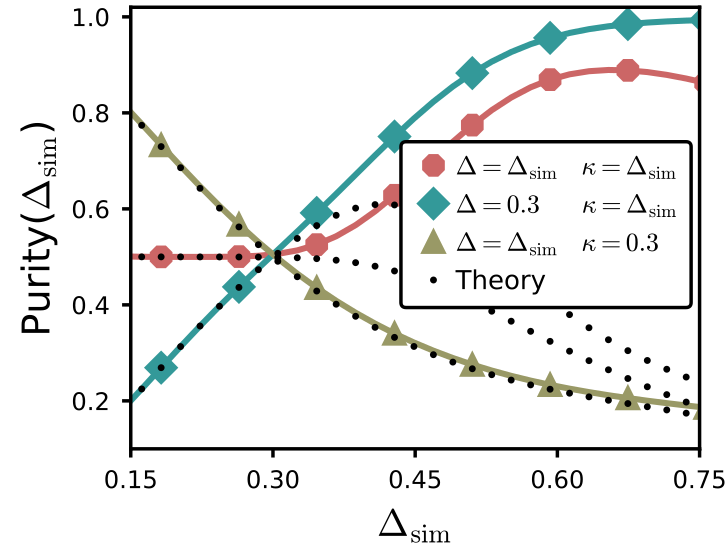}
    \caption{Purity $\mathsf{P}(\rho)$ of the first subsystem after the application of the $\CZ$ gate. The input states are $\ket{00}_{\Delta,\kappa}$ (cf. Eq~\eqref{eq:finite_com_basis}) with different $\Delta$ and $\kappa$ parameters. The dotted curve follow the analytical approximation given in~Eq.~\eqref{eq:purity_formula}.}
    \label{fig:purity_vs_delta_kappa}
\end{SCfigure}

\iftoggle{arXiv}{ \PRLsec{Gates decompositions}{sec:gate_decompositions}
}{ \section{Gates decompositions}\label{sec:gate_decompositions} }

In this section, we now focus on the decomposition of the presented entangling operations in terms of well established linear optical elements such as squeezers and beamsplitters. We distinguish two type of constructions: exact and approximate ones. The former realizations implement the desired gates identically and are based on the so-called Bloch-Messiah reduction~\cite{bloch_canonical_1962,braunstein_squeezing_2005,cariolaro_bloch-messiah_2016}. The approximate decomposition gives an alternative perspective where the two-qubit gate is realized for certain limiting parameters. In the last subsection we discuss the advantages and disadvantages of both methods.

\subsec{Exact gate decompositions}{sec:exact_gate_decomp}

\citet{braunstein_squeezing_2005} established that any linear multimode unitary Bogoliubov transformation can be realized identically using two multiport linear interferometers (a.k.a.~beamsplitters) enclosing a layer of single-mode squeezers. The decomposition of the original transformation is done using the Bloch-Messiah reduction theorem~\cite{bloch_canonical_1962} which in sum corresponds to a joint singular value decomposition of the transformation matrices~\cite{cariolaro_bloch-messiah_2016}.

Given the fact that the $\CNOT$ and $\CZ$ gates are both linear Bogoliubov transformations, we can then determine their exact decompositions. Regarding the $\CNOT$, we report here the result from \citet{terhal_towards_2020}. After performing the Bloch-Messiah reduction, the authors obtain the following expression (written in the symplectic form)
\begin{equation}\label{eq:bloch_messiah_CNOT}
    S_\CNOT =
    \frac{1}{\sqrt{2}}
    \begin{pmatrix}
        s_\theta & c_\theta & s_\theta & -c_\theta \\
        -c_\theta & s_\theta & c_\theta & s_\theta \\
        c_\theta & -s_\theta & c_\theta & s_\theta \\
        s_\theta & c_\theta & -s_\theta & c_\theta 
    \end{pmatrix}
    \begin{pmatrix}
        e^{r} & 0 & 0 & 0  \\
        0 &  e^{-r} & 0 & 0  \\
        0 &  0 & e^{r} & 0 \\
        0 &  0 &  0 & e^{-r}
    \end{pmatrix}
    \frac{1}{\sqrt{2}}
    \begin{pmatrix}
        c_\theta & -s_\theta & c_\theta & s_\theta \\
        s_\theta & c_\theta & -s_\theta & c_\theta \\
        s_\theta & c_\theta & s_\theta & -c_\theta \\
        -c_\theta & s_\theta & c_\theta & s_\theta 
    \end{pmatrix}
\end{equation}
where $c_\theta$ and $s_\theta$ are $\cos(\theta)$ and $\sin(\theta)$, respectively. The parameter values for which this equality holds follow from the method and are equal to $\theta = 1/2\,\n{arcsin}\!\left(2/\sqrt{5}\right)$ and $r = \n{arccosh}\!\left(\sqrt{5}/2\right)$. They can also be obtained from the symplectic form of the $\CNOT$ gate given in Eq.~\eqref{eq:CNOT_phase_space}. An alternative trigonometrically equivalent expression for these parameters is given by $\theta=1/2\,\n{arccot}\!\left(1/2\right)$ and $r=\n{arcsinh}\!\left(1/2\right)$.

We note that the decomposition given in Eq.~\eqref{eq:bloch_messiah_CNOT} can be rewritten using a layer of squeezers $\hat{S}(-r)\otimes \hat{S}(-r)$ and two 50:50 beamsplitters of the form~\cite{braunstein_squeezing_2005,terhal_towards_2020}
\begin{equation}
    \begin{pmatrix} \h{a}_c \\ \h{a}_t \end{pmatrix}
    \mapsto \frac{1}{\sqrt{2}}
    \begin{pmatrix}
        -i e^{i\theta} & i e^{-i\theta} \\
           e^{i\theta} &   e^{-i\theta} \\
    \end{pmatrix}
    \begin{pmatrix} \h{a}_c \\ \h{a}_t \end{pmatrix}
    \qquad\textnormal{and}\qquad
    \begin{pmatrix} \h{a}_c \\ \h{a}_t \end{pmatrix}
    \mapsto \frac{1}{\sqrt{2}}
    \begin{pmatrix}
           e^{i\theta} &   e^{-i\theta} \\
        -i e^{i\theta} & i e^{-i\theta} \\
    \end{pmatrix}
    \begin{pmatrix} \h{a}_c \\ \h{a}_t \end{pmatrix}\,,
\end{equation}
for the left and right hand-side interferometers, respectively. Here, $\h{a}_c$ and $\h{a}_t$ indicate the creation operators of the control and target modes. This form of interactions is common for photonic architectures. However, we can also reformulate it in terms of phase space rotations $\h{R}$ and anti-symmetric beamsplitters $\hat{B}_A$ introduced in Eqs.~\eqref{eq:rotation_operator} and~\eqref{eq:beamsplitter_operators}. The whole scheme would thus have the following representation:
\begin{equation}\label{eq:bloch_messiah_plus_rot_CNOT}
    \CNOT = 
    \hat{B}_A\!\left(\frac{\pi}{4}\right)
    \left(\hat{R}(\theta)\otimes \hat{R}\left(\frac{\pi}{2}-\theta\right)\right)
    \left(\hat{S}(-r)\otimes \hat{S}(-r)\right)
    \hat{B}_A\!\left(-\frac{\pi}{4}\right)
    \left(\hat{R}\left(\theta-\frac{\pi}{2}\right)\otimes \hat{R}(-\theta)\right).
\end{equation}
Since the rotations are not strictly speaking operations but only changes of the oscillator's reference frame, the $\CNOT$ gate can in practice be realized using only three main unitaries. If one absorbs these rotations into the beamsplitter interaction and use asymmetric squeezing of the control and target modes, one would then obtain the decomposition suggested in Ref.~\cite{tzitrin_progress_2020}
\begin{equation}\label{eq:CNOT_tzitrin}
    \CNOT = \hat{B}_S\!\left(\theta + \frac{\pi}{2}\right)\left(\hat{S}(-r)\otimes \hat{S}(r)\right)\hat{B}_S(\theta)\,.
\end{equation}

\begin{figure*}[b]
    \centering
    \includegraphics[scale=1.2]{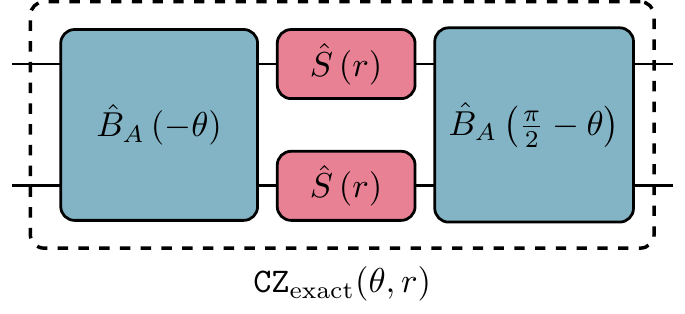}
    \qquad
    \includegraphics[scale=1.2]{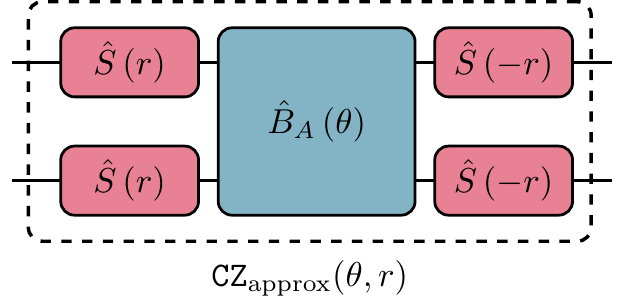}
    \caption{Gate based representation of the exact (left) and approximate (right) decompositions of the $\CZ$ gate as defined in Eqs.~\eqref{eq:CNOT_tzitrin} and~\eqref{eq:CZ_CNOT_approx_decomp}, respectively. The elements composing both decompositions are the squeezing $\h{S}(r)$ and anti-symmetric beamsplitter $\h{B}_A(\theta)$ operations (cf. Eqs.~\eqref{eq:squeezing_operator} and~\eqref{eq:beamsplitter_operators}). The right parameters $(r,\theta)$ for realizing the desired gate are given in the text. }
    \label{fig:decompositions_gates_diagram}
\end{figure*}

Regarding the $\CZ$ gate, we perform again the Bloch-Messiah reduction and obtain the following symplectic expression
\begin{equation}\label{eq:bloch_messiah_CZ}
    S_\CZ =
    \begin{pmatrix}
        0 & -s_\theta & c_\theta & 0 \\
        s_\theta & 0 & 0 & c_\theta  \\
        c_\theta & 0 & 0 & -s_\theta \\
        0 & c_\theta & s_\theta & 0  
    \end{pmatrix}
    \begin{pmatrix}
        e^{-r} & 0 & 0 & 0  \\
        0 &  e^{r} & 0 & 0  \\
        0 &  0 & e^{-r} & 0 \\
        0 &  0 &  0 & e^{r}
    \end{pmatrix}
    \begin{pmatrix}
        0 & -s_{\theta-\frac{\pi}{2}} & c_{\theta-\frac{\pi}{2}} & 0 \\
        s_{\theta-\frac{\pi}{2}} & 0 & 0 & c_{\theta-\frac{\pi}{2}}  \\
        c_{\theta-\frac{\pi}{2}} & 0 & 0 & -s_{\theta-\frac{\pi}{2}} \\
        0 & c_{\theta-\frac{\pi}{2}} & s_{\theta-\frac{\pi}{2}} & 0  
    \end{pmatrix}\,,
\end{equation}
where the two parameters take the same values as for the $\CNOT$ decompositions which is not surprising since as stated previously $\CNOT$ and $\CZ$ are unitarily equivalent. Contrastingly, we have here an identical multiport interferometer before and after the layer of single-mode squeezers. 

The more compact form of this decomposition reads
\begin{equation}\label{eq:CZ_exact_decomp}
    \CZ = 
    \hat{B}_A\!\left(-\theta\right)
    \left(\hat{S}(r)\otimes \hat{S}(r)\right)
    \hat{B}_A\!\left(-\theta+\frac{\pi}{2}\right)\,.
\end{equation}

The decompositions of the $\CNOT$ and $\CZ$ gates given in Eqs.~\eqref{eq:CNOT_tzitrin} and~\eqref{eq:CZ_exact_decomp} will be from now referred as $\CNOT_\n{exact}(\theta,r)$ and $\CZ_\n{exact}(\theta,r)$ as opposed to their approximate decompositions derived in the next subsection.

\subsec{Approximate gate decompositions}{sec:approximate_gate_decomposition}

To derive the the approximate decompositions we make use of the BCH formula by enclosing a beamsplitter between two single-mode squeezers. For the $\CNOT$ and $\CZ$ gates we choose the symmetric and anti-symmetric beamsplitter operators, respectively, and obtain
\begin{equation}\label{eq:CZ_CNOT_approx_decomp}
    \begin{split}
    \CNOT_\n{approx}(\theta,r) &= \left(\hat{S}(r)\otimes \hat{S}(-r)\right) \hat{B}_S(\theta) \left(\hat{S}(-r)\otimes \hat{S}(r)\right) = 
    e^{i\theta(e^{2r}\q_1\p_2 - e^{-2r} \p_1\q_2)}, \\
    \CZ_\n{approx}(\theta,r) &= \left(\hat{S}(r)\otimes \hat{S}(r)\right) \hat{B}_A(\theta) \left(\hat{S}(-r)\otimes \hat{S}(-r)\right) =
    e^{i\theta(e^{2r}\q_1\q_2 + e^{-2r} \p_1\p_2)}.
    \end{split}
\end{equation}
Notice that the squeezers are here selected as in the exact decompositions given in Eqs.~\eqref{eq:CNOT_tzitrin} and~\eqref{eq:CZ_exact_decomp}. It can be understood as following: the first pair of squeezers enhances the position or momentum probability distribution of the input state. This effectively ensures that the state becomes an eigenstate of $\q_1\p_2$ (for the $\CNOT$ gate) or $\q_1\q_2$ (for the $\CZ$ gate) operators. In the second step, the beamsplitter operators $\hat{B}_S(\theta)$ or $\hat{B}_A(\theta)$ realize the desired gates. Finally, the second layer of squeezers defocuses the quadrature distributions and brings them back into their initial configuration. From Eq.~\eqref{eq:CZ_CNOT_approx_decomp}, it is thus straightforward to determine the parameters $(\theta,r)$ that would realize our $\CNOT$ and $\CZ$ gates, namely we would like $\theta=e^{-2r}$ and $r\gg1$ such that the undesired operators are exponentially suppressed compared to the wanted ones.

\begin{figure*}[b]
    \centering
    \includegraphics{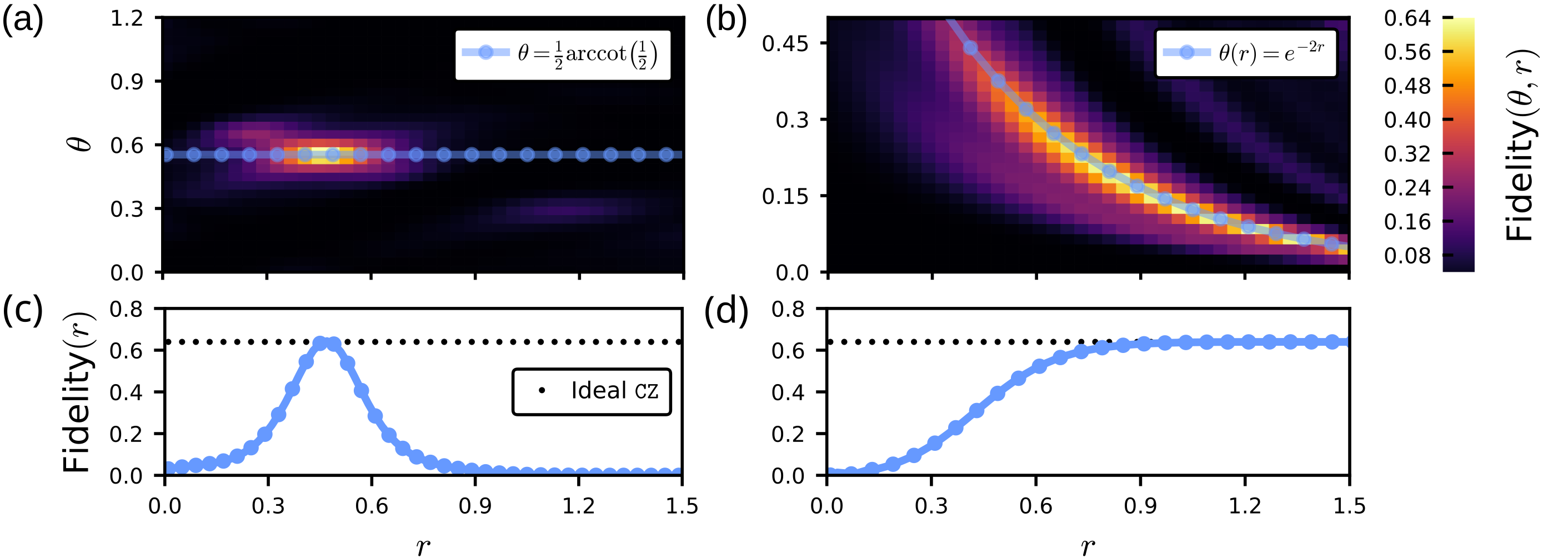}
    {\phantomsubcaption\label{fig:heatmap_exact}}
    {\phantomsubcaption\label{fig:heatmap_approximate}}
    {\phantomsubcaption\label{fig:theta_slice_exact}}
    {\phantomsubcaption\label{fig:theta_slice_approximate}}
    \caption{Exact (left) and approximate (right) decompositions of the $\CZ$ gate fidelity. Plots (a) and (b) illustrate the overlap fidelity between the input state subjected to $\CZ(\theta,r)$ and the desired output state in function of the beamsplitter $\theta$ and  squeezing $r$ parameters. The simulations have been performed with the Bell state $\ket{\Phi_{+}}_\Delta$ in input and  $\ket{\Phi_1}_\Delta$ as the desired one, both with an finite-energy parameter $\Delta=0.3$. Plots (c) and (d) show cross-sections of the respective heatmap for specific $\theta$.}
    \label{fig:decompositions_fidelity_heatmap}
\end{figure*}

This condition for the optimal beamsplitter and squeezing parameters can also be obtained using the symplectic representation of the two optical linear elements. The symplectic forms of the products squeezer--beamsplitter--squeezer given above read (for readability issue we dropped the subscript ``approx'')
\[
    S_{\CNOT(\theta,r)} =
    \begin{pmatrix}
        \cos{\theta} & 0 & e^{-2r}\sin{\theta} & 0 \\
        0 & \cos{\theta} & 0 & e^{2r}\sin{\theta} \\
        -e^{2r}\sin{\theta} & 0 & \cos{\theta} & 0 \\
        0 & -e^{-2r}\sin{\theta} & 0 & \cos{\theta} \\
    \end{pmatrix} \qquad\!\!
    S_{\CZ(\theta,r)} =
    \begin{pmatrix}
        \cos{\theta} & 0 & 0 & e^{-2r}\sin{\theta} \\
        0 & \cos{\theta} & -e^{2r}\sin{\theta} & 0 \\
        0 & e^{-2r}\sin{\theta} & \cos{\theta} & 0 \\
        -e^{2r}\sin{\theta} & 0 & 0 & \cos{\theta} \\
    \end{pmatrix}
\]
which when compared to the symplectic form of the $\CZ$ and $\CNOT$ gates (cf. Eqs.~\eqref{eq:CZ_phase_space} and~\eqref{eq:CNOT_phase_space}) allows us to conclude that the beamsplitter and squeezer parameters have to satisfy the following system of equations
\be 
    \Bigg\{
    \begin{matrix}
        \,e^{2r}\sin{\left(\theta\right)} = 1\,,\\
        \cos{\left(\theta\right)}=1\,.
    \end{matrix}
\ee
This system has no exact solutions, but it can nevertheless be approximated by $\theta(r)=e^{-2r}$ and $r\rightarrow\infty$ since the limits $\lim_{r\rightarrow\infty}e^{2r}\sin(e^{-2r})$ and $\lim_{r\rightarrow\infty}\cos(e^{-2r})$ exist and are both equal to $1$. Thus, the desired $\CNOT$ and $\CZ$ gates are realized in the infinite squeezing limit. However, considering a beamsplitter parameter $\theta(r)=e^{-2r}$ and a sufficiently large $r$ allows to approximate enough the desired gates. We can evaluate the distance between the desired symplectic matrices and the approximated ones using the Hilbert-Schmidt norm of their differences
\be \label{eq:distance_approx_and_ideal}
    \mathrm{D}=\left\lVert S_{\CZ} - S_{\CZ_\n{approx}(\theta(r),r)}\right\rVert_\mathrm{H.S.} =
    \sqrt{-8 \sqrt{1-e^{-4r}}+2 e^{-8r}-4 e^{-4r}+8} \approx \sqrt{3}\,e^{-4r}\,.
\ee
Note that the distance decreases exponentially with the squeezing parameter. For comparison, for $r=0.75$ and $r=1.2$ the distance evaluates to $\mathrm{D}\approx0.09$ and $\mathrm{D}\approx0.01$, respectively. This result is identical for the $\CNOT$ gate.

Fig.~\ref{fig:decompositions_fidelity_heatmap} shows the comparison of the two presented decompositions in terms of their physical overlap fidelity for various parameter pairs $(r,\theta)$ and the symmetric (i.e., $\Delta=\kappa$) finite-energy GKP Bell state ${\ket{\Phi_{+}}_{\Delta}:=1/\sqrt{2}(\ket{00}_{\Delta}+\ket{11}_{\Delta})}$ (arbitrary choice). In particular, the comparison presented here is between the exact (Fig.~\ref{fig:heatmap_exact}) and approximate (Fig.~\ref{fig:heatmap_approximate}) decompositions of the $\CZ$ gate (similar results for the $\CNOT$). One can notice that both schemes reach in some regions of the parameter space the maximum value imposed by Eq.~\eqref{eq:phys_fidelity_full}. Yet, for the exact decomposition the maximum is attained at the unique optimal $\theta$ and $r$ which are as stated previously given by $\theta=1/2\,\n{arccot}\!\left(1/2\right)$ and $r=\n{arcsinh}\!\left(1/2\right)$. In the approximate decomposition, on the other hand, we observe that the fidelity increases monotonically for increasing $r$ and a beamsplitter parameter that follows $\theta(r)=e^{-2r}$. It reaches the maximum success probability for $r\geq0.75$. Figs.~\ref{fig:theta_slice_exact} and~\ref{fig:theta_slice_approximate} illustrate this difference between the two schemes on specific cross-section of the fidelity heatmaps. Concisely, this analysis suggests that ideal $\CNOT$ and $\CZ$ operations between two GKP qubits can besides the finite-energy effects be realized using common linear optical elements.

Most importantly, in the right parameters' regime both decompositions work similarly and implement the two gates identically or with great precision. Indeed, the variance of the fidelity with respect to the variance of the parameters is comparable in both cases. This can be observed on Fig.~\ref{fig:decompositions_fidelity_heatmap_plus_derivative} where we numerically evaluate the partial derivatives $\partial_\theta$ and $\partial_r$ of the overlap fidelity along selected direction in the parameter space. With this we can conclude that both decomposition (disregarding their fault-tolerance aspect that we study in the next Section) would be suitable for implementing an entangling gate for GKP codes and should thus be equally considered for experiments.

\begin{figure*}[t]
    \centering
    \includegraphics{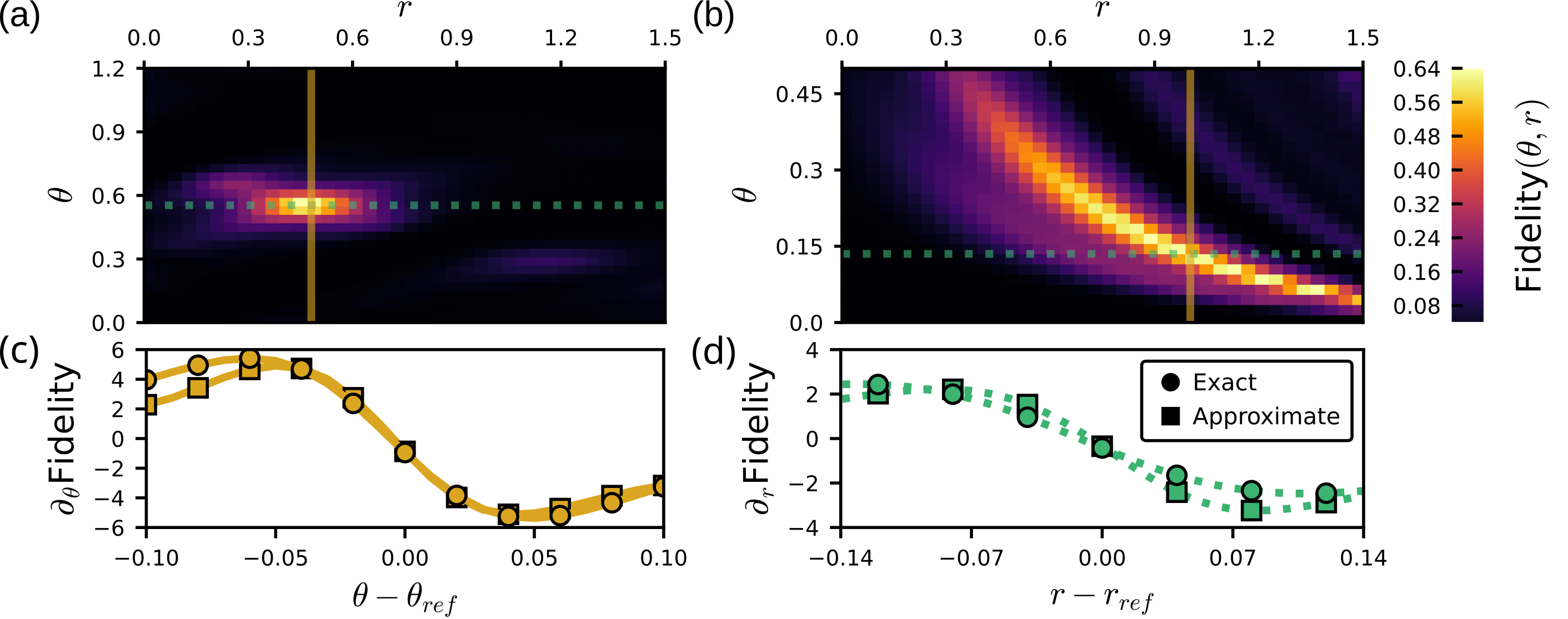}
    {\phantomsubcaption\label{fig:dummy1}}
    {\phantomsubcaption\label{fig:dummy2}}
    {\phantomsubcaption\label{fig:partial_theta}}
    {\phantomsubcaption\label{fig:partial_r}}
    \caption{Parameters stability of the two exact and approximate decompositions of the $\CZ$ gate. Plots (a) and (b) are copies from Fig.~\ref{fig:decompositions_fidelity_heatmap}. Plots (c) and (d) show partial derivatives of the overlap fidelity along the $\theta$ and $r$ directions, respectively, in parameter space. These derivatives were evaluated numerically using the finite difference formula. The reference values $\theta_{ref}$ and $r_{ref}$ for both decompositions are illustrated as dotted and solid lines on the respective heatmap.}
    \label{fig:decompositions_fidelity_heatmap_plus_derivative}
\end{figure*}

\subsec{Experimental implementation in trapped ions}{subsec:trapped_ion_experiment}

In order to realize the gate between two GKP qubits, we consider the necessary beamsplitter and squeezing operations in trapped ions architectures. Furthermore, the introduction of the operation should not be at the expense of the qubits, that is one can turn off the interaction sufficiently.

In trapped ion systems, the Coulomb repulsion can be utilized to achieve a beamsplitter interaction between two modes \cite{brown_coupled_2011}. Consider for example two trapped ions at positions $\h{\ve{r}}_1$, $\h{\ve{r}}_2$. They interact via the potential given by 
\be
\begin{split}
    U(\h{\ve{r}}_1,\h{\ve{r}}_2)&= \frac{q_1 q_2}{4\pi\epsilon_0}\frac{1}{|\h{\ve{r}}_1-\h{\ve{r}}_2|}
    = \frac{q_1 q_2}{4\pi\epsilon_0}\frac{1}{|\h{\ve{x}}_1-\h{\ve{x}}_2-\ve{s}_0|} \approx \\[7pt]
    &\approx \frac{q_1 q_2}{4\pi\epsilon_0}
    \left( 
        \frac{1}{s_0} + 
        \cos(\gamma)\frac{\h{x}_1  - \h{x}_2 }{s_0^2} +
        \left(1-3\cos^2(\gamma)\right)\frac{- \h{x}_1^2- \h{x}_2^2 + 2 \h{x}_1 \h{x}_2}{2s_0^3}
    \right),
\end{split}
\ee
where $\h{\ve{r}}_1$ and $\h{\ve{r}}_2$ are the position vectors of the two ions,  $\h{\ve{x}}_1$ and $\h{\ve{x}}_2$ are the displacements from their equilibrium positions in the trap, $\ve{s}_0$ is the distance between the two equilibrium positions, and $|\ve{s}_0|=s_0$, $|\h{\ve{x}}_{1,2}|=\h{x}_{1,2}$. We assume that the two ions are displaced along the same direction, i.e., the motional modes we consider are parallel, and that the angle between $\ve{x}_1$ and $\ve{s}_0$ is given by $\gamma$.

The term of interest is the coupling $\h{x}_1 \h{x}_2$ which is equivalent to the beamsplitter interaction when one considers the position operators in terms of ladder operators $\hat{x}_i = \sqrt{\frac{\hbar}{2m_i\omega_i}}(\hat{a}_i + \hat{a}_i^\dagger)$ where $m_i$ and $\omega_i$ are the masses of the ions and their mode frequencies, respectively. In particular, the desired unitary reads
\be
    \label{eq:beamsplitter_derivation}
    \h{H}_B =  (1-3\cos^2(\gamma))\frac{q_1 q_2}{4\pi\epsilon_0} \frac{\h{x}_1 \h{x}_2}{s_0^3} 
    = -\hbar\,\Omega_B(\gamma) \left(\h{a}_1+\h{a}_1^\dagger\right)\left(\h{a}_2+\h{a}_2^\dagger\right)
    = -\hbar\,\Omega_B(\gamma) \left(\h{a}_1 \h{a}_2^\dagger+\h{a}_1^\dagger \h{a}_2\right)\,.
\ee
The second equality is obtained using the rotating wave approximation. The coupling coefficient is parameterized by
\be
    \Omega_B(\gamma)= \left(1-3\cos^2(\gamma)\right)\frac{q_1 q_2}{8\pi\epsilon_0 s_0^3\sqrt{m_1m_2}\sqrt{\omega_1\omega_2}}\,.
\ee

For two $^{40}\text{Ca} ^+$ ions with mode frequencies of $\omega_1=\omega_2= 2\pi \times \SI{3}{\mega\hertz}$ and a equilibrium distance of $s_0=\SI{10}{\micro\meter}$ and $\gamma=\pi/2$ the interaction rate becomes $\Omega_B(\pi/2)=2\pi\times\SI{15}{\kilo\hertz}$. The beamsplitter interaction can be suppressed by a combination of detuning the two modes and increasing the inter-ion distance.

Squeezing can be achieved by parametric modulation of the trapping potential \cite{burd_quantum_2019}. Consider an ion sitting in a trap that is modulated with amplitude $k_1$ at frequency $2\omega$, then the time-dependent Hamiltonian describing its movement~is
\begin{equation}\label{eq:h_modulated_trap}
    \hat{H}(t) = \frac{\hat{p}^2}{2m} + \frac 1 2\left(k_0 + k_1 \cos(2\omega t) \right) \hat{x}^2 =
    \hbar \omega_0 \left(\hat{a}^\dagger \hat{a} + \frac{1}{2}\right) + \hbar\omega \frac{k_1}{k_0}\cos(2\omega t) 
    \left( \frac{1}{2}\left(\hat{a}^{\dagger2} + \hat{a}^2 \right) + \hat{a}^\dagger \hat{a} + \frac{1}{2}\right). 
\end{equation}
Here, the wanted squeezing Hamiltonian resides in the parenthesis proportional to the cosine function. In a rotating wave approximation and the oscillator's frame of reference, the interaction term becomes
\begin{equation}
  \h{H}_{S} = \hbar \omega \frac{k_1}{2k_0} \frac{1}{2} \left( \h{a}^{\dagger 2} + \h{a}^2 \right)\,,
\end{equation}
which for a trap frequency of $\omega = 2\pi \times 3 \si{MHz}$ and a modulation of 1\% of the trap potential yields a squeezing interaction strength of 
$\Omega_S = \omega \frac{k_1}{2k_0}= 2\pi\times \SI{15}{\kilo\hertz}$.
Note that since squeezing is an active operation, it can be completely turned off. 

To realize the approximate $\CZ$ gate as defined in Section~\ref{sec:approximate_gate_decomposition} we can choose $r=1$ and $\theta=e^{-2r}=0.14$. Using the interaction strengths calculated above this yields a time for the beamsplitter of $t_B=\theta/\Omega_B\approx\SI{1.5}{\micro s}$ and for the squeezing of $t_S=r/\Omega_S\approx\SI{11}{\micro s}$ and therefore a total gate time of approximately $\SI{24}{\micro s}$.

Analogously, for the exact gate as defined in Section~\ref{sec:exact_gate_decomp} we will assume ${r=\n{arcsinh}\!\left(1/2\right)\approx0.48}$ and ${\theta=1/2\,\n{arccot}\!\left(1/2\right)\approx0.55}$ which yields $t_B=\theta/\Omega_B\approx\SI{6}{\micro s}$, $t_S=r/\Omega_S\approx\SI{5}{\micro s}$ and a total gate time of approximately~$\SI{17}{\micro s}$.

\subsec{Experimental implementation in superconducting cavities}{subsec:supercond_experiment}

Superconducting cavities are also a promising platform for the experimental realization of GKP qubits \cite{campagne-ibarcq_quantum_2020, eickbusch_fast_2022, sivak_real_time_2023}. Here, we give an overview of the recent advances in the implementation of the desired beamsplitter and squeezing operations.

Several implementations of bilinear couplings have been realized in the past ten year~\cite{baust_tunable_2015,pfaff_controlled_2017,collodo_observation_2019}. The first realization of a beamsplitter operation between two cavities storing some quantum information was reported in Ref.~\cite{gao_programmable_2018} where the authors coupled two superconducting cavities using a single transmon. An improved version of this operation was realized recently by \citet{chapman_high_2023} using instead a SNAIL resonator to couple the two bosonic systems. The effective interaction between two cavities is given by 
\begin{equation}
    \h{H}_{B} = g_B(t) \left( e^{i\varphi_B}\,\hat{a}_1^\dagger \hat{a}_2 + e^{-i\varphi_B}\,\hat{a}_1 \hat{a}_2^\dagger  \right)
\end{equation}
where the coupling strength $g_B(t)$ and the phase $\varphi_Bi$ are both obtained from the microwave drives used to activate the respective four- or three-wave mixing interactions.

The SNAIL resonator is also capable of producing squeezing, as was demonstrated in \cite{frattini_optimizing_2018,sivak_kerr-free_2019}. In this case, following the derivation from Appendix F in Ref.~\cite{chapman_high_2023}, it is apparent that one can get a squeezing operation by pumping the snail at twice the target cavity's frequency. The squeezing interaction Hamiltonian is then given by 
\begin{equation}
    \h{H}_{S} = g_S(t) \left( e^{i\varphi_S}\,\hat{a}^{\dagger2} + e^{-i\varphi_S}\,\hat{a}^2 \right)
\end{equation}
Importantly, the strength of the squeezing and of the beamsplitter are comparable in amplitude and realizable using the same circuit element. 

\iftoggle{arXiv}{ \PRLsec{Fault-tolerance aspect of two-qubit gates}{sec:ft_threshold} 
}{ \section{Fault-tolerance aspect of two-qubit gates}\label{sec:ft_threshold} }

In this section, we discuss the fault-tolerance aspect of the two decompositions presented in Section~\ref{sec:gate_decompositions}. For that we consider the value of the error threshold for a fault-tolerant concatenation of GKP and surface codes determined in Ref.~\cite{noh_fault-tolerant_2020} $\frac{\kappa}{G} \leq 0.69\%$ with $\kappa$ being the rate of the error processes and $G$ denoting the coupling strength of the ideal $\CNOT$ gate
\begin{equation}\label{eq:CNOT_with_t_dep_coupling}
    \CNOT(G,t) = \exp\left(-i \int_0^t \,G\,\q_1\p_2 \right)\,.
\end{equation}
If $G=t^{-1}$ this would then realize the desired ideal gate given in Eq.~\eqref{eq:CNOT_operator}. The error processes are here assumed to be Markovian and dominated by the photon loss and heating, i.e., the dynamics of the system $\rho$ under these effects is described by the following master equation
\be \label{eq:error_master_eq}
    \dot{\rho} = \kappa \left(\cl{D}[\,\h{a}_1\,](\rho) + \cl{D}[\,\h{a}_1^\dagger\,](\rho) + 
                 \cl{D}[\,\h{a}_2\,](\rho) + \cl{D}[\,\h{a}_2^\dagger\,](\rho) \right)
\ee
where $\cl{D}[\,L_k\,](\rho)\coloneqq  L_k\,\rho\,L_k^{\dagger} - \frac{1}{2}\left\{L_k^{\dagger} L_k\,,\rho\right\}$ is the so-called dissipator superoperator for the jump operator $L_k$.

\begin{SCfigure}[][b!]
    \centering
    \includegraphics{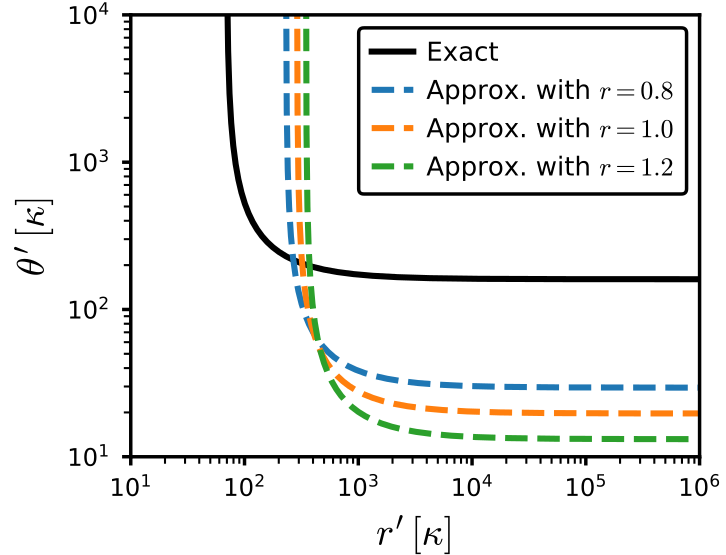}
    \caption{Values of the beamsplitter $\theta'$ and squeezer $r'$ interaction strengths in units of the photon loss rate $\kappa$ that allow to reach the fault-tolerance threshold of the Surface-GKP code as defined in Ref.~\cite{noh_fault-tolerant_2020}. These values are given for both exact (full line) and approximate (dashed lines) decompositions of the $\CZ$ gate (cf. Section~\ref{sec:gate_decompositions}). Then, for a given parameter pair $(r',\theta')$ the threshold is reached when it lays above the corresponding curve.}
    \label{fig:fault_tol}
\end{SCfigure}

In Section~\ref{sec:gate_decompositions}, we determined that the exact gate decompositions is realized when the beamsplitter and squeezing parameters are equal to $\theta=1/2\,\n{arccot}\!\left(\frac{1}{2}\right)$ and $r=\n{arcsinh}\!\left(1/2\right)$. These are unitless quantities that can be related in a similar manner as in Eq.~\eqref{eq:CNOT_with_t_dep_coupling} to an interaction time and strength, $\theta = \theta'\,t_B$ and $r = r'\,t_{S}$. Assuming that the total gate time $t$ equals to the sum of the interactions that compose it, we obtain the following inequality
\be \label{eq:inequality_exact_decomp}
    t = 2t_B + t_{S} = 
    \frac{1}{\theta'} \mathrm{cot}^{-1}\!\!\left(\frac{1}{2}\right) + 
    \frac{1}{r'} \sinh^{-1}\!\!\left(\frac{1}{2}\right)
    \equiv \frac{1}{G} \leq \frac{0.69\%}{\kappa}
\ee

Regarding the approximate decompositions, the condition that the beamsplitter and squeezing parameter have to fullfil is $\theta = \theta'\,t_B = \n{arcsin}\!\left(e^{-2r}\right)$ for a sufficiently large $r=r'\,t_{S}$. Then, the inequality relating the error rate to the interaction strength is
\be \label{eq:inequality_approx_decomp}
    t = t_B + 2t_{S} = 
    \frac{1}{\theta'} \sin^{-1}\!\!\left(e^{-2r}\right) + 
    2\frac{r}{r'}
    \equiv \frac{1}{G} \leq \frac{0.69\%}{\kappa}\,.
\ee
We notice that now the inequality depends on three parameters. As we have seen previously, the approximate decomposition of the two-qubit gates is sufficiently close to the ideal gate for squeezing parameters larger than $0.75$, we will thus consider in the following discussion the values of $r\in\{0.8,1.0,1.2\}$.

Eqs.~\eqref{eq:inequality_exact_decomp} and~\eqref{eq:inequality_approx_decomp} allow us to determine the range of the interaction strengths $\theta'$ and $r'$ that would satisfy the fault-tolerance threshold of a surface-GKP code given the system's error rate $\kappa$. Fig.~\ref{fig:fault_tol} show these ranges in units of noise rate for both the exact and the approximate decompositions. We notice that the exact decompositions allow for a wider range of values for the squeezing parameters, whereas the approximate schemes offer a more flexible selection of the beamsplitter coupling coefficient. This behavior is due to the difference in the number of beamsplitters and squeezers between both decompositions. 

Using these expressions we can also roughly evaluate the desired interaction strengths for the two architectures that have demonstrated the preparation and stabilization of GKP states. In the case of superconducting devices, we can evaluate from recent works~\cite{,campagne-ibarcq_quantum_2020,gao_programmable_2018} the photon loss rate and the beamsplitter strength to be $\kappa\sim 1.5\,\n{kHz}$ and $\theta'\sim0.3\,\n{MHz}$. Thus, the squeezing strength required for a fault-tolerant surface-GKP code is at least $r'\geq0.1\,\n{MHz}$. If the ratio $\theta'/\kappa$ ends up be lower than $160$ then the use of the approximate scheme would be needed, in which case $r'\geq0.3\,\n{MHz}$. A similar observation can be done in trapped ion systems where the heating rate is of the order of $\sim 10\,\n{Hz}$ and squeezing with strengths larger than $0.1\,\n{MHz}$ can be reached~\cite{burd_quantum_2019}. Then, the beamsplitter interaction strength can be set to be above $160\kappa=1.6\,\n{kHz}$ if one desires to implement the exact scheme or lower if one tries to realize the approximate decomposition. These values for both architectures are a priori achievable in current systems. Crucially, our rough estimations do not include any imperfections arising in the implementation of the $\h{B}_{S/A}(\theta)$ and $\h{S}(r)$ Hamiltonians which would add some additional requirements on the values of $\theta'$ and $r'$. We leave that analysis for further studies.

\iftoggle{arXiv}{ \PRLsec{Error-corrected two-qubit gates}{sec:error_correction}
}{ \section{Error-corrected two-qubit gates} \label{sec:error_correction} }

As mentioned in the main text the first method that could help to deal with finite-energy effects arising from ideal two GKP qubit gates is the so-called quantum error correction~(QEC). In this section, we will briefly introduce the recent developments along the stabilization and error correction of GKP codes. Then, we demonstrate numerically that they indeed eliminate the undesired effects bringing the states back into the logical subspace. However, due to an increase of asymmetry between stabilizers' expectation values we observe a logical decoherence too. 

\subsec{Introduction to GKP code stabilization}{subsec:intro_to_stabil}

In the case of ideal grid states, the correction procedure would consist of measuring the stabilizers $\h{S}_x$ and $\h{S}_z$ (cf.~Eq.~\eqref{eq:square_stab_ops}) and then applying a corresponding displacement operator $\h{D}(\alpha)$. In practice this procedure is tedious since it requires a fault-tolerant and non-destructive readout of the quadratures $\q$ and $\p$. The alternative and most common method is to use an auxiliary system and perform a so-called phase-estimation protocol. The additional system is ideally another GKP state which together with a high-efficiency homodyne detection can realize a bosonic version of Steane or Knill error correction (we refer the reader to the review in Ref.~\cite{grimsmo_quantum_2021} for more details). However, to this day only protocols with additional two-level systems has been experimentally demonstrated~\cite{fluhmann_encoding_2019,campagne-ibarcq_quantum_2020,de_neeve_error_2022}. Nonetheless, these schemes have enabled to reach the largest gain in logical coherence time compared to the physical one~\cite{sivak_real_time_2023}. Thus, we will only concentrate on QEC and stabilization protocols that rely on phase estimation with an auxiliary qubit.

\begin{SCfigure}[1.6][t!]
    \centering
    \includegraphics[scale=1.2]{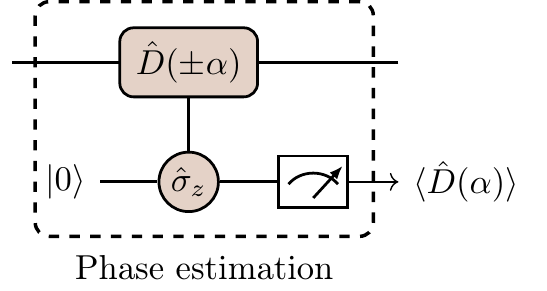}
    \caption{Gate based representations of the phase estimation protocol. The circuit takes in input a bosonic state and uses one auxiliary qubit which is initialized in $\ket{0}$ and measured at the end of the protocol. The unitary operation used in the circuit represent a displacement operators (rectangle) that is conditioned on the state $\ket{0}$ or $\ket{1}$ (i.e., eigenbasis of the Pauli Z operator $\s_z$) of the auxiliary system (circles). The mathematical form of the operator is given in Eq.~\eqref{eq:spin_dependent_displ}. The measurement gives one bit of information about the expectation value of the desired displacement.}
    \label{fig:phase_estimation}
\end{SCfigure}

The phase estimation protocol in question is illustrated in Fig.~\ref{fig:phase_estimation}. Its central operation is a conditional displacement (a.k.a. spin-dependent force~\cite{haljan_spin-dependent_2005}) which reads as
\be \label{eq:spin_dependent_displ}
    C\h{D}(\alpha) = e^{-i\,\sqrt{2}\,\big(\n{Re}(\alpha)\,\p - \n{Im}(\alpha)\,\q\big)\,\s_z}
\ee
with $\alpha$ being the magnitude of the controlled displacement. We can easily check that if the auxiliary system is in the state $\ket{0}$ or $\ket{1}$ this operation results in displacements $\h{D}(\alpha)$ or $\h{D}(-\alpha)$, respectively. Together with qubit rotations one can then implement an operation controlled on arbitrary spin states. Thus, if we consider that the spin is initially in the state $\ket{+}$ and that after the conditional operation we measure it in the $\s_x$ basis, we effectively obtain a single bit of information about $\langle\h{D}(\alpha)\rangle$. This expectation value can then be retrieved with large enough statistics. Replacing $\alpha$ with some appropriate values would thus allows us to measure the syndromes of the GKP code, $\langle\h{S}_x\rangle$ and $\langle\h{S}_z\rangle$, which in turn can be used to eliminate the errors from the bosonic system.

More concretely, the probability of measuring $\ket{+}$ or $\ket{-}$ as well as the post-measurement state of the bosonic system (after one measurement) are given by
\be \label{eq:phase_est_proba}
    P_{\pm}\left(\h{S}_{x/z}\right) = 
    \frac{1}{2}\left[1\pm\frac{1}{2}\left(\left\langle \h{S}_{x/z} \right\rangle + \left\langle \h{S}^\dagger_{x/z} \right\rangle \right)\right] 
    \qquad\text{and}\qquad 
    \ket{\psi_\pm} = \frac{1}{2}\,\h{S}_{x/z}\,\ket{\psi} \pm \frac{1}{2}\,\h{S}^\dagger_{x/z}\,\ket{\psi}\,.
\ee
For infinite-energy grid states, $\h{S}^\dagger_{x/z}$ is equivalent to $\h{S}_{x/z}$. Thus, if the system's state is in the ideal logical subspace (i.e.,~$\ket{\psi}=\ket{\psi}_\n{I}$) one would measure the spin in $\ket{+}$ with unit probability. Moreover, the bosonic system would in this case remain unchanged since $\h{S}_{x/z}\,\ket{\psi}_\n{I}\equiv\ket{\psi}_\n{I}$. 

However, for finite-energy GKP states, $\h{S}_{x}$ and $\h{S}_z$ as defined in Eq.~\eqref{eq:square_stab_ops} are not energy preserving operations as they displace both the peaks and the envelope of the states. Hence, the stabilizers' expectation values which effectively corresponds to the overlap between $\h{S}_{x/z}\ket{\psi}_{\Delta,\kappa}$ and $\ket{\psi}_{\Delta,\kappa}$ are strictly lower than one (id. in Section~\ref{sec:eff_squeez_param})
\be \label{eq:ideal_stab_expect_val}
    \left\langle \h{S}_{x} \right\rangle_{\Delta,\kappa} =
    \prescript{}{{\Delta,\kappa}}{\bra{\psi}}\,\h{S}_{x}\ket{\psi}_{\Delta,\kappa} 
    \approx e^{-\pi\kappa^2} \qquad\Rightarrow\qquad P_{+}\left(\h{S}_{x}\right) \approx \frac{\pi}{2}\,\kappa^2\,e^{-\frac{\pi}{2}\kappa^2}
\ee
(similarly for $\h{S}_z$ but with $\Delta^2$ instead of $\kappa^2$). Beside a reduced probability $P_\pm$ the state of the oscillator is also distorted after every round of the phase-estimation protocol. Indeed, the post-measurement state $\ket{\psi_{+}}_{\Delta,\kappa}$ although being centered around the phase space's origin has a larger envelope than the one of $\ket{\psi}_{\Delta,\kappa}$. All in all, the energy of the system increases with every measurement and should then be corrected too.

\citet{campagne-ibarcq_quantum_2020} determined that the correction of the envelope's expansion can be done with an additional phase-estimation round that consists of the same conditional displacement $C\h{D}(\alpha)$ but with a lower amplitude $\alpha$ and along an axis orthogonal to the first displacement. Namely, if one desires to measure the syndrome associated to $\h{S}_{x}$ one would need to interleave $C\h{D}(\sqrt{2\pi})$ with $C\h{D}(i\epsilon)$ where $\epsilon\ll\sqrt{2\pi}$ is related to the size of the envelope. The authors referred to these two measurement processes as ``Sharpen'' and ``Trim'', respectively. \citet{royer_stabilization_2020} and \citet{de_neeve_error_2022} then unveiled that these two distinct steps can be realized dissipatively and correspond to the 1st order Trotter decomposition of the finite-energy stabilizers
\be \label{eq:finite_energy_stab}
    \h{S}_{x,(\Delta,\kappa)} = e^{i\,2\sqrt{\pi}\n{cosh}(\Delta\kappa)\,\p+2\sqrt{\pi}\frac{\Delta}{\kappa}\n{sinh}(\Delta\kappa)\,\q}
    \qquad\text{and}\qquad
    \h{S}_{z,(\Delta,\kappa)} = e^{-i\,2\sqrt{\pi}\n{cosh}(\Delta\kappa)\,\q+2\sqrt{\pi}\frac{\kappa}{\Delta}\n{sinh}(\Delta\kappa)\,\p}\,.
\ee

\begin{figure*}[t!]
    \centering
    
    \includegraphics{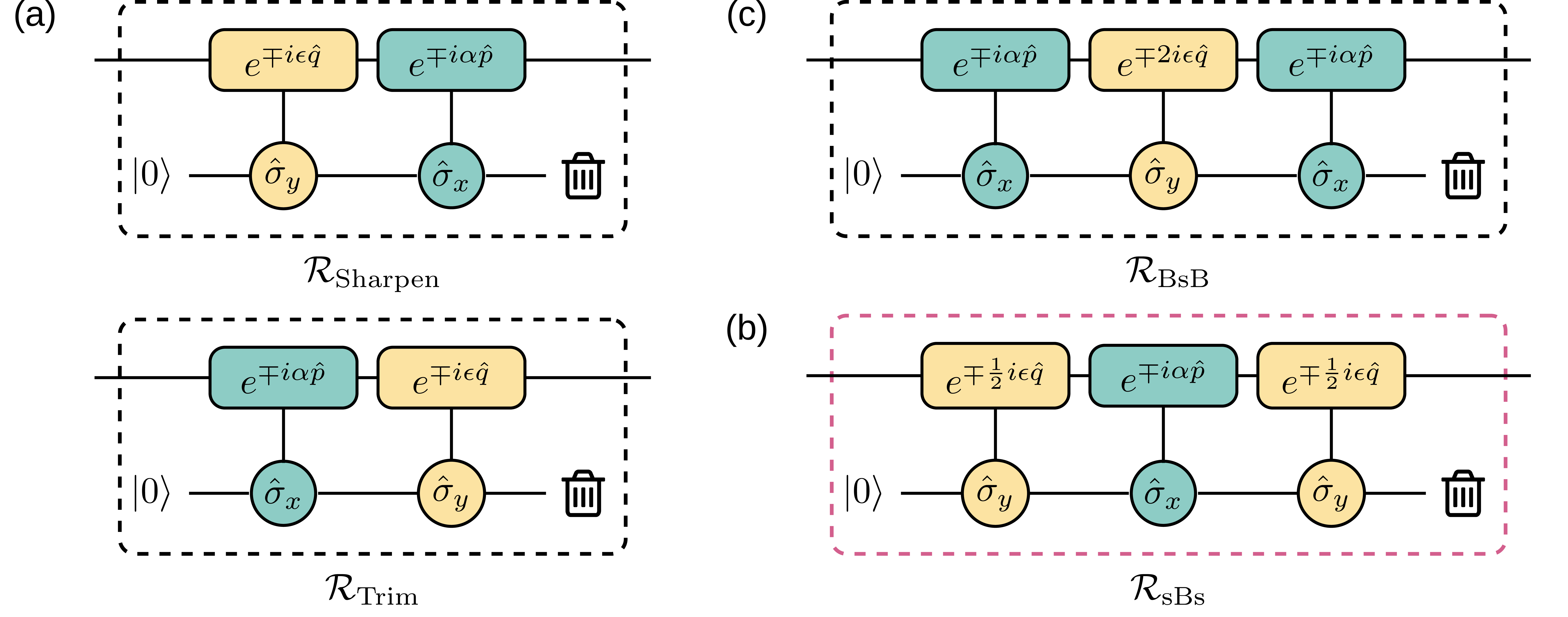}
    \caption{Gate based representations of one dissipative stabilization cycle for the $\q$ quadrature of the GKP code. The diagrams represent three different Trotter decompositions of the finite-energy stabilizer $\h{S}_{x,(\Delta,\kappa)}$ (cf. Eq.~\eqref{eq:finite_energy_stab}), namely the (a) Sharpen-Trim, (b) Big-small-Big and (c) small-Big-small schemes. The latter one will be used in the next subsection to correct for finite-energy effects of the $\CZ$ gate. Each circuit takes in input a bosonic state and uses one auxiliary qubit which is initialized in $\ket{0}$ and reset at the end of the process (illustrated here by a dustbin). The unitary operations in each schemes are given in Eq.~\eqref{eq:trotter_stabilization_gates} and represent displacement operators (rectangles) that are conditioned on a specific Pauli basis of the auxiliary system (circles). The parameters of the circuit follow $\alpha=\sqrt{\pi}\n{cosh}(\Delta\kappa)$ and $\epsilon=\sqrt{\pi}\frac{\Delta}{\kappa}\n{sinh}(\Delta\kappa)$. The schemes for $\h{S}_{z,(\Delta,\kappa)}$ are obtained after the transformations $(\q,\p)\rightarrow(-\p,\q)$ and $(\Delta,\kappa)\rightarrow(\kappa,\Delta)$.}
    \label{fig:circuits_QEC_schemes}
\end{figure*}

\noindent
The authors also derived two 2nd order approximations of the discretized unitary that stabilize the logical subspace. Overall, the three unitary operations stabilizing the $\q$-quadrature read
\be \label{eq:trotter_stabilization_gates}
\begin{split}
    \textnormal{1st order Trotter:} \quad 
    & \h{U}_\n{Sharpen} = 
    e^{-i\sqrt{\pi} \frac{\Delta}{\kappa}s_{\Delta,\kappa} \,\q\,\s_y }\,\,
    e^{-i\sqrt{\pi} c_{\Delta,\kappa} \,\p\,\s_x } 
    \quad\textnormal{and}\quad
    \h{U}_\n{Trim} = 
    e^{-i\sqrt{\pi} c_{\Delta,\kappa} \,\p\,\s_x } \,\,
    e^{-i\sqrt{\pi} \frac{\Delta}{\kappa}s_{\Delta,\kappa} \,\q\,\s_y }    
    \\
    \textnormal{2nd order Trotter:} \quad
    &\h{U}_\n{sBs} = 
    e^{-i\frac{1}{2}\sqrt{\pi} \frac{\Delta}{\kappa}s_{\Delta,\kappa} \,\q\,\s_y }\,\,
    e^{-i\sqrt{\pi} c_{\Delta,\kappa} \,\p\,\s_x }\,\,
    e^{-i\frac{1}{2}\sqrt{\pi} \frac{\Delta}{\kappa}s_{\Delta,\kappa} \,\q\,\s_y }\, \\
    &\h{U}_\n{BsB} = 
    e^{-i\sqrt{\pi} c_{\Delta,\kappa} \,\p\,\s_x }\,\,
    e^{-i2\sqrt{\pi} \frac{\Delta}{\kappa}s_{\Delta,\kappa} \,\q\,\s_y }\,\,
    e^{-i\sqrt{\pi} c_{\Delta,\kappa} \,\p\,\s_x }
\end{split}
\ee
with $c_{\Delta,\kappa}$ and $s_{\Delta,\kappa}$ abbreviating $\n{cosh}(\Delta\kappa)$ and $\n{sinh}(\Delta\kappa)$. To obtain the unitaries which approximate $\h{S}_{z,(\Delta,\kappa)}$ we can make use of the Fourier transform and replace $(\q,\p)\rightarrow(-\p,\q)$ as well as $(\Delta,\kappa)\rightarrow(\kappa,\Delta)$. These three protocols are illustrated in Fig.\ref{fig:circuits_QEC_schemes}. The \textit{stabilization cycle} for one of the quadratures then consists from the application of one of these unitaries followed by a reset of the two-level system. In the absence of noise, $\h{U}_\n{sBs}$ has been seen to outperform the other protocols~\cite{de_neeve_error_2022,royer_stabilization_2020}. In the next subsection, we will thus concatenate this stabilization procedure together with the ideal $\CZ$ gate.

In order to make the simulations tractable, we can rewrite this dissipative stabilization procedures in terms of completely positive trace preserving maps $\cl{R}_{x,(\Delta,\kappa)}$ and $\cl{R}_{z,(\Delta,\kappa)}$ for the stabilization of the $\p$ and $\q$ quadratures, respectively. The Kraus operators of $\cl{R}_{x,(\Delta,\kappa)}$ are given by~\cite{de_neeve_error_2022}
\be \label{eq:stabilization_kraus_operators}
\begin{split}
    K_{x,(\Delta,\kappa)}^{(0)} &=
    \frac{1}{\sqrt{2}}\,
    e^{i \frac{1}{2}\sqrt{\pi} \frac{\Delta}{\kappa}s_{\Delta,\kappa}\,\q}
    \left( \cos\!\left(\sqrt{\pi} c_{\Delta,\kappa}\,\p\right) \, 
    e^{i \frac{1}{2}\sqrt{\pi} \frac{\Delta}{\kappa}s_{\Delta,\kappa}\,\q} -
    \sin\!\left(\sqrt{\pi} c_{\Delta,\kappa}\,\p\right) \, 
    e^{-i \frac{1}{2}\sqrt{\pi} \frac{\Delta}{\kappa}s_{\Delta,\kappa}\,\q}
    \right)
    \\[6pt] 
    K_{x,(\Delta,\kappa)}^{(1)} &= 
    \frac{1}{\sqrt{2}}\,
    e^{-i \frac{1}{2}\sqrt{\pi} \frac{\Delta}{\kappa}s_{\Delta,\kappa}\,\p}
    \left( \cos\!\left(\sqrt{\pi} c_{\Delta,\kappa}\,\p\right) \, 
    e^{-i \frac{1}{2}\sqrt{\pi} \frac{\Delta}{\kappa}s_{\Delta,\kappa}\,\q} +
    \sin\!\left(\sqrt{\pi} c_{\Delta,\kappa}\,\p\right) \, 
    e^{i \frac{1}{2}\sqrt{\pi} \frac{\Delta}{\kappa}s_{\Delta,\kappa}\,\q}
    \right)
\end{split}
\ee
where $\ket{0/1}$ are the states of the auxiliary qubit. The operators for $\cl{R}_{z,(\Delta,\kappa)}$ can be obtained as above by replacing $(\q,\p)\rightarrow(-\p,\q)$ as well as $(\Delta,\kappa)\rightarrow(\kappa,\Delta)$.

Importantly, $\h{U}_\n{Sharpen}/\h{U}_\n{Trim}$ and $\h{U}_\n{sBs}$ protocols have a peculiar behavior compared to the $\h{U}_\n{BsB}$ one which is that every stabilization cycle (i.e., application of the unitary + reset) corresponds as well to a logical Pauli operation~\cite{de_neeve_error_2022,royer_stabilization_2020}. This is due to the fact that the three procedures stabilize simultaneously two lattices where the first one is the desired one and the second is a superlattice appearing due to our requirement of having a low excitation transfer to the bath (i.e., throughout the process the orthogonal spin state is weakly populated). The difference between the three protocols is then the lattice constant of the superlattice that they stabilize. Although the Pauli transformations can be monitored using Pauli frame tracking, we will instead consider in what follows that one \textit{round of stabilization} consists of two stabilization cycles.

\subsec{Stabilization after a two GKP qubit gate}{subsec:stabilized_cz}

\begin{figure}[t!]
    \centering
    \includegraphics{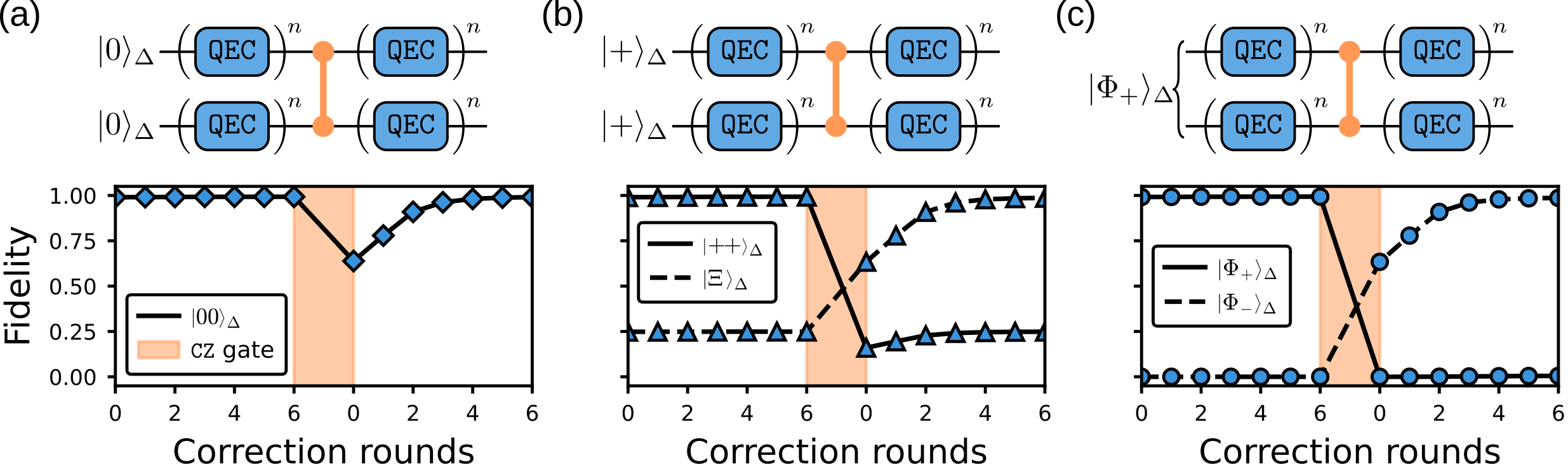}
    {\phantomsubcaption\label{fig:qec_and_cz_various_states_00}}
    {\phantomsubcaption\label{fig:qec_and_cz_various_states_p0}}
    {\phantomsubcaption\label{fig:qec_and_cz_various_states_Phi}}
    \caption{Fidelity of system's state with the input (full) and desired (dashed) states throughout stabilization rounds before and after the $\CZ$ gate. The finite-energy parameter $\Delta$ of the input states is equal to $0.3$. The simulations have been performed using the density matrix representation of the system's state with stabilization implemented as completely positive trace preserving maps with Kraus operators given in Eq.~\eqref{eq:stabilization_kraus_operators}.}
    \label{fig:qec_and_cz_various_states}
\end{figure}

As mentioned above we remedy the finite-energy effects of the two-qubit GKP gates by concatenating some stabilization rounds before and after the entangling gate. The purpose of these rounds is to preserve the logical information in the system as well as to constrain the energy of the physical states. As we noted in the previous sections the energy of the encodings is measurably altered, however from empirical observations the logical information seemed to be unaffected by the physical effects. Thus, our initial hypothesis is that in a noiseless situation the stabilization that follows the $\CNOT$ or $\CZ$ gates would solely reduce the energy of physical states to their initial value.

As a starting point we test this hypothesis by evaluating the fidelity of system's state with the desired ones throughout the stabilization rounds. The fidelity is defined as $F(\ket{\psi_\n{des}},\rho_\n{targ})=\bra{\psi_\n{des}}\rho_\n{sys}\ket{\psi_\n{des}}$. It is worth mentioning that this quantity was evaluated using the ideal $\psi_\n{des}$ and not the stabilized one which would be computationally intractable. Fig.~\ref{fig:qec_and_cz_various_states} illustrates the dynamics of this measure for three different initial states that are $\ket{00}_{\Delta}$, $\ket{++}_{\Delta}$ and the Bell state ${\ket{\Phi_{+}}_{\Delta}}$ (for simplicity we chose $\kappa=\Delta$). In all the three cases, we note that the fidelity is initially lowered. This behavior comes from the approximate aspect of the $\h{U}_\n{sBs}$ stabilization. Nevertheless, during the rounds before the entangling gate, the fidelity converges quickly to a finite value that is close to 1. After the gate, the value drops drastically according to our analysis from Section~\ref{sec:physical_fidelity}. Finally, the stabilization rounds that follow improve as expected the fidelity with the desired two-qubit state and converges qualitatively to a similar value than before the gate. 

To test quantitatively the agreement with our assumption we compare the end fidelity with its value before the gate. Fig.~\ref{fig:diff_fidelity_stab_CZ} shows this difference for three initial states as well as for the $\ket{+0}_{\Delta}$ as a function of the energy parameter $\Delta$. These plots show a clear distinction between computational states and those with logical coherence. Indeed, when the system is in $\ket{00}_{\Delta}$ the fidelity is recovered identically (provided some approximation error due to the sBs correction scheme), whereas in the other case the difference increases exponentially with increasing $\Delta$ (cf. Sec.~\ref{subsec:log_info_decoherence} for the analytical expression of the scaling). This unexpected behavior exhibits a potential limit to quantum information processing with GKP encodings as any algorithm will be exponentially limited by the number of ideal $\CZ$ and $\CNOT$ gates that it contains. 

We now investigate this phenomenon in more details by performing logical state tomography throughout the stabilization rounds that we define as
\be \label{eq:logic_state_tomography}
    \big[\rho^\n{logic}\big]_{(i,j)} := \bra{\psi_i}\rho^\n{phys}\ket{\psi_j} 
    \qquad \textnormal{with} \qquad
    \ket{\psi_i},\,\ket{\psi_j}\,\in\,\big\{\ket{00}_{\Delta},\ket{01}_{\Delta},\ket{10}_{\Delta},\ket{11}_{\Delta}\big\}
\ee
where $\rho^\n{logic}$ is a two-qubit logical density matrix. We verify that the resulting matrix already satisfies the positivity condition expected from a density operator as its eigenvalues are strictly positive. Regarding the unit trace requirement, we find that it follows the same behavior as the fidelity curves presented in Fig.~\ref{fig:qec_and_cz_various_states}, but with the difference that after the stabilization $\n{Tr}\left[\rho^\n{logic}\right]$ converges to nearly the same value as before the application of the $\CZ$ gate. Indeed, the difference between the two values is of the order of $10^{-5}$ (which can be associated to simulation precision as we simulated with numbers in single-precision floating-point format, $\mathtt{Float32}$). This observation points out to the fact that with the stabilization rounds the system is brought back to the initial logical subspace with energy $\Delta$. Finally, we normalize $\rho^\n{logic}$ by its trace to obtain the logical density matrix.

\begin{SCfigure}[][t!]
    \centering
    \includegraphics{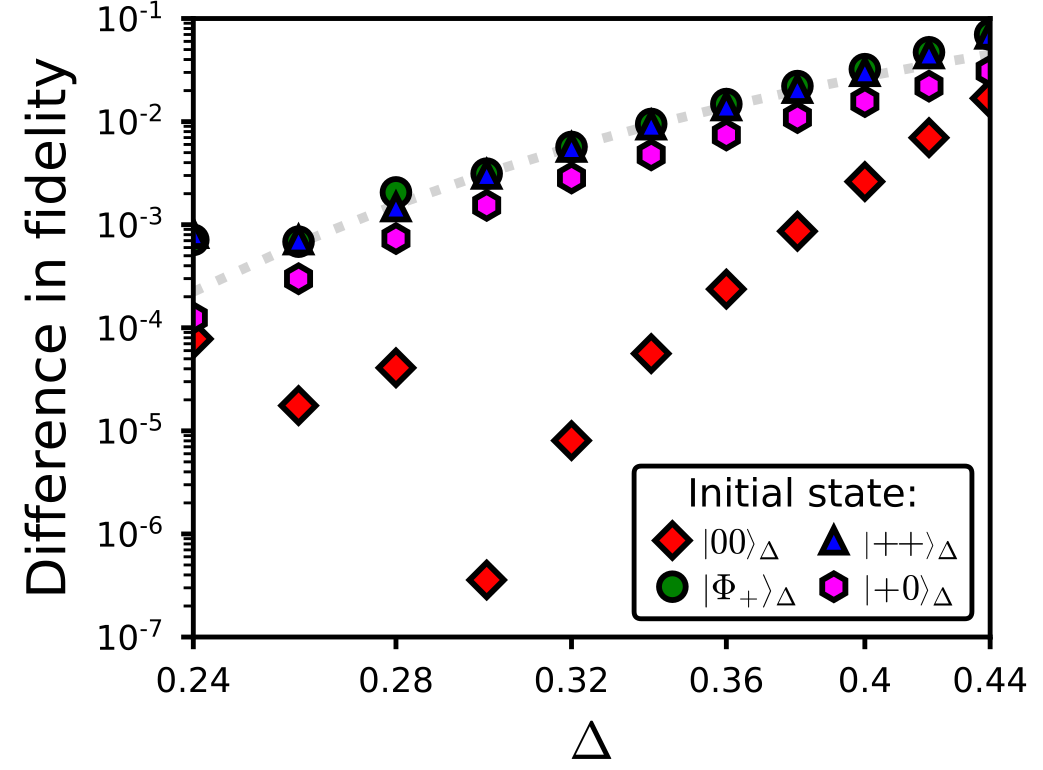}
    \caption{Difference in fidelity of the system's state with the desired one before and after the stabilized $\CZ$ or $\CNOT$ gates ($18$ stabilization rounds) for the three states used in Fig.~\ref{fig:qec_and_cz_various_states}. The dotted line follows the analytical logical infidelity given in Eq.\eqref{eq:diff_logical_fidelity_full}. Here, due to simulation constraints data points for ${\Delta\leq0.28}$ are more susceptible to rounding errors.}
    \label{fig:diff_fidelity_stab_CZ}
\end{SCfigure}

Having access to the logical state of the system allows us to study the dynamics of various properties such as the purity, the entanglement entropy and concurrence of the logical state. The entropy of entanglement of a bipartite system corresponds to the entropy of either of the reduced subsystems
\be  \label{eq:entanglement_entropy}
    \cl{S}(\rho_{AB}) = - \n{Tr}\big[\rho_A\,\log\left(\rho_A\right)\big] 
    = - \n{Tr}\big[\rho_B\,\log\left(\rho_B\right)\big]
    \qquad \textnormal{with} \qquad 
    \rho_{A/B} = \n{Tr}_{B/A}\big[\rho_{AB}\big]
\ee
and ranges from 0 for separable states to $\log(N)$ if the two $N$-dimensional subsystems are entangled. While for pure states $\rho_{AB}$ this quantity is a good measure of entanglement, for an entangled mixture it is not and can only signal the presence of entanglement~\cite{bennett_mixed-state_1996}. We will thus use the concurrence to quantify the amount of entanglement in the system. For a two-qubit state, it reads as~\cite{wootters_entanglement_1998}
\be \label{eq:concurrence}
    \cl{C}(\rho_{AB}) = \n{Max}\{0,\,\sqrt{\lambda_1}-\sqrt{\lambda_2}-\sqrt{\lambda_3}-\sqrt{\lambda_4}\}
\ee
where $\lambda_1\geq\lambda_2\geq\lambda_3\geq\lambda_4$ are eigenvalues of the operator
\be
    \h{\Lambda} = \rho_{AB}\,\h{\Sigma}\,\rho_{AB}^{\intercal}\,\h{\Sigma} 
    \qquad \text{with} \qquad
    \hat{\Sigma} = \left(\begin{array}{cccc}
        0 & 0 & 0 & -1 \\
        0 & 0 & 1 & 0 \\
        0 & 1 & 0 & 0 \\
        -1 & 0 & 0 & 0
    \end{array}\right)
\ee
known as the spin flip matrix. $\cl{C}$ maps the quantity of entanglement to a unit scale where the zero value is attained by any pure separable state, i.e., $\rho_{AB}=\ket{\psi}\!\!\bra{\psi}\otimes\ket{\phi}\!\!\bra{\phi}$, as the spin-flip matrix $\h{\Sigma}$ will map it to an orthogonal state. On the other hand, $\cl{C}(\rho_{AB})=1$ arises only when the bipartite state is pure and maximally entangled. Moreover, thanks to the convexity of the concurrence formula for pure states, there exist no mixture state that could reach a larger or equal value than one.

\begin{figure}[t!]
    \centering
    \includegraphics{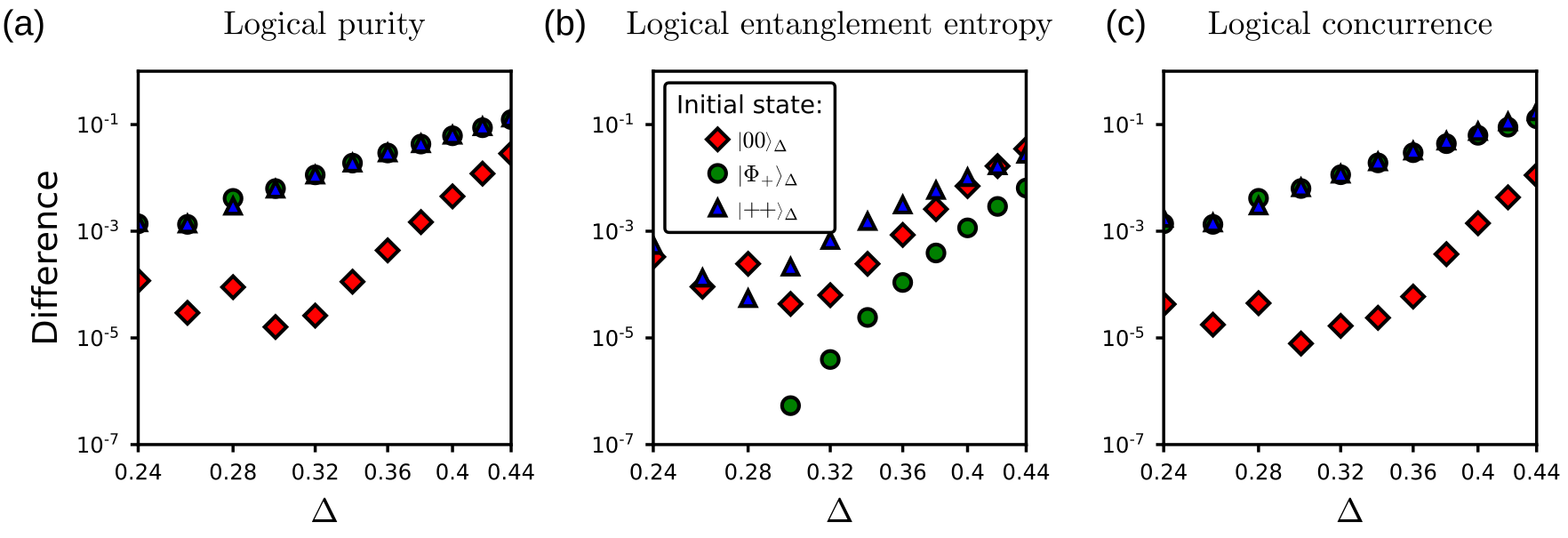}
    {\phantomsubcaption\label{fig:diff_logical_purity}}
    {\phantomsubcaption\label{fig:diff_logical_entropy}}
    {\phantomsubcaption\label{fig:diff_logical_concurrence}}
    \caption{Difference of logical purity, entanglement entropy~(cf. Eq.~\eqref{eq:entanglement_entropy}) and concurrence~(cf. Eq.~\eqref{eq:concurrence}) of the system's state before and after the stabilized $\CZ$ or $\CNOT$ gates ($18$ stabilization rounds) for the three states used in Fig.~\ref{fig:qec_and_cz_various_states}. These measures were performed after performing a logical state tomography as defined in Eq.~\eqref{eq:logic_state_tomography} and appropriately renormalizing it. We observe that when the gate involves entanglement it also increases the decoherence of the information encoded in the system. Here, due to simulation constraints data points for $\Delta\leq0.28$ are more susceptible to rounding errors. }
    \label{fig:diff_logical_quantities_stab_CZ}
\end{figure}

Fig.~\ref{fig:diff_logical_quantities_stab_CZ} illustrate the difference of these three measures for the stabilized states before and after the $\CZ$ gate. We notice that the difference in logical purity shows the same behavior as in the physical fidelity (cf. Fig.~\ref{fig:diff_fidelity_stab_CZ}). This observation together with the knowledge that the final state is back in the logical subspace highlights a decoherence of the logical information that is faster for states involving logical coherences. Despite that, entanglement entropy demonstrates that the stabilization preserves the presence of entanglement. Indeed, for separable states such as $\ket{00}_{\Delta}$ and $\ket{+0}_{\Delta}$, $\cl{S}(\rho_{AB})$ is brought back to the desired $0$ value, whereas for initial states that are maximally entangled like $\ket{\Phi_+}_{\Delta}$ this value still converges to the $\log(2)$ after the stabilized $\CZ$. Thus, the conclusion that we draw from Figs.~\ref{fig:diff_logical_purity} and~\ref{fig:diff_logical_entropy} is that the operation is indeed mixing the information encoded in the states but at the same time does not completely disentangle the desired final state. Last in order, the logical concurrence is presented in~Fig.~\ref{fig:diff_logical_concurrence}. It follows the same trend as the logical purity, i.e., entangled states show a decrease in their amount of entanglement while $\cl{C}(\rho_{AB})$ for the unentangled state converge after the stabilized $\CZ$ to the ideal $0$ value.

\subsec{Explanation of the logical information decoherence}{subsec:log_info_decoherence}

This decoherence of the logical information can be easily explained from our analysis of the finite-energy effects. Indeed, we have seen in Section~\ref{sec:marginal_distributions} that the peaks' and envelope's widths of both subsystem states increases (either in position or momentum space) which consequently increases their overlap with the opposite parity states and thus the probability of misinterpreting $\ket{0}_{\Delta}$ for $\ket{1}_{\Delta}$ (and vice-versa) during the error correction procedure. Similar conclusions are drawn from the stabilizers' expectation value (cf. Eq.~\eqref{eq:ideal_stab_expect_val} or Section~\ref{sec:eff_squeez_param}). Since either $\langle\h{S}_{x}\rangle$ or $\langle\h{S}_{z}\rangle$ increase by a factor of $\sqrt{2}$, it increases the probability of a logical error during the correction procedure. Finally, the purity studied in Section~\ref{sec:purity} also adverts this logical decoherence. Indeed, a local stabilization of each subsystem disregards the state of the other subsystem which in some sense can be seen as its partial trace. Therefore, states with logical coherence suffer the most from this decoherence.

Altogether, this analysis indicates that the state of the system after the entangling operation is no longer a proper~+1 eigenstate of the finite-energy stabilizers $\h{S}_{x,(\Delta,\kappa)}$ and $\h{S}_{z,(\Delta,\kappa)}$ defined in Eq.~\eqref{eq:finite_energy_stab}. Omitting the fact that these operators are neither unitary nor Hermitian, we can evaluate their expectation value after the $\CZ$ gate. First, we note 
\be 
\begin{split}
    \CZ^{\dagger}\!\left(\h{S}_{x,(\Delta,\kappa)} \otimes \h{I}\right)\!\CZ 
    &= 
    \CZ^{\dagger}\!\left(\h{E}_{\Delta,\kappa} \otimes \h{I}\right)\!\CZ\,\, 
    \CZ^{\dagger}\!\left(\h{S}_{x} \otimes \h{I}\right)\!\CZ\,\,
    \CZ^{\dagger}\!\left(\h{E}_{\Delta,\kappa}^{-1} \otimes \h{I}\right)\!\CZ = \\    
    &= 
    \CZ^{\dagger}\!\left(\h{E}_{\Delta,\kappa} \otimes \h{I}\right)\!\CZ \,\,
    \left(\h{S}_x \otimes \h{S}_z\right)\,\,
    \CZ^{\dagger}\!\left(\h{E}_{\Delta,\kappa}^{-1} \otimes \h{I}\right)\!\CZ
\end{split}
\ee
where the first equality comes from the definition of a finite-energy operator in Eq.~\eqref{eq:finite_energy_operation_general} and the last one from the property of the stabilizers derived in Section~\ref{sec:eff_squeez_param}. Then, commuting $\CZ$ and the energy operator $\h{E}_{\Delta,\kappa}$ results in 
\be \label{eq:commut_CZ_and_envl}
    \CZ^{\dagger}\!\left(\h{E}_{\Delta,\kappa}^{\pm1} \otimes \h{I}\right)\!\CZ = \h{E}_{\Delta,\kappa}^{\pm1}\,e^{\pm\kappa^2\,\p_1\,\q_2}\,\,e^{\mp\frac{1}{2}\kappa^2\,\q_2} + \cl{O}(\Delta^2\kappa^2) \,.
\ee
The last two exponential terms commute with $\h{S}_x \otimes \h{S}_z$. Thus the expectation value of $\h{S}_{x,(\Delta,\kappa)}$ after the gate reads
\be \label{eq:expect_value_FE_stab_x_after_CZ}
    \left\langle \CZ^{\dagger}\!\left(\h{S}_{x,(\Delta,\kappa)} \otimes \h{I}\right)\!\CZ \right\rangle_{\Delta,\kappa} =
    \left\langle \h{S}_{x,(\Delta,\kappa)} \otimes \h{S}_z \right\rangle_{\Delta,\kappa} \equiv
    \left\langle \h{S}_z \right\rangle_{\Delta,\kappa} \approx e^{-\pi\Delta^2}
\ee
with the expectation value taken over the finite-energy states $\ket{\psi}_{\Delta,\kappa}$. The expectation value for the complementary stabilizer is obtained in a similar manner except that now $\CZ^{\dagger}(\h{S}_{z} \otimes \h{I})\CZ=\h{S}_{z} \otimes \h{I}$ does not commute with the additional exponential terms from Eq.~\eqref{eq:commut_CZ_and_envl}. It thus reads as
\be
    \left\langle \CZ^{\dagger}\!\left(\h{S}_{z,(\Delta,\kappa)} \otimes \h{I}\right)\!\CZ \right\rangle_{\Delta,\kappa} =
    \left\langle \h{S}_{z,(\Delta,\kappa)} \otimes e^{-2\sqrt{\pi}\kappa^2\,\q_2} \right\rangle_{\Delta,\kappa} \equiv
    \left\langle e^{-2\sqrt{\pi}\kappa^2\,\q_2} \right\rangle_{\Delta,\kappa} \,.
\ee
Using the shifted grid state representation of $\ket{0}_{{\Delta},{\kappa}}$, we can evaluate this (pseudo-)expectation value to
\be \label{eq:expect_value_FE_stab_z_after_CZ}
    \left\langle e^{-2\sqrt{\pi}\kappa^2\,\q_2} \right\rangle_{{\Delta},{\kappa}} = 
    \prescript{}{{{\Delta},{\kappa}}}{\bra{0}}\,e^{-2\sqrt{\pi}\kappa^2\,\q_2} \ket{0}_{{\Delta},{\kappa}} 
    \approx e^{\pi\,\kappa^2}\,.
\ee
We omitted here the result for the situation when the $\kappa$ in the operator's exponent differs from the state's parameter. We observe that the obtained function is strictly increasing with $\kappa$ and it would thus be beneficial to choose $\kappa\ll1$ during the stabilization of the $\p$ quadrature (cf. next subsection). The formula in Eqs.~\eqref{eq:expect_value_FE_stab_x_after_CZ} and~\eqref{eq:expect_value_FE_stab_z_after_CZ} are numerically verified in Fig.~\ref{fig:expect_value_FE_stab}.

\begin{SCfigure}[][t!]
    \centering
    \includegraphics{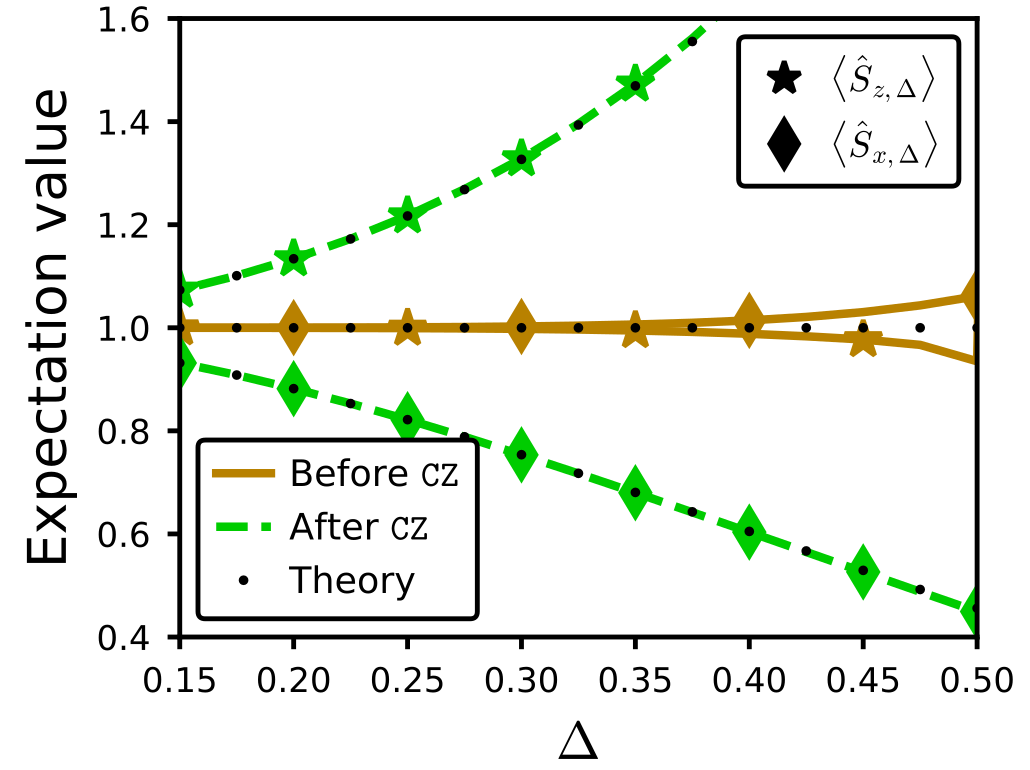}
    \caption{Expectation value of the finite energy stabilizers $\h{S}_{z,(\Delta,\kappa)}$ and $\h{S}_{x,(\Delta,\kappa)}$ (cf. Eq.~\eqref{eq:finite_energy_stab}) for symmetric finite-energy GKP states (i.e.,~${\kappa=\Delta}$) before and after the $\CZ$ gate (no stabilization involved). The theoretical (dotted) curves show that before the entangling gate their value is as expected equal to $1$, whereas after the gate they diverge exponentially with increasing $\Delta$. These curves are described by Eqs.~\eqref{eq:expect_value_FE_stab_x_after_CZ} and~\eqref{eq:expect_value_FE_stab_z_after_CZ}. These numerical results were obtained with the Bell state $\ket{\Phi_{+}}_\Delta$ which suggest that the theoretically derived expressions are state independent. }
    \label{fig:expect_value_FE_stab}
\end{SCfigure}

The scaling of the fidelity difference for states containing coherences (cf. Fig.~\ref{fig:diff_fidelity_stab_CZ}), which can be associated to the logical infidelity, can be obtained using our results from Sec.~\ref{sec:marginal_distributions} regarding the marginal distributions. Indeed, let us assume the initial state to be $\ket{+0}_{\Delta,\kappa}$. In the last subsection, we have numerically verified that all the undesired entanglement between both subsystems is erased after few rounds of QEC. Therefore, the logical infidelity should to a large extent come from the first subsystem only. The momentum marginal distribution $P(p_1)$ of the first subsystem after the $\CZ$ gate is given by the same expression as in Eq.~\eqref{eq:marginal_final} to the only difference that the periodicity is now $2\sqrt{\pi}$ instead of simply $\sqrt{\pi}$. The most naive evaluation of the logical information stored in this subsystem is to proceed to a simple binning of the marginals, i.e. integrating $P(p_1)$ over $[-\sqrt{\pi}/2,\sqrt{\pi}/2)+2\sqrt{\pi}\bb{Z}$. This calculation leads to a logical infidelity 
\be \label{eq:diff_logical_fidelity_full}
    1-F_\n{logic} = 1 - \n{erf}\left(\frac{\sqrt{\pi}}{2} \sqrt{\frac{\Delta ^2+\kappa ^2}{\left(\Delta ^2+\kappa ^2\right)^2 + \Delta ^4 \kappa ^4}}\right)
\ee
where $\n{erf}(\cdot)$ denotes the error function. For symmetric GKP states, i.e. $\kappa=\Delta$, we get the expression from the main text by taking the series expansion of the $\n{erf}(\cdot)$ functions up to $\cl{O}(\Delta^4)$. This expression is illustrated by the dashed curve in Fig.~\ref{fig:diff_logical_quantities_stab_CZ}. We observe that it describes well the general trend for states involving logical coherence, namely $\ket{+0}_\Delta$, $\ket{\Phi_+}_\Delta$ and $\ket{++}_\Delta$ states. Yet, it does not align perfectly with the $\ket{+0}_\Delta$ data points because the stabilization procedure used in simulations is more sophisticated than the simple binning of the marginals. We notice however that they lay on the curve described by $0.5(1-F_\n{logic})$. The agreement in the trend is accurate up to $\Delta\approx0.38$. Past this point the data diverge from Eq.\eqref{eq:diff_logical_fidelity_full} because of the approximate aspect of the finite-energy error-correction process (cf. Sec.~\ref{subsec:intro_to_stabil}). This is also the reason why we see a non-zero difference in fidelity for the state $\ket{00}_\Delta$.

We note that this result can also be obtained using an ideal error correction. The recovery process can in general be represented by a map $\cl{R}(\rho)=\sum_i U_i P_i \rho P_i U_i^\dagger$ where $P_i$ are projectors onto some error subspace and $U_i$ are the recovery operators from this subspace to the logical one. In the continuous variable case the sum becomes an integral over the space of errors which for the ideal GKP codes are spanned by the shifted grid states $\ket{u,v}$. The projectors read
\be
    P_{u,v} = \ket{u,v}\!\!\bra{u,v} + \ket{1,u,v}\!\!\bra{1,u,v} = \ket{u,v}\!\!\bra{u,v} + e^{i\sqrt{\pi}v} \ket{1,u,v}\!\!\bra{u+\sqrt{\pi},v}
\ee
with $\ket{\psi,u,v} := e^{iu\p}e^{iv\q}\ket{\psi}_\n{I}$ as already defined in Sec.~\ref{sec:shifted_grid_basis}. The correctability condition is that the shifts $u,\,v$ remain below $\sqrt{\pi}/2$. The correction operation is then given by the inverse displacements, namely $U_{u,v}=e^{-iv\q}e^{-iu\p}$, which bring the states $\ket{u,v}$ and $\ket{1,u,v}$ back to the ideal computational states $\ket{0}_\n{I}$ and $\ket{1}_\n{I}$, respectively. Thus, the recovery map for infinite-energy GKP states reads
\be
    \cl{R}_\n{I}(\rho) = \sum_{\mu,\mu'\in\{0,1\}} \,\,
    \int\displaylimits_{-\sqrt{\pi}/2}^{\sqrt{\pi}/2}\!\n{d}u\!\int\displaylimits_{-\sqrt{\pi}/2}^{\sqrt{\pi}/2}\!\n{d}v \,\,
    e^{i\sqrt{\pi}v(\mu'-\mu)} 
    \bra{u+\mu\sqrt{\pi},v}\rho\ket{u+\mu'\sqrt{\pi},v} \,\, \ket{\mu}_\n{I}\prescript{}{\n{I}}{\bra{\mu'}} \,.
\ee
To obtain the expression in Eq.~\eqref{eq:diff_logical_fidelity_full}, one can use $\cl{R}_\n{I}$ to get the decoded fidelity of $\ket{\psi}_{\Delta,\kappa}$ and $\CZ\ket{\psi}_{\Delta,\kappa}$ with their respective ideal states. The recovery map for the a multi-qubit state is easily generalizable in a similar fashion. It is however important to note that this map would remain local and correspond to a decoder for square GKP codes. Hence, it would produce exactly the same effects as the finite-energy correction process used in simulations. Moreover, this is a naive recovery procedure, the projector and the correctability domain of the shifts can be adapted to various situations corresponding to different QEC schemes (cf. Ref.~\cite{shaw_stabilizer_2024}).

\subsec{Improve the performance of stabilized two GKP qubit gates}{subsec:postprocessing_stabil}
As seen above, the experimentally demonstrated QEC schemes for GKP codes can help to fight against the finite-energy effects induced by an ideal entangling operation on the code but it also slightly mixes the logical information present in the system. In this way, the coherence of the information during a computation will exponentially decay with the number of $\CZ$ that an algorithm requires. This decay is illustrated in Fig.~\ref{fig:logical_fidelity_repet_CZ} for the states $\ket{00}_\Delta$ and $\ket{\Phi_{+}}_\Delta$ with an energy $\Delta\in\{0.3,0.32,0.34\}$. In particular it represents the logical infidelity with the desired state as a function of the number of $\CZ\,+\,18\,$ stabilization rounds. We can approximate the trend of these curves using $p_{\n{err}}\,N+p_{\n{err}}^0$ where $p_{\n{err}}$ would be the error per stabilized two-qubit gate and $p_{\n{err}}^0$ the initial error related to the finite-energy aspect of the states. Their values are given in the Table~\ref{tab:logical_fidelity_repet_CZ}.

\begin{table}[b!]
    \centering
    \begin{minipage}{0.5\linewidth}
		\centering
		\includegraphics{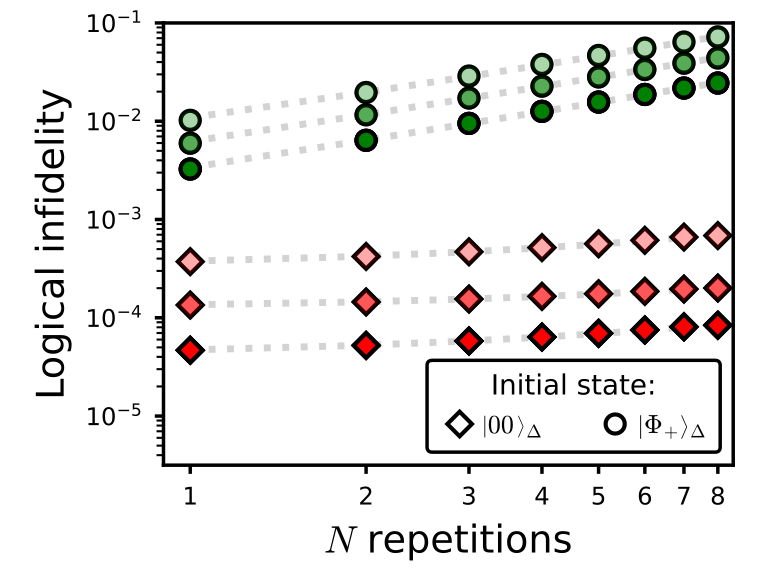}
		\captionof{figure}{Logical infidelity of system's state with the desired one as a function of the number of repetition of a stabilized $\CZ$ gate ($18$ stabilization rounds). The initial states are $\ket{00}_\Delta$ and $\ket{\Phi_{+}}_\Delta$ of energy $\Delta\in\{0.3,0.32,0.34\}$ (bottom to top in the plot). The difference in the trend between the two states is qualitatively identical to the results presented in Figs.~\ref{fig:diff_fidelity_stab_CZ} and~\ref{fig:diff_logical_quantities_stab_CZ}. The dotted lines correspond to linear fits with parameters presented in Table~\ref{tab:logical_fidelity_repet_CZ}.}
		\label{fig:logical_fidelity_repet_CZ}
	\end{minipage}
    \hfill
	\begin{minipage}{.45\linewidth}
		\centering
		\begin{tabular}{C{50pt}C{30pt}C{60pt}C{60pt}}
            Initial state   & $\Delta$ & $p_\n{err}$ & $p_\n{err}^0$ \\ \hline\hline
            \multicolumn{1}{c|}{\multirow{3}{*}{$\ket{00}_\Delta$}} &  0.34   &   $4.62\times10^{-5}$     &   $3.30\times10^{-4}$   \\
            \multicolumn{1}{c|}{}      &  0.32    &   $9.75\times10^{-6}$     &   $1.26\times10^{-4}$    \\
            \multicolumn{1}{c|}{}      &  0.30    &   $5.48\times10^{-6}$     &   $4.16\times10^{-5}$    \\ \hline
            \multicolumn{1}{c|}{\multirow{3}{*}{$\ket{\Phi_{+}}_\Delta$}} &  0.34   &  $8.85\times10^{-3}$  &   $1.99\times10^{-3}$   \\
            \multicolumn{1}{c|}{}      &  0.32    &   $5.45\times10^{-3}$     &   $7.83\times10^{-4}$    \\
            \multicolumn{1}{c|}{}      &  0.30    &   $3.05\times10^{-3}$     &   $3.32\times10^{-4}$    
        \end{tabular}
        \caption{Parameter of the linear fits from Fig.~\ref{fig:logical_fidelity_repet_CZ}. The fitting formula is given by $p_{\n{err}}\,N+p_{\n{err}}^0$ where $p_{\n{err}}$ is the error per stabilized two-qubit gate and $p_{\n{err}}^0$ the initial error related to the finite-energy aspect of the states.}
		\label{tab:logical_fidelity_repet_CZ}
	\end{minipage}
\end{table}

The question is now to determine if it is possible to improve the performances of the error-corrected two GKP qubit gate by modifying only the QEC procedure. The answer is yes and we will show below how it can be done and a first improvement to it.

\begin{figure}[b!]
    \centering
    \includegraphics{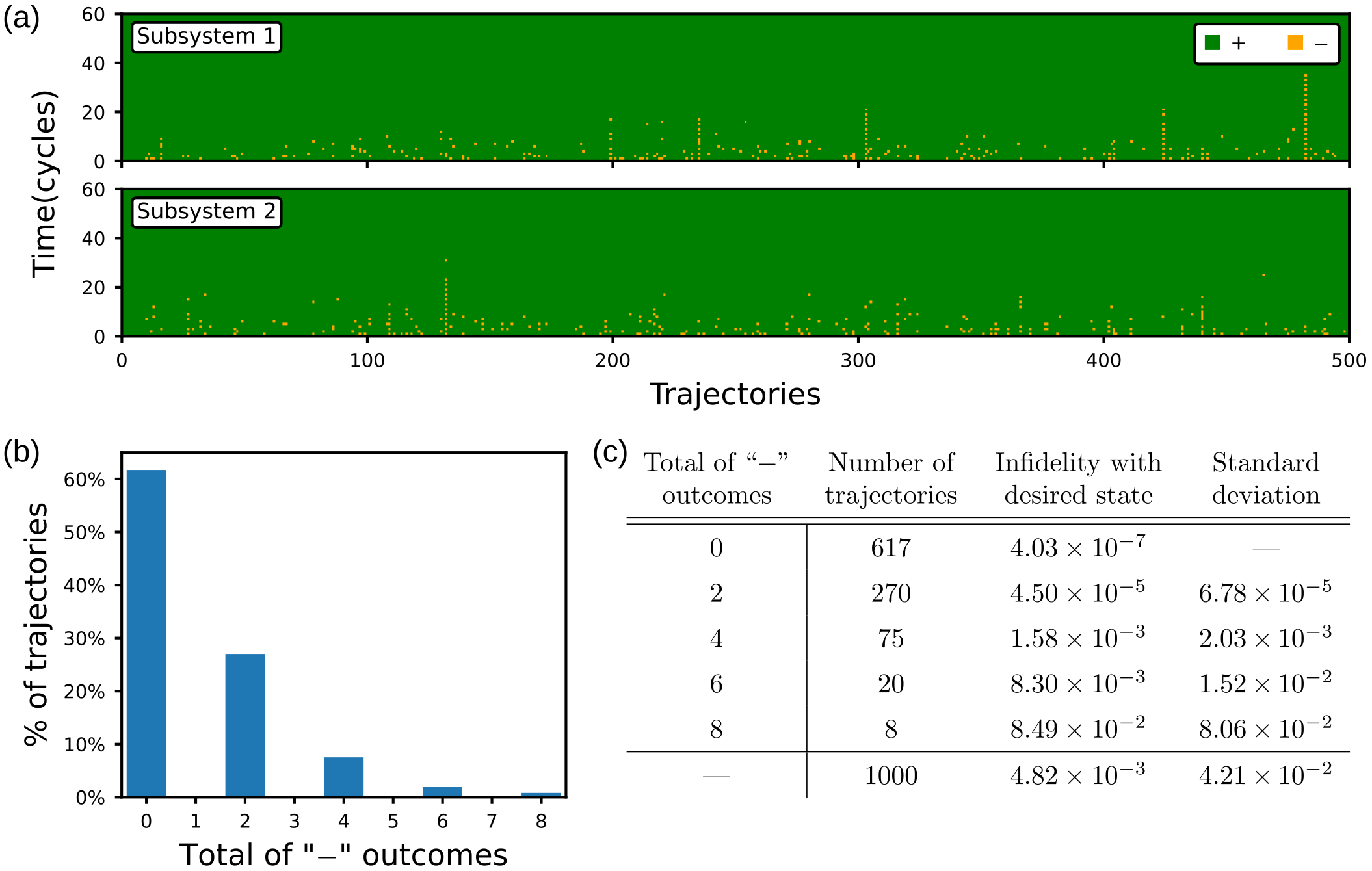}
    {\phantomsubcaption\label{fig:trajectories_outcomes}}
    {\phantomsubcaption\label{fig:trajectories_histogram}}
    {\phantomsubcaption\label{fig:trajectories_infidelity}}
    \caption{Trajectories of system through the error correction cycles after a $\CZ$ gate. The simulations were done with the Bell state $\ket{\Phi_{+}}_{\Delta}$ that has been beforehand stabilized for $15$ rounds and subjected to a $\CZ$ gate. The error correction cycles correspond were represented in terms of Kraus maps with operators $K_{j,(\Delta,\Delta)}^{(\pm)}$ where $j\in\{x,z\}$ (cf. Eq.~\eqref{eq:outcome_stab_kraus_op}). Every outcome has been drawn according to the probabilities calculated using Eq.~\eqref{eq:outcomes_probability} and the post-measurement state was then obtained following Eq.~\eqref{eq:post_measurement_state}.}
    \label{fig:trajectories}
\end{figure}

The stabilization Kraus operators presented in Eq.~\eqref{eq:stabilization_kraus_operators} were obtained using the states $\ket{0}$ and $\ket{1}$ of the auxiliary qubit. If we were to measure this qubit instead of resetting it after each cycle we would be able to make use of the outcome information in post-processing to determine if the stabilization procedure went well or not. This technique was in fact recently employed in~\citet{sivak_real_time_2023} for demonstrating QEC beyond break-even using a single GKP qubit. However, in the chosen basis $\{\ket{0},\ket{1}\}$ both outcomes have an equal chance to appear which makes it more difficult to work with. To make it more intuitive, we measure the auxiliary system in a rotated basis $\{\ket{\pm}\propto\ket{0}\pm\ket{1}\}$. In this basis, the dissipative stabilization channel $\cl{R}_{x,(\Delta,\kappa)}$ would thus be given by the Kraus operators
\be \label{eq:outcome_stab_kraus_op}
    K_{x,(\Delta,\kappa)}^{(\pm)}=\frac{1}{\sqrt{2}}\left(K_{x,(\Delta,\kappa)}^{(0)}\pm K_{x,(\Delta,\kappa)}^{(1)}\right)
\ee
(similarly for $\cl{R}_{z,(\Delta,\kappa)}$). In the measurement-based framework every stabilization cycle $\cl{R}_{x/z,(\Delta,\kappa)}$ would give an outcome $+$ with a probability of almost $1$ if the system is an eigenstate of the finite-energy stabilizer $\h{S}_{x/z,(\Delta,\kappa)}$. The probability to observe the opposite outcome is then nearly $0$ when one are in the computational subspace. If the system leaves this subspace, every observation of a $\ket{-}$ state heralds a process in which the bosonic state was transferred one level down the error hierarchy. We refer the reader to the work of~\citet{sivak_real_time_2023} for a further discussion about the physical meaning of these measurements (N.B.: the authors use the notation $g/e$ for the outcomes $+/-$). In our noiseless case, every outcome would correspond to a false-positive error, meaning that the GKP qubit has been stabilized towards an opposite parity state. Mathematically speaking, the outcome probabilities are given by
\be \label{eq:outcomes_probability}
    P_{x/z,(\Delta,\kappa)}^{(\pm)}\big(\rho\big) = 
    \n{Tr}\left(K_{x/z,(\Delta,\kappa)}^{(\pm)\,\dagger}K_{x/z,(\Delta,\kappa)}^{(\pm)}\,\rho \right) 
    \qquad \text{or} \qquad
    P_{x/z,(\Delta,\kappa)}^{(\pm)}\big(\ket{\psi}\big) = 
    \bra{\psi}K_{x/z,(\Delta,\kappa)}^{(\pm)\,\dagger}K_{x/z,(\Delta,\kappa)}^{(\pm)}\,\ket{\psi}
\ee
for a pure state $\ket{\psi}$ and the post-measurement state reads
\be \label{eq:post_measurement_state}
    \ket{\psi_\pm} = \frac{1}{\sqrt{P_{x/z,(\Delta,\kappa)}^{(\pm)}\big(\ket{\psi}\big)}} \, 
    K_{x/z,(\Delta,\kappa)}^{(\pm)\,\dagger}K_{x/z,(\Delta,\kappa)}^{(\pm)}\,\ket{\psi}\,.
\ee
We can therefore analyze the trajectories that the system's state takes during a stabilized entangling gate. Fig.~\ref{fig:trajectories} illustrates the outcome strings for $10^3$ trajectories. The simulations were performed using the state vector representation of the system and updated probabilistically using expressions in Eqs.~\eqref{eq:outcomes_probability} and~\eqref{eq:post_measurement_state}. The state is first initialized in $\ket{\Phi_{+}}_\Delta$ with $\Delta=0.30$, stabilized for $15$ rounds and finally subjected to a $\CZ$ gate. Then, the state is stabilized for some additional $15$ rounds consisting as mentioned previously of two cycles of $\cl{R}_{\Delta}=\cl{R}_{x,(\Delta,\Delta)}\circ\cl{R}_{z,(\Delta,\Delta)}$. The measurement outcomes for the stabilization after the gate are organized accordingly in Fig.~\ref{fig:trajectories_outcomes}.

As a first observation we note that at the end of $15$ rounds the system always converges to the computational subspace which confirms what we already concluded in the previous subsection. Filtering every run in terms of the number of ``$-$'' outcomes that the joint subsystems contain features that most of the trajectories are errorless or comprise of two errors at most. This filtering also highlights the peculiar behavior of the stabilized $\CZ$ gate that the ``$-$'' outcomes are strongly correlated as only an even number of them is permitted. Table.~\ref{fig:trajectories_infidelity} presents the average infidelity with the desired stabilized $\ket{\Phi_-}_\Delta$ state for different categories of trajectories. We see that the $0$ and $2$ error trajectories show a several order of magnitude lower infidelity than what we observed using dissipative stabilization (cf.~Fig.~\ref{fig:diff_fidelity_stab_CZ}). The rest of the trajectories show higher values such that when the number of ``$-$'' outcomes is larger or equal to $10$ the fidelity drops below $90\%$ (happens in $\sim1\%$ of all trajectories). All together we recover in average the infidelity obtained with dissipative stabilization. The capability to record the outcome of every stabilization cycle could offer the ability to discard in post-processing the runs that were problematic and thus improve algorithms' performances.

However, this technique can also help to improve the performance of the gate in average. Indeed, we empirically determined that the first $\cl{R}_{z,(\Delta,\kappa)}$ cycles just after the $\CZ$ gate show a higher probability $P_{z,(\Delta,\kappa)}^{(+)}$ if one fixes ${(\Delta,\kappa)\rightarrow\left(\Delta/\sqrt{2},0\right)}$. This choice of parameters can be explained by two analysis made previously. The $\sqrt{2}$ comes from our observation that for symmetric states (i.e.,~$\Delta=\kappa$) the envelope's width of each subsystem will decrease by this factor (cf.~Eq.~(1)), it is thus expected that a $\h{S}_{z,(\Delta,\kappa)}$ stabilizer with $\Delta/\sqrt{2}$ should be more appropriate for the correction procedure. The setting $\kappa=0$ on the other side comes from our derivation of the expectation value of the finite-energy stabilizer $\h{S}_{z,(\Delta,\kappa)}$ in Eq.~\eqref{eq:expect_value_FE_stab_z_after_CZ}. Since the quantity grows exponentially with $\kappa$, it is natural to set this parameter as low as possible. In practice, $\kappa=\Delta/\sqrt{2}$ is sufficient enough. Fig~\ref{fig:proba_outcome_plus_after_CZ} compares the probability $P_{z,(\Delta,\kappa)}^{(+)}$ for the parameters $\Delta=\kappa=\Delta_\n{sim}/\sqrt{2}$ with the initial naive ones $\Delta=\kappa=\Delta_\n{sim}$ used so far. As above the input state is $\ket{\Phi_{+}}_{\Delta_\n{sim}}$ with $\Delta_\n{sim}\in(0.1,0.4)$ that has been beforehand stabilized for $15$ rounds and subjected to a $\CZ$ gate. 

\begin{SCfigure}[][t!]
    \centering
    \includegraphics{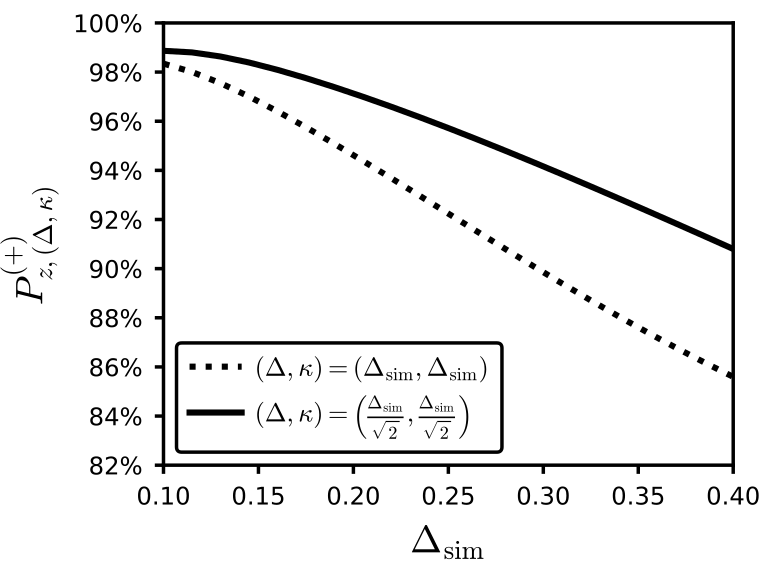}
    \caption{Probability $P_{z,(\Delta,\kappa)}^{(+)}$ of outcome ``$+$'' for a stabilization cycle $\cl{R}_{z,(\Delta,\kappa)}$ after the $\CZ$ gate for two different regimes. In the first one (dotted line) the stabilization is performed as with the same parameters $(\Delta,\kappa)$ that correspond to the input state ones. In the second case (full line) the parameters of the stabilization cycle are reduced by a factor of $\sqrt{2}$. The simulations were done with the input state $\ket{\Phi_{+}}_{\Delta}$ that has been beforehand stabilized for $15$ rounds and subjected to a $\CZ$ gate. }
    \label{fig:proba_outcome_plus_after_CZ}
\end{SCfigure}

Fig.~\ref{fig:trajectories_improved_QEC} presents how this optimization would modify the trajectories of the system's state and their statistics. In these simulations, the first round of stabilization consists of two cycles of ${\cl{R}_{x,(\Delta,\Delta)}\circ\cl{R}_{z,(\Delta/\sqrt{2},\Delta/\sqrt{2})}}$. Since the stabilized state has now a larger energy parameter in the momentum space we adiabatically reduce it during the next nine rounds by applying twice $\cl{R}_{\Delta,\epsilon}=\cl{R}_{x,(\Delta,\Delta)}\circ\cl{R}_{z,(\Delta/\sqrt{\epsilon},\Delta/\sqrt{\epsilon})}$ with $\epsilon\in\{1.9,\,1.8,\,\ldots,\,1.1\}$. These steps correspond to the lattice reshaping suggested in Ref.~\cite{royer_stabilization_2020} which can also be seen as a code deformation technique. The remaining stabilization rounds were performed as described previously using $\cl{R}_{\Delta}$. These new simulations show first of all that the trajectories qualitatively differ from the the ones in Fig.~\ref{fig:trajectories_outcomes}. Indeed, we notice that long sequences of ``$-$'' outcomes are now less probable. This is due to the fact that these outlier sequences were mostly starting due to an error in the first cycle. Since the probability of having the wrong outcome using $\cl{R}_{z,(\Delta/\sqrt{2},\Delta/\sqrt{2})}$ is $\sim5\%$ lower (for $\Delta=0.3$) than with $\cl{R}_{z,(\Delta,\Delta)}$ we reduced by almost $10\%$ the probability of a false positive error in the first cycle in either of the two subsystems. Moreover, in this setting the fidelity with the desired state also increased compared to the naive correction procedure used above. Fig.~\ref{fig:trajectories_infidelity_improved} shows that the infidelity for trajectories with $0$ or $2$ outcomes is in average $\sim3-4$ times lower than in the situation depicted in Fig.~\ref{fig:trajectories} and the total average infidelity over all the shots is twice as low as before.

This was only the improvement of the first two cycles of QEC after a two GKP qubit gate. We believe that the correction cycles can be optimized even further using for instance reinforced learning techniques already demonstrated in Refs.~\cite{eickbusch_fast_2022,sivak_real_time_2023}. However, we leave this aspect for future investigations.

\begin{figure}[t!]
    \centering
    \includegraphics{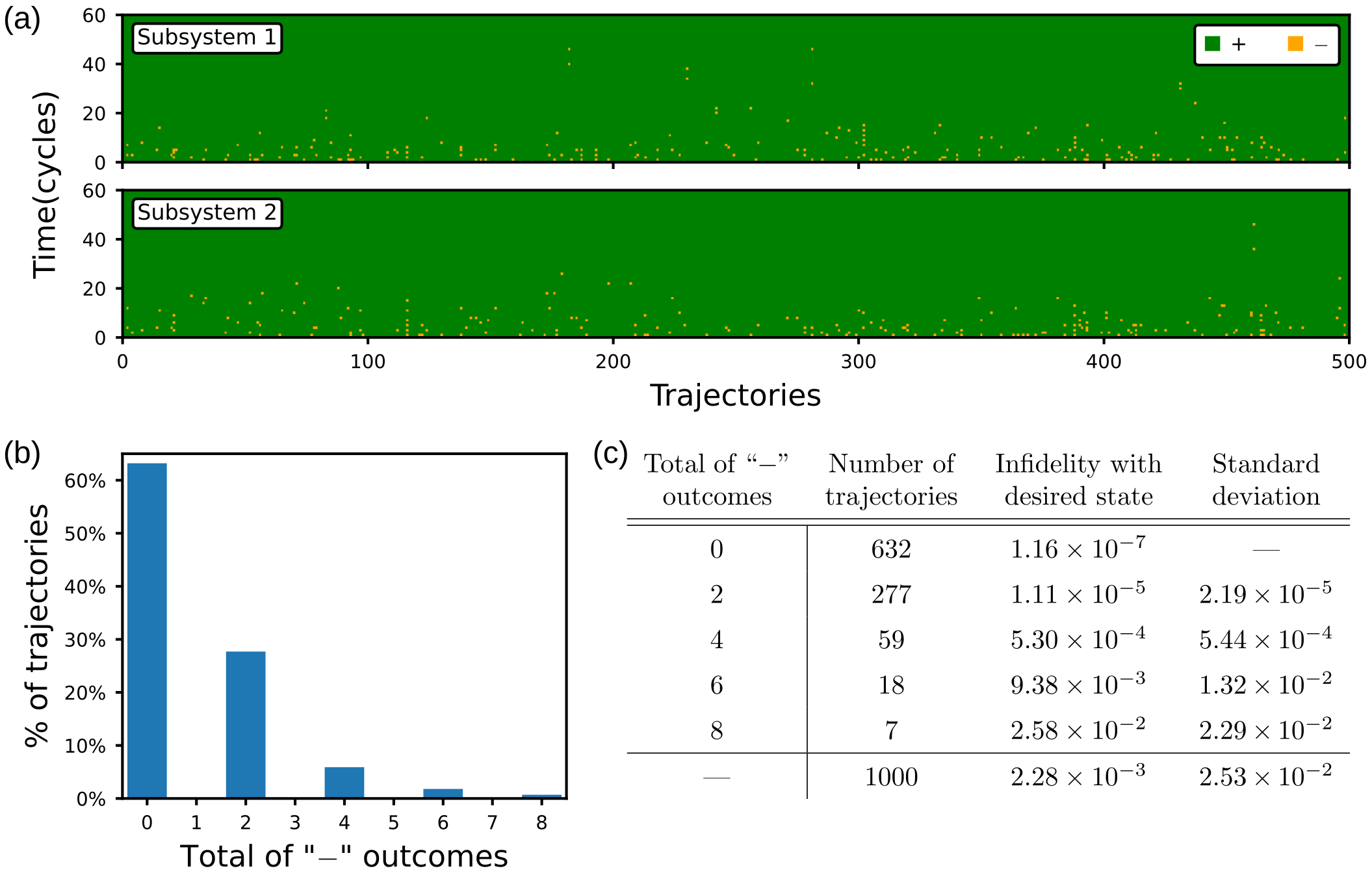}
    {\phantomsubcaption\label{fig:trajectories_outcomes_improved}}
    {\phantomsubcaption\label{fig:trajectories_histogram_improved}}
    {\phantomsubcaption\label{fig:trajectories_infidelity_improved}}
    \caption{Trajectories of system through some improved error correction cycles after a $\CZ$ gate. The simulations were done in a similar manner as for Fig.~\ref{fig:trajectories} but with a modified the error correction scheme. The first stabilization cycle corresponds to the map ${\cl{R}_{x,(\Delta,\Delta)}\circ\cl{R}_{z,(\Delta/\sqrt{2},\Delta/\sqrt{2})}}$ where stabilization of the $p$ quadrature is using the Kraus operators $K_{z,(\Delta/\sqrt{2},\Delta/\sqrt{2})}^{(\pm)}$. In the nine subsequent rounds of QEC the parameter of the first map was adiabatically (in a linear fashion) brought to the desired value $\Delta$.}
    \label{fig:trajectories_improved_QEC}
\end{figure}

\iftoggle{arXiv}{ \PRLsec{Finite-energy gates}{sec:FE_gates}
}{ \section{Finite-energy gates} \label{sec:FE_gates}}

So far we only considered ideal unitary entangling gates acting on finite-energy GKP states and have seen that this results in some undesired effects that ultimately can to some extent be corrected using local error-correction schemes. An alternative approach is to try to realize the finite-energy version of these gates which would then preserve the intrinsic energy of the bosonic states and would thus lead to higher gate fidelities. To obtain the finite-energy version $\h{U}_{\Delta,\kappa}$ of a GKP gate $\h{U}_\n{I}$ using the energy envelope operator $\h{E}_{\Delta,\kappa}$ (cf. Eq.~\eqref{eq:energy_op})
\be \label{eq:finite_energy_operation_general}
    \h{U}_{\Delta,\kappa} := \h{E}_{\Delta,\kappa}\,\h{U}_\n{I}\,\h{E}_{\Delta,\kappa}^{-1} \, .
\ee
This definition is motivated by the fact that finite-energy states are defined as $\ket{\psi}_{\Delta,\kappa}=\cl{N}_{\Delta,\kappa}\h{E}_{\Delta,\kappa}\ket{\psi}_\n{I}$. Therefore, under $\h{U}_{\Delta,\kappa}\ket{\psi}_{\Delta,\kappa}$ the envelope operator $\h{E}_{\Delta,\kappa}$ would cancel out with its inverse leaving room for the ideal interaction to happen and thus bypassing eventual undesired effects due to the finite size of the input state. For gates $\h{U}_\n{I}$ involving multiple qubits at the same time the transformation that leads to $\h{U}_{\Delta,\kappa}$ is given by tensor product of the single qubit energy operator $\h{E}_{\Delta,\kappa}$.

Applying Eq.~\eqref{eq:finite_energy_operation_general} to the ideal $\CZ$ and $\CNOT$ gates defined in Eqs.~\eqref{eq:CZ_operator} and~\eqref{eq:CNOT_operator} yields to
\be \label{eq:finite_energy_gates_full_with_kappa}
\begin{split}
    \CZ_{\Delta,\kappa} &= \exp\!\Bigg[ i 
    \left(\cosh^2(\Delta\kappa)\,\q_1\q_2 - \frac{\kappa^2}{\Delta^2}\sinh^2(\Delta\kappa)\,\p_1\p_2 \right) -
    \frac{\kappa}{\Delta}\sinh(\Delta\kappa)\cosh(\Delta\kappa) \bigg(\q_1\p_2 + \p_1\q_2\bigg) \Bigg] \,, \\
    \CNOT_{\Delta,\kappa} &= \exp\!\Bigg[ i 
    \bigg(\cosh^2(\Delta\kappa)\,\q_1\p_2 + \sinh^2(\Delta\kappa)\,\p_1\q_2 \bigg) +
    \sinh(\Delta\kappa)\cosh(\Delta\kappa) \bigg( \frac{\Delta}{\kappa}\,\q_1\q_2 - \frac{\kappa}{\Delta}\,\p_1\p_2\bigg) \Bigg] \,.
\end{split} 
\ee
The finite-energy version of the $\CNOT$ gate can also be obtained from $\CZ_{\Delta,\kappa}$ by simply replacing $\q_2\rightarrow\p_2$ as well as $\p_2\rightarrow-(\Delta^2/\kappa^2)\q_2$ which is equivalent to perform two Hadamard transform on the target mode before and after the $\CZ$. Note that the obtained operators have both a unitary and a non-unitary components. For the $\CZ_{\Delta,\kappa}$ for instance, the unitary part is non-surprisingly dominated by the $\q_1\q_2$ term which corresponds to the desired ideal interaction. The non-unitary component is identified by the $\q_1\p_2 + \p_1\q_2$ term which in turn coincides with the Hamiltonian of an ideal $\CNOT$. This behavior is similar to what has been observed in finite-energy stabilizers analyzed by~\citet{royer_stabilization_2020} and realized by~\citet{de_neeve_error_2022}. Indeed, $\h{S}_{x}$ ($\h{S}_{z}$) which in the ideal case is parallel to $\p$ ($\q$) operator requires a small amount of the orthogonal quadrature $\q$ ($\p$) in order to preserve the envelope of the input state (cf. Eq.~\eqref{eq:finite_energy_stab}).

In the limit of $\Delta^2,\kappa^2\ll1$, expressions in Eq.~\eqref{eq:finite_energy_gates_full_with_kappa} can be approximated by
\be \label{eq:finite_energy_gates_with_kappa}
    \CNOT_{\Delta,\kappa} \approx e^{i\q_1\p_2 + \Delta^2\,\q_1\q_2 - \kappa^2\,\p_1\p_2 }\qquad\text{and}\qquad
    \CZ_{\Delta,\kappa}   \approx e^{i\q_1\q_2 - \kappa^2\,\big( \q_1\p_2 + \p_1\q_2 \big) }\,,
\ee
where only the terms of the order $\cl{O}(\Delta^2)$ and $\cl{O}(\kappa^2)$ were kept. Moreover, these formula are exact up to the second order in the product $\Delta\kappa$. Eq.~(5) from the main text is then obtained by setting $\kappa=\Delta$. Finally, here we assumed that both states have the same energy parameters, but similar conclusions can be drawn if we perform the same analysis with $\h{E}_{\Delta_1,\kappa_1}\otimes\h{E}_{\Delta_2,\kappa_2}$ as the transformation operator.

\subsec{Implementation of finite-energy gates}{sec:FE_gate_implement}

The main challenge is now to realize these finite-energy entangling operations using experimentally accessible resources. We can lift the non-unitarity of these gates by going into a higher dimensional Hilbert space (i.e., using the Stinespring dilation theorem) that we then trace out. This technique called reservoir engineering has been proposed in Refs.~\cite{de_neeve_error_2022,royer_stabilization_2020} for the implementation of finite-energy stabilizers. In this subsection, we follow the same approach and present below some first strategies for the implementation of $\CZ_{\Delta,\kappa}$.

We can realize dissipatively and in a continuous fashion the desired interactions by coupling the two oscillators to a common lossy bosonic bath 
\be \label{eq:dissip_gate}
    \dot{\rho} = -i \left[\h{H}_{\Delta,\kappa}\,,\,\rho\right]+\gamma\,\cl{D}\big[\,\h{b}\,\big](\rho) \qquad\text{with}\qquad 
    \h{H}_{\Delta,\kappa} = \sqrt{\Gamma}\left( \h{O}_{\Delta,\kappa}\,\h{b}^\dagger + \h{O}_{\Delta,\kappa}^\dagger\,\h{b} \right)\,.
\ee
Here, $\h{b}$ is the bath's annihilation operator and $\h{H}_{\Delta,\kappa}$ represents the system-bath interaction of strength $\sqrt{\Gamma}$ where $\h{O}_{\Delta,\kappa}$ is logarithm of the desired operation, e.g. $\h{O}_{\Delta,\kappa} \approx -\q_1\q_2 + i\,\kappa^2\,(\q_1\p_2 + \p_1\q_2)$ for the $\CZ_{\Delta,\kappa}$ gate. Finally, $\gamma$ corresponds to the dissipation rate of the auxiliary mode's energy. For a sufficiently large $\gamma$, we can adiabatically eliminate the bath~\cite{gardiner_quantum_2004} and the dynamics of the system becomes $\dot{\rho} = \Gamma\,\cl{D}[\,\h{O}_{\Delta,\kappa}\,](\rho)$. The bosonic bath can also be replaced by a two-level system which is continuously cooled down, in this case $\h{b}\rightarrow\s_{-}$. However, as pointed out in Ref.~\cite{royer_stabilization_2020}, engineering an interaction with an auxiliary system described by $\h{H}_{\Delta,\kappa}$ is challenging in practice. {\color{\mycolor} On top of the experimental challenge of realizing such an interaction Hamiltonian, the continuous and dissipative implementation of $\CZ_{\Delta,\kappa}$ poses a fundamental problem which is that $\h{O}_{\Delta,\kappa}\ket{\psi}_\Delta$ is not necessarily trivial for all states since some of them must aquire a phase of $\pi$.}

To solve the experimental implementation challenge we take the alternative approach of discretizing the dynamics of the system using a collisional model of dissipation~\cite{ciccarello_collision_2017}. In this framework, the unitary evolution given by $\h{H}_{\Delta,\kappa}$ is interleaved by some active reset/cooling of the auxiliary system. Importantly the evolution must be sufficiently short such that the bath stays at zero temperature (i.e., on average it contains less than one excitation). Overall, the dynamics can be written as
\be \label{eq:collisional_model_gate}
    \rho_{n+1} = \n{Tr}_\n{B}\Big( \h{U}_\n{S+B}\,\big(\rho_{n}\otimes\ket{0}\!\!\bra{0}\!\big)\,\h{U}_\n{S+B}^\dagger \Big) \qquad\text{where}\qquad 
    \h{U}_\n{S+B} = \exp\!\Big(\!-i\delta t\h{H}_{\Delta,\kappa}\Big) \,,
\ee
with indices $\n{S}$ and $\n{B}$ denoting the system of interest and the bath, respectively. {\color{\mycolor} To address the fundamental issue, we apply this dissipative step only once, effectively embodying a block-encoding of the nonunitary operation $\CZ_{\Delta,\kappa}$.  For additional insights, readers are directed to Ref.~\cite{gilyen_quantum_2019}}. Separating the unitary and dissipative parts, allows us to approximate the desired interaction $\h{H}_{\Delta,\kappa}$ with some experimentally realizable coupling Hamiltonians. A common method to proceed with is the so-called Suzuki-Trotter decomposition (a.k.a. the exponential product formula) which in the general case reads~\cite{hatano_finding_2005} 
\be \label{eq:trotter_decomp_formula}
    e^{x(A+B)} = e^{p_0xA}\,e^{p_1xB}\,e^{p_2xA}\,e^{p_3xB}\,\cdots\,e^{p_MxB}+\cl{O}\left(x^{M+1}\right)
\ee
where $x\in\bb{R}$, $A$ and $B$ are arbitrary operators and $p_k\in\bb{R}$ are some appropriately chosen parameters that depend on the correction order $M$. The first- and second-order approximations are parametrized by ${\{p_0=1,\,p_1=1\}}$ and ${\{p_0=1/2,\,p_1=1,\,p_2=1/2\}}$. We apply these formula to $\h{U}_\n{S+B}$ with $\h{H}_{\Delta,\kappa}$ realizing the finite-energy $\CZ$ gate by coupling to a two-level system, i.e.,
\be \label{eq:unitary_FE_cz_gate}
    \h{U}_\n{S+B} = \exp\!
    \Big(\!-i\delta t\sqrt{\Gamma}
    \left( \h{O}_{\Delta,\kappa}\,\s_+ + \h{O}_{\Delta,\kappa}^\dagger\,\s_- \right)
    \Big) = \exp\!
    \Big( i\delta t\sqrt{\Gamma}
    \big(\q_1\q_2\s_x - \kappa^2\,(\q_1\p_2 + \p_1\q_2)\s_y \big)
    \Big)\,
\ee
and get
\be \label{eq:trotter_cz_gate}
\begin{split}
    \textnormal{1st order Trotter:} \qquad 
    &\h{U}_\n{S+B} \approx 
    e^{i\delta t\sqrt{\Gamma}\,\q_1\q_2\s_x }\,\,
    e^{-i\delta t\sqrt{\Gamma}\kappa^2(\q_1\p_2 + \p_1\q_2)\s_y}\,, \\
    \textnormal{2nd order Trotter:} \qquad
    &\h{U}_\n{S+B} \approx 
    e^{-i\frac{1}{2}\delta t\sqrt{\Gamma} \kappa^2(\q_1\p_2 + \p_1\q_2)\s_y}\,\,
    e^{i\delta t\sqrt{\Gamma}\,\q_1\q_2\s_x }\,\,
    e^{-i\frac{1}{2}\delta t\sqrt{\Gamma} \kappa^2(\q_1\p_2 + \p_1\q_2)\s_y}\,.
\end{split}
\ee
We must now implement two distinct spin-dependent Hamiltonians, namely $\h{H}_1 \propto \q_1\q_2\s_x$ and $\h{H}_2 \propto \kappa^2(\q_1\p_2 + \p_1\q_2)\s_y$. Using the additional assumption that $\kappa^2\ll1$, we can split the unitaries involving $\h{H}_2$ into two exponential operators with the terms $\q_1\p_2\s_y$ and $\p_1\q_2\s_y$, respectively. Hence, this derivation suggests that we can approximate a finite-energy $\CZ_{\Delta,\kappa}$ operation using some ideal $\CNOT$ and $\CZ$ interactions that are conditioned on the states of an auxiliary two-level system's state. Despite seeming difficult to implement, we can think of several ways to realize these spin-dependent unitaries (N.B. we always assume that the auxiliary two-level system is fully controllable, meaning that arbitrary spin rotations are available):

As a first example, one can for instance consider to have access to some conditional anti-symmetric and symmetric beamsplitter operations given by 
\be \label{eq:conditional_beamsplitters}
    C\hat{B}_A(\theta) = e^{i\theta(\q_1\q_2 + \p_1\p_2)\s_z} 
    \qquad \text{and} \qquad
    C\hat{B}_S(\theta) = e^{i\theta(\q_1\p_2 - \p_1\q_2)\s_z}.
\ee
Indeed, it is straightforward to see that using $C\hat{B}_A(\theta)$ in the approximate decomposition $\CZ_\n{approx}(\theta,\kappa)$ (cf. Eq.~\eqref{eq:CZ_CNOT_approx_decomp}) we can realize Hamiltonians of type $\q_1\q_2\s_z$ (provided that we are in the right $(\theta,r)$ regime). Then, $\h{H}_1$ can be implemented with two additional spin rotations (or Hadamard gates) before and after the conditional beamsplitter interaction. Terms of the form $\kappa^2\q_1\p_2\s_y$ and $\kappa^2\p_1\q_2\s_y$ can be similarly obtained using instead $C\hat{B}_S(\theta)$ and change the beamsplitter and squeezing parameters accordingly. Thus, all the necessary elements for $\h{U}_\n{S+B}$ can be realized using this method. A gate-based representation of this schemes is shown in Fig.~\ref{fig:circuit_cond_BS_and_Sq}(left). In fact, this experimentally demonstrated spin-dependent interaction~\cite{gao_programmable_2018,gan_hybrid_2020,chapman_high_2023} together with single-mode squeezing and arbitrary spin rotations can be straightforwardly used for implementing arbitrary two-mode operators with linear Bogoliubov transformations that are conditioned on the state of the auxiliary system. This property (that we are not going to prove here) follows directly from the Bloch-Messiah theorem (employed in Section~\ref{sec:gate_decompositions}) and the expressivity of single-qubit rotations. Note that the first order Trotter decomposition of $\h{U}_\n{S+B}$ would amount to three conditional beamsplitters whereas the second order one to five, i.e., circuit lengths that could eventually be detrimental for architectures with low-coherence auxiliary qubits.

\begin{figure*}[t!]
    \centering
    \includegraphics[scale=1.1]{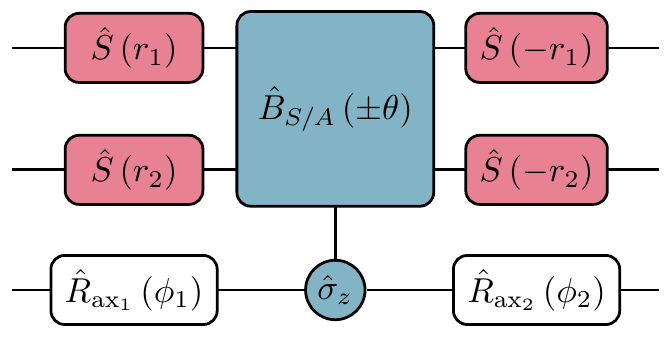}
    \qquad
    \includegraphics[scale=1.1]{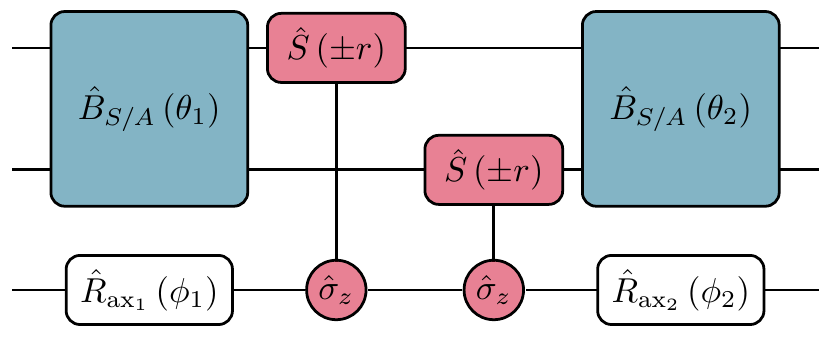}
    \caption{Quantum circuits realizing the conditional operations required for the implementation of unitary $\h{U}_\n{S+B}$ (cf.~Eq.~\eqref{eq:trotter_cz_gate}). Both circuits take in input two bosonic mode (top wires) and one auxiliary qubit (bottom wire). Provided an appropriate choice of parameters the two schemes would allow the realization of Hamiltonians of types $\q_1\q_2\s_z$, $\q_1\p_2\s_z$ and $\p_1\q_2\s_z$. Then, the exponential operators composing the unitary $\h{U}_\n{S+B}$ are achieved with the help of specific spin rotations $\h{R}_{\n{ax}_j}(\phi_j)$ along the axes $\n{ax}_j$ and angles $\phi_j$. The left circuit uses the conditional beamsplitter interaction given in Eq.~\eqref{eq:conditional_beamsplitters}, whereas the right one utilizes the spin-dependent squeezers in Eq.~\eqref{eq:conditional_squeezers}. The sign of both operations is conditioned on the states $\ket{0}$ or $\ket{1}$ (i.e., eigenbasis of the Pauli Z operator $\s_z$) of the auxiliary system (circles). }
    \label{fig:circuit_cond_BS_and_Sq}
\end{figure*}

Alternatively, we can make use of conditional displacements $C\h{D}(\alpha)$ (cf. Eq.~\eqref{eq:spin_dependent_displ}). Indeed, these operations available in both trapped ions and superconducting cavities architectures have been proven to be a sufficient resource (together with spin rotations) for a universal control of a single oscillator~\cite{eickbusch_fast_2022}. We can thus replace the standard single-mode squeezing operators by conditional squeezers 
\be \label{eq:conditional_squeezers}
    C\hat{S}(r) = e^{i \frac{1}{2}r(\q\p + \p\q)\s_z}
\ee
which would be realized with a specific combination of $C\h{D}(\alpha)$ and spin rotations. Hence, by the same argument as above we can claim that a universal control of a pair of finite-energy GKP state can also be reached with standard beamsplitters (i.e., $\h{B}_A(\theta)$ and $\h{B}_S(\theta)$), spin-dependent displacements and arbitrary rotations of the auxiliary qubit (cf.~Fig.~\ref{fig:circuit_cond_BS_and_Sq}(right)). In this situation, however, $\h{U}_\n{S+B}$ would require the spin to couple to both oscillators as the squeezing operations were always considered local. It would then imply the existence of a non-zero probability of an error propagating from one bosonic mode to the other one through the auxiliary system. Moreover, $C\hat{S}(r)$ might require a large amount of $C\h{D}(\alpha)$ and spin rotations leading to potentially long circuits. Note that a direct implementation of spin-dependent squeezing operators would also be possible. Such operations have recently been proposed and realized in trapped ion systems using higher order terms in the laser-ion Lamb-Dicke expansion~\cite{sutherland_universal_2021,katz_n-body_2022,katz_programmable_2023,shapira_robust_2023}. {\color{\mycolor}For superconducting microwave cavities, the conditional version of squeezing and beamsplitter interactions can be obtained straightforwardly using the same superconducting elements as their non-conditional counterparts~\cite{pietikainen_controlled_2022} -- the conditional beamsplitter was realized in Refs.~\cite{gao_programmable_2018,chapman_high_2023}.}

We can however think of a third situation where both spin-dependent displacements and beamsplitters are available. In this case we believe that some operation efficient schemes that directly implement the unitary in Eq.~\eqref{eq:unitary_FE_cz_gate} can be developed. Similarly to the universal control of a single oscillator~\cite{eickbusch_fast_2022}, these schemes could be found using some numerical optimization strategies (e.g. reinforced learning) and their performance would depend on the number of spin-dependent operations as well as the coherence time of the auxiliary two-level system.

Crucially, these three solutions employ an auxiliary system acting solely as a bath to implement approximations of the finite-energy version of the desired entangling gates (as given in Eq.~\eqref{eq:finite_energy_gates_with_kappa}). In other words, no logical information is ever stored in the auxiliary spin. In fact, the non-unitary terms in $\CZ_{\Delta,\kappa}$ or alternatively the Hamiltonian $\h{H}_2$ in $\h{U}_\n{S+B}$ correspond effectively to a modified simultaneous error-correction procedure of both oscillators' states. This collective recovery map has been recently suggested by \citet{shaw_stabilizer_2024}. The authors derived it using an ideal (non-physical) error correction scheme as well as their new stabilizer subsystem decomposition for GKP codes. They observed that this modified recovery map can indeed correct identically the finite-energy effects. In what we presented above, we went further by describing a first possible implementation on current experimental platforms.

Finally, a two GKP qubit gate that would preserve the energy of both bosonic states can be performed through the auxiliary qubit itself. To put it another way, one could use the auxiliary spin to store the logical information of a bosonic state. Then, using the beamsplitter interaction we would implement a $\mathtt{SWAP}$ operation which would exchange both GKP qubits. This way the controlled NOT and phase gates would require at most two beamsplitters which can be seen as a much better construction than the previously proposed ones. {\color{\mycolor}This solutions is proposed in Section~\ref{sec:qutrit_scheme}.}

\subsec{Performance of the approximate finite-energy gates}{sec:FE_gate_performance}

We now evaluate the performance of the Trotter approximations of $\CZ_{\Delta,\kappa}$ given in Eq.~\eqref{eq:trotter_cz_gate} using as before the three input states $\ket{00}_\Delta$, $\ket{++}_\Delta$ and $\ket{\Phi_{+}}_\Delta$. In order to limit the number of two-modes operations we fix $\delta t\sqrt{\Gamma}=1$ and are thus compare the channels $\cl{E}_k(\rho) = \n{Tr}_\n{S}\Big( \h{U}_k\,\big(\rho\otimes\!\ket{0}\!\!\bra{0}\!\big)\,\h{U}_k^\dagger \Big)$ with $\rho=\ket{\psi}_\Delta\!\prescript{}{\Delta\!}{\bra{\psi}}$ and $\h{U}_k$ defined as
\be \label{eq:FE_CZ_trotter_schemes}
\begin{split}
    \textnormal{1st order Trotter:} \qquad 
    &\h{U}_1 = e^{i\q_1\q_2\s_x }\,\, e^{-i\kappa^2\q_1\p_2\s_y} \,\, e^{-i\kappa^2\p_1\q_2\s_y}\,, \\
    \textnormal{1st order Trotter modified:} \qquad 
    &{\color{\mycolor} \h{U}_2 =  e^{-i\frac{1}{2}\kappa^2\q_1\p_2\s_y} \,\, e^{i\q_1\q_2\s_x }\,\, e^{-i\frac{1}{2}\kappa^2\p_1\q_2\s_y} } \,, \\
    \textnormal{1st order Trotter (x2):} \qquad 
    &\h{U}_3 = \Big[ e^{i\frac{1}{2} \q_1\q_2\s_x }\,\, e^{-i\frac{1}{2}\kappa^2\q_1\p_2\s_y} \,\, e^{-i\frac{1}{2}\kappa^2\p_1\q_2\s_y}\Big]^2\,, \\
    \textnormal{1st order Trotter modified (x2):} \qquad 
    &\h{U}_4 =  \Big[e^{-i\frac{1}{4}\kappa^2\q_1\p_2\s_y} \,\, e^{i\frac{1}{2}\q_1\q_2\s_x }\,\, e^{-i\frac{1}{4}\kappa^2\p_1\q_2\s_y}\Big]^2\,, \\
    \textnormal{2nd order Trotter:} \qquad
    &\h{U}_5 = 
    e^{-i\frac{1}{2} \kappa^2\q_1\p_2\s_y} \,\,  e^{-i\frac{1}{2} \kappa^2 \p_1\q_2\s_y}\,\,
    e^{i\,\q_1\q_2\s_x }\,\,
    e^{-i\frac{1}{2} \kappa^2\q_1\p_2\s_y} \,\,  e^{-i\frac{1}{2} \kappa^2 \p_1\q_2\s_y}\,.
\end{split}
\ee

\begin{figure*}[t!]
    \centering
    {\phantomsubcaption\label{fig:FE_CZ_fidelity_00}}
    {\phantomsubcaption\label{fig:FE_CZ_fidelity_pp}}
    {\phantomsubcaption\label{fig:FE_CZ_fidelity_Phi0}}
    \includegraphics{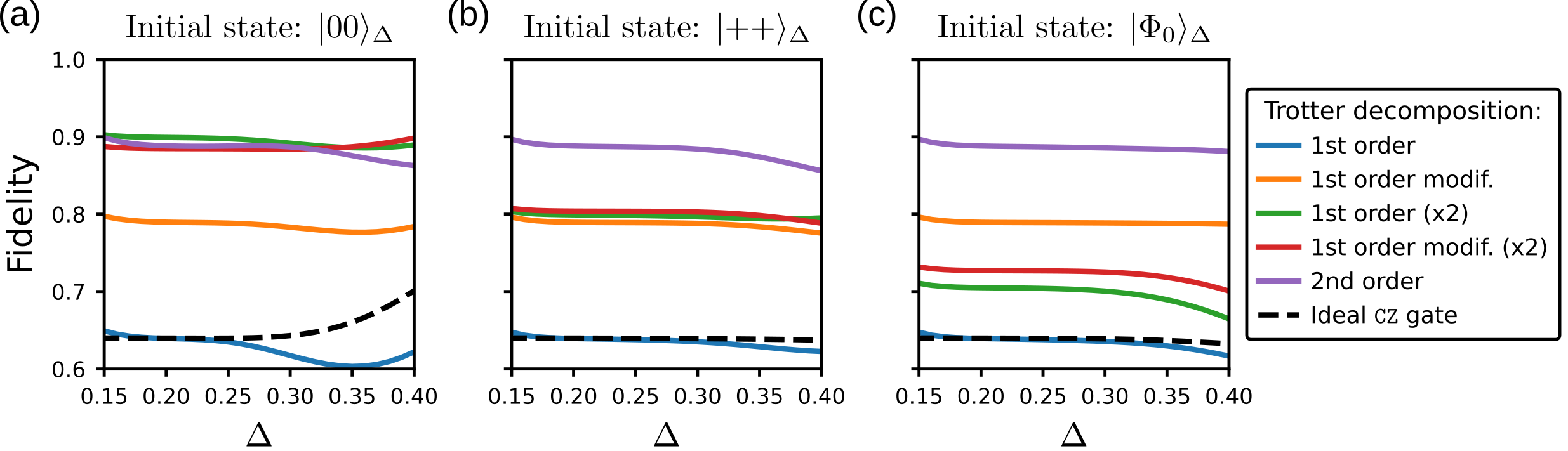}
    \caption{Performance of different approximations of the finite-energy $\CZ_{\Delta,\kappa}$ gate. The five approximations correspond to different Trotter decomposition listed in Eq.~\eqref{eq:FE_CZ_trotter_schemes}. Their performance is evaluated using the overlap fidelity between the output state and the desired one and compared to the fidelity of the ideal $\CZ$ gate (dashed). (a), (b) and (c) represent this quantity for three different input states.}
    \label{fig:FE_CZ_fidelity}
\end{figure*}

\noindent
In addition to the 1st and 2nd order Trotter decompositions, $\h{U}_1$ and $\h{U}_5$, we added here a modified 1st order as well as two-step 1st order approximations. We justify $\h{U}_2$ by the fact that the unitaries with $\q_1\p_2\s_y$ and $\p_1\q_2\s_y$ terms correspond to some perturbation of the main $\CZ$ interaction; thus, placing them symmetrically with respect to the $\q_1\q_2\s_x$ term mimics a 2nd order Trotter formula that then reduces the approximation error. 

The performance of these channels have been measured in term of the physical overlap fidelity between the output and desired states. The simulation results are presented in Fig.~\ref{fig:FE_CZ_fidelity}. We observe that the bare 1st order formula performs identically to the ideal $\CZ$ gate, whereas $\h{U}_2$ on the other hand show for the three states an improvement of ${\sim5\%}$. The two two-step Trotter approximations improve the fidelity even further, but their performance depends heavily on the input state. Finally, the fidelity of the 2nd order unitary saturate at $\sim89\%$. We believe that the enhanced performances of $\h{U}_2$ and $\h{U}_5$ compared to $\h{U}_{1}$ and $\h{U}_{3/4}$, respectively, is the result of some interference of approximation errors that have been recently studied in Ref.~\cite{layden_first-order_2022}.

Although these approximations of the desired $\CZ_{\Delta,\kappa}$ do not reach unit fidelity, we conceive that they remove from the system a substantial part of undesired entanglement and that their infidelity comes from approximation errors which are local. This hypothesis is supported by observations from Ref.~\cite{shaw_stabilizer_2024} where the authors argue that only a single ``error correction on a modified patch'' is needed for eliminating the unwanted entanglement. We therefore believe that the fidelity after any approximate $\CZ_{\Delta,\kappa}$ described in Eq.~\eqref{eq:FE_CZ_trotter_schemes} can recover a unit value using local error correction procedures. We leave this demonstration for future works.

After publishing, we became aware of ongoing work that generalizes the construction of hybrid GKP-qubit gates using quantum signal processing techniques. An initial presentation of this work can be found in Ref.~\cite{singh2023composite}.

{ \color{\mycolor}
\iftoggle{arXiv}{ \PRLsec{Logical entanglement mediated by an auxiliary three-level system}{sec:qutrit_scheme}} 
{ \section{Logical entanglement mediated by an auxiliary three-level system} \label{sec:qutrit_scheme} }

\begin{figure}[t]
    \centering
    {\phantomsubcaption\label{fig:qutrit_scheme}}
    {\phantomsubcaption\label{fig:alternative_grid_state_wigner}}
    {\phantomsubcaption\label{fig:qutrit_infidelity}}
    \includegraphics{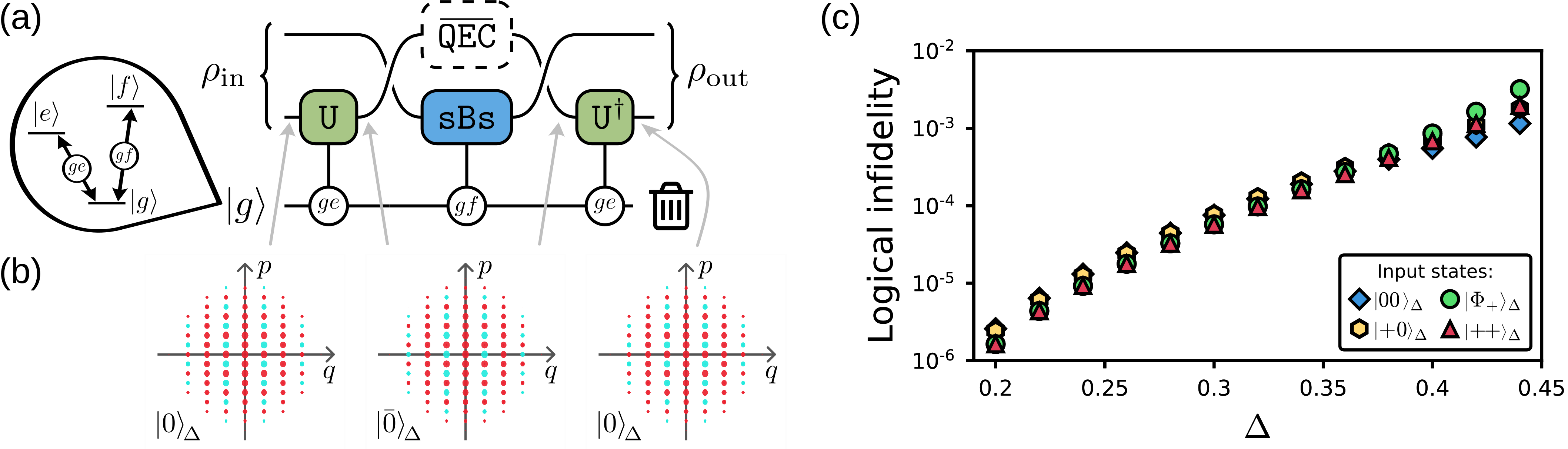}
    \caption{Two-qubit gate mediated by an auxiliary three-level system. (a) Circuit representation of the protocol. The system is initialized in a state $\rho_\n{in}$. The auxiliary system starting in $\ket{g}$ is used to retrieve the logical information of one of the oscillator using $\mathtt{CU}$ after what the two oscillators are swapped using a 50:50 beamsplitter. $\mathtt{sBs}$ represents a finite-energy stabilization which also effects a Pauli $Z$ operation if the qutrit is in $\ket{g}$. The circuit is then reversed in order to disentangle the oscillators from the auxiliary system. (b) Example of Wigner quasiprobability of the lower oscillator state before, between and after the unitaries $\mathtt{CU}$ and $\mathtt{CU}^\dagger$. The state in between is an eigenstate of the stabilizers $(-\hat{S}_{x,\Delta},\hat{S}_{z,\Delta})$. (c) The logical infidelity evaluated in a similar fashion as in Fig.~2(c).}
    \label{fig:qutrit_protocol_and_result}
\end{figure}

As mentioned earlier, it is possible to achieve a finite-energy version of an entangling gate between two GKP states by employing an auxiliary system to facilitate logical information exchange between oscillators. In the subsequent discussion, we will show that all the necessary components for executing this protocol are readily accessible and have been recently demonstrated. Nevertheless, the protocol presents some non-trivial insights.

To accomplish the information transfer, we require an auxillary three-level system (composed of levels $\ket{g}$, $\ket{e}$ and $\ket{f}$) and three operations: a conditional unitary suggested in Refs.~\cite{de_neeve_error_2022,hastrup_improved_2021} for high-fidelity readout of finite-energy GKP states, a standard 50:50 beamsplitter operation, and the small-Big-small quantum error correction scheme proposed in Refs.~\cite{de_neeve_error_2022,royer_stabilization_2020}. The entire protocol is illustrated in Fig.~\ref{fig:qutrit_scheme}. 

The first unitary operation~\cite{de_neeve_error_2022,hastrup_improved_2021}
\be \label{eq:GKP_readout}
    \mathtt{CU}=e^{-i \frac{1}{2} \sqrt{\pi} \q \s_y^{(ge)}}\,e^{-i \frac{1}{2} \sqrt{\pi} \Delta^2 \p \s_x^{(ge)}}
\ee
effectively realizes a control NOT gate between a logical qubit encoded in a GKP state (control) and a bare two-level system (target) when the latter is prepared in its ground state. Operators $\s_y^{(ge)}$ and $\s_x^{(ge)}$ are the standard Pauli operators defined using the states $\ket{g}$ and $\ket{e}$. This scheme has first been demostrated in \citet{de_neeve_error_2022} for high-fidelity readout of finite-energy GKP states in the logical $Z$ basis. The readout in the logical $X$ basis is performed similarly after replacing $(\q,\p)\rightarrow(-\p,\q)$. One can note the similarity of this operation with the first two unitaries of $\h{U}_\n{sBs}$ given in Eq.~\eqref{eq:trotter_stabilization_gates}. Indeed, in the dissipative stabilization of GKP states the unitaries perfom first a readout of the finite-energy stabilizers (as opposed to logical Pauli $Z$ and $X$) before a coherent controlled feedback on the oscillator. A specificity of this circuit, which has not been highlighted in prior works, is that, although being in a grid state, the oscillator state is now a joint eigenstate of the stabilizers $(-\hat{S}_{x,\Delta},\hat{S}_{z,\Delta})$. In other words, the action of $\mathtt{CU}$ can be summarized by
\[
    \mathtt{CU}\,\big[\alpha \ket{0}_\Delta + \beta \ket{1}_\Delta\big]\ket{g} = 
    \alpha \ket{\overbar{0}}_\Delta \ket{g} + \beta \ket{\overbar{1}}_\Delta \ket{e}
    \qquad \text{where} \qquad
    \forall \ket{\overbar{\psi}}_\Delta, \quad 
    -\hat{S}_{x,\Delta}\ket{\overbar{\psi}}_\Delta = \hat{S}_{z,\Delta}\ket{\overbar{\psi}}_\Delta = \ket{\overbar{\psi}}_\Delta
\]
In the Wigner quasiprobability picture these states differ from the standard ones by the presence of negative peaks at ${q=0}$ and ${p=(2k+1)\sqrt{\pi},}\,\,k\in\mathbb{Z}$ (see Fig.~\ref{fig:alternative_grid_state_wigner} for an example of $\ket{\overbar{0}}_\Delta$). These alternative grid states can fortunately be stabilized in a similar way (cf. Fig.~\ref{fig:circuits_QEC_schemes}) as the standard ones up to some phase factors. As an example the equivalent of the small-Big-small operator $\h{U}_\n{sBs}$ for states $\ket{\overbar{\psi}}_\Delta$ is 
\be \label{eq:sBs_for_alt_grid_states}
    \overbar{\mathtt{sBs}}_x = 
    e^{i\frac{1}{2}\sqrt{\pi} \Delta^2 \,\q\,\s_y }\,\,
    e^{i\sqrt{\pi} \,\p\,\s_x }\,\,
    e^{-i\frac{1}{2}\sqrt{\pi} \Delta^2 \,\q\,\s_y }\,
    \qquad \text{and} \qquad
    \overbar{\mathtt{sBs}}_z = 
    e^{i\frac{1}{2}\sqrt{\pi} \Delta^2 \,\p\,\s_y }\,\,
    e^{i\sqrt{\pi} \,\q\,\s_x }\,\,
    e^{i\frac{1}{2}\sqrt{\pi} \Delta^2 \,\p\,\s_y }\,
\ee
(for compactness we will in this section refer to $\h{U}_\n{sBs}$ as $\mathtt{sBs}_x$ and $\mathtt{sBs}_z$ for the stabilization of $q$ and $p$ quadratures, respectively). Note that here $\overbar{\mathtt{sBs}}_z\equiv\mathtt{sBs}_z$ since both the standard and alternative grid states are $+1$ eigenstates of the stabilizer $\hat{S}_{z,\Delta}$. Thus, after applying $\mathtt{CU}$ the oscillator state can be error corrected independently of the state of the qutrit using an additional auxiliary qubit. We will refer to this error correcting process as $\overbar{\mathtt{QEC}}$.

The next operation necessary for our approach is two 50:50 interferometer, described in terms of the symmetric beamsplitter interaction (see Eq.~\eqref{eq:beamsplitter_operators}) as $\hat{B}_S(\pi/2)$. This operation conserves energy inherently and is analogous to exchanging states between the two oscillators. Consequently, we will refer to this operator as $\mathtt{SWAP}$.

Finally, as mentioned above, the last essential operation we need is the small-Big-small unitary, chosen for its ability to deterministically apply a logical Pauli operator to the oscillator state (see Section~\ref{sec:error_correction}). Specifically, for realizing a $\CZ$ gate, we utilize the operation 
\be \label{eq:sBs_qutrit}
    \mathtt{sBs}_z = 
    e^{-i\frac{1}{2}\sqrt{\pi} \Delta^2 \,\p\,\s_y^{(gf)} }\,\,
    e^{i\sqrt{\pi} \,\q\,\s_x^{(gf)} }\,\,
    e^{-i\frac{1}{2}\sqrt{\pi} \Delta^2 \,\p\,\s_y^{(gf)} }\,
\ee
where the operators $\s_y^{(gf)}$ and $\s_x^{(gf)}$ are the standard Pauli operators defined using the states $\ket{g}$ and $\ket{f}$. We utilize $\mathtt{sBs}_z$ as it implements a logical Pauli $Z$ allowing us to introduce a phase to the oscillator state when the auxiliary system is in $\ket{g}$. Alternatively, if one aims to implement a $\CNOT$, the necessary operation would be $\mathtt{sBs}_x$ (using the same qutrit states), as it applies a logical Pauli $X$. After such small-Big-small operations the state of the auxilliary system is almost entirely in $\ket{g}$ when the oscillator state starts in the codespace. If the latter starts in an erroneous state, the auxilliary system will have some population in $\ket{f}$ that one can reset using a non-unitary process realizing $\ket{f}\mapsto\ket{g}$.

With the three operations outlined above, we are now equipped to demonstrate the protocol's mechanism. Let us consider the two bosonic systems to start in arbitrary GKP states $\ket{\psi}_\Delta$ and $\alpha\ket{0}_\Delta+\beta\ket{1}_\Delta$ with $\alpha^2+\beta^2=1$, respectively. Moreover, we consider the three-level system to be coupled exclusively to the second oscillator which is characteristic to hybrid discrete-continuous variable systems~\cite{andersen_hybrid_2015}. We initialize this auxiliary system in $\ket{g}$ and perform the protocol described in Fig.~\ref{fig:qutrit_scheme}. Analytically the state of the entire system evolves as
\[
\begin{split}
    \ket{\psi}_\Delta \Big[\alpha\ket{0}_\Delta+\beta\ket{1}_\Delta\Big]\ket{g}
    &\xmapsto[\qquad\qquad]{\mathtt{CU}} 
    \ket{\psi}_\Delta \Big[\alpha \ket{\overbar{0}}_\Delta \ket{g} + \beta \ket{\overbar{1}}_\Delta \ket{e}\Big] \\[3pt]
    &\xmapsto[\qquad\qquad]{\mathtt{SWAP}}
    \alpha \ket{\overbar{0}}_\Delta \ket{\psi}_\Delta \ket{g} + \beta \ket{\overbar{1}}_\Delta \ket{\psi}_\Delta \ket{e} \\
    &\xmapsto[\qquad\qquad]{\mathtt{sBs}_z}
    \alpha \ket{\overbar{0}}_\Delta \left( \h{Z}_\Delta \ket{\psi}_\Delta \right) \ket{g} + \beta \ket{\overbar{1}}_\Delta \ket{\psi}_\Delta \ket{e}\\
    &\xmapsto[\qquad\qquad]{\mathtt{SWAP}}
    \alpha \left( \h{Z}_\Delta \ket{\psi}_\Delta \right) \ket{\overbar{0}}_\Delta \ket{g} + \beta \ket{\psi}_\Delta \ket{\overbar{1}}_\Delta \ket{e} \\
    &\xmapsto[\qquad\qquad]{\mathtt{CU}^\dagger}
    \Big[\alpha \left( \h{Z}_\Delta \ket{\psi}_\Delta \right) \ket{0}_\Delta + \beta \ket{\psi}_\Delta \ket{1}_\Delta \Big] \ket{g} 
\end{split}
\]
where $\h{Z}_\Delta$ represents a logical finite-energy Pauli $Z$ operation. It is worth noting that the operation achieved corresponds to a $\CZ$ gate up to a $\hat{Z}_\Delta$ transformation applied to the first grid state. This transformation can be monitored through Pauli frame tracking or actively implemented using another small-Big-small operation with an additional auxiliary qubit. The last step of the protocol consists of a complete reset of the three-level system as it is no longer entangled with the bosonic systems. 

We evaluate the fidelity of the protocol in a similar way as the fidelity of the error-corrected gate GKP presented in Section~\ref{subsec:stabilized_cz}. Specifically, we prepare a two-qubit stabilized GKP state, execute the proposed protocol, and then determine the fidelity of the resulting output state with the desired stabilized state. The resulting infidelity is depicted in Fig.~\ref{fig:qutrit_infidelity}. Notably, we observe that  the protocol's performance surpasses in average that of both the error-corrected two-qubit gate and the finite-energy gate detailed in Section~\ref{sec:FE_gates}. This high performance stems from the entanglement propagation through a discrete-variable system, which prevents undesired entanglement between different parts of the bosonic systems. As seen previously, the efficacy of entanglement propagation is facilitated by $\mathtt{CU}$, which transfers logical information from the oscillator to a two-level system with high fidelity. By employing $\overbar{\mathtt{QEC}}$ to correct the state of the first oscillator between the two $\mathtt{SWAP}$ operations, the infidelity of the protocol can be further reduced. This correction addresses approximation errors inherent in the $\mathtt{CU}$ unitary while also providing protection against external disturbances during this idle period.

}

\iftoggle{arXiv}{ \PRLsec{Numerical methods}{sec:num_method}} 
{ \section{Numerical methods} \label{sec:num_method} }

In this last section we briefly discuss the numerical methods used in this work for simulating GKP states and the operations involving them. All the results were obtained using a package for continuous variable systems simulation \textsf{CVsim.jl}~\cite{cvsim_rojkov_2024} that we wrote in the programming language Julia~\cite{bezanson2017julia} and which uses principally the \textsf{QuantumOptics.jl} package~\cite{kramer2018quantumoptics}. Our simulations differ from the usual ones by the choice of basis used to perform the simulations. Indeed, instead of using the Fock basis (i.e. eigenstates of $\h{a}^\dagger\h{a}$ operator) to represent the states and operators, \textsf{CVsim.jl} uses the position or momentum bases. In other words, we use here the first quantization of quantum mechanics instead of the second one. There exist several advantages for this approach: First, some continuous variable states such as GKP ones require a large number of Fock states in order to be accurately represented which consequently increases the complexity of the calculations with them (especially if one treats two-qubit systems). Second, some of the desired bosonic operators such as the $\CZ$ gate or some conditional operators $C\h{D}$ are diagonal in the $q$ or $p$ basis. Lastly, one can easily transition between the position and momentum representations of a state or operator using discrete Fourier transforms for which efficient numerical algorithms are known. Thus, the marginal distributions $P(q)$ and $P(p)$ for example are easily accessible as they correspond simply to the diagonal of the density matrix in the position and momentum basis, respectively.
\end{SuppMat}}{}

\end{document}